\documentclass[journal]{IEEEtran}
\usepackage[utf8]{inputenc}
\usepackage[caption=false,font=normalsize,labelfont=sf,textfont=sf]{subfig}
\usepackage{rotating,graphicx}
\graphicspath{{./fig/}}
\usepackage{array,multirow}
\usepackage{wrapfig}
\usepackage{amsmath,amsfonts}
\usepackage{lipsum}
\usepackage{mathtools}
\usepackage{float}
\usepackage{xcolor}
\usepackage{enumitem}
\usepackage{csquotes}
\usepackage{tabularx}
\usepackage{colortbl,booktabs}

\usepackage{cite}

\usepackage[%
    acronym=true,
    nonumberlist,%
    nopostdot,%
    seeautonumberlist,%
    shortcuts,%
    section=chapter,%
    toc=false,%
    nogroupskip=true,%
    prefix,
]{glossaries-extra}
\setabbreviationstyle[acronym]{long-short}
\newacronym{3GPP}{3GPP}{3rd Generation Partnership Project}
\newacronym{4G}{4G}{fourth-generation}
\newacronym{5G}{5G}{fifth-generation}
\newacronym{6G}{6G}{sixth-generation}
\newacronym{AI}{AI}{artificial intelligence}
\newacronym{API}{API}{application programming interface}
\newacronym{BPMN}{BPMN}{business process management notation}
\newacronym{BWP}{BWP}{bandwidth partitioning}
\newacronym{CD}{CD}{continuous deployment}
\newacronym{CI}{CI}{continuous integration}
\newacronym{CM}{CM}{continuous monitoring}
\newacronym{CT}{CT}{continuous testing}
\newacronym{CL}{CL}{control loop}
\newacronym{CNF}{CNF}{containerized network function}
\newacronym{CSI}{CSI}{communication service instance}
\newacronym{CSMF}{CSMF}{communication service management function}
\newacronym{CST}{CST}{communication service template}
\newacronym{NST}{NST}{network slice template}
\newacronym{NSST}{NSST}{network slice subnet template}
\newacronym{E2E}{E2E}{end-to-end}
\newacronym{eMBB}{eMBB}{enhanced mobile broadband}
\newacronym{IL}{IL}{infrastructure layer}
\newacronym{KPI}{KPI}{key performance indicator}
\newacronym{MANO}{MANO}{management and orchestration}
\newacronym{ML}{ML}{machine learning}
\newacronym{MNO}{MNO}{mobile network operator}
\newacronym{NF}{NF}{network function}
\newacronym{NFL}{NFL}{network function layer}
\newacronym{NFMF}{NFMF}{network function management function}
\newacronym{NFVO}{NFVO}{network function virtualization orchestrator}
\newacronym{NPN}{NPN}{non-public network}
\newacronym{VNFM}{VNFM}{virtual network functions manager}
\newacronym{RO}{RO}{resource orchestration}
\newacronym{SL}{SL}{service layer}
\newacronym{NSO}{NSO}{network service orchestration}
\newacronym{NFO}{NFO}{network function orchestration}
\newacronym{NFVI}{NFVI}{NFV infrastructure}
\newacronym{VIM}{VIM}{virtualized infrastructure manager}
\newacronym{NSI}{NSI}{network slice instance}
\newacronym{NSSI}{NSSI}{network slice subnet instance}
\newacronym{SLA}{SLA}{service level agreement}
\newacronym{QoS}{QoS}{quality of service}
\newacronym{QoE}{QoE}{quality of experience}
\newacronym{UPF}{UPF}{user plane function}
\newacronym{WP}{WP}{work package}
\newacronym{NSMF}{NSMF}{network slice management function}
\newacronym{NSSMF}{NSSMF}{network slice subnet management function}
\newacronym{URLLC}{URLLC}{ultra reliable low latency communication}
\newacronym{mMTC}{mMTC}{massive machine type communication}
\newacronym{MIoT}{MIoT}{massive IoT}
\newacronym{IoT}{IoT}{internet of things}
\newacronym{V2X}{V2X}{vehicle to everything}
\newacronym{O-RAN}{O-RAN}{Open RAN}
\newacronym{RAN}{RAN}{radio access network}
\newacronym{IQ}{IQ}{in-phase and quadrature}
\newacronym{SON}{SON}{self-organizing network}
\newacronym{RIC}{RIC}{RAN intelligent controller}
\newacronym{NR}{NR}{new radio}
\newacronym{PRB}{PRB}{physical resource block}
\newacronym{NFV}{NFV}{network function virtualization}
\newacronym{SDN}{SDN}{software-defined networking}
\newacronym{O-DU}{O-DU}{O-RAN DU}
\newacronym{DU}{DU}{distributed unit}
\newacronym{vO-DU}{vO-DU}{virtualized O-DU}
\newacronym{vO-CU}{vO-CU}{virtualize O-CU}
\newacronym{O-RU}{O-RU}{O-RAN RU}
\newacronym{FB}{FB}{functional block}
\newacronym{O-CU}{O-CU}{O-RAN CU}
\newacronym{O-CU-CP}{O-CU-CP}{O-CU control plane}
\newacronym{O-CU-UP}{O-CU-UP}{O-CU user plane}
\newacronym{CU-UP}{CU-UP}{CU user plane}
\newacronym{CU-CP}{CU-CP}{CU control plane}
\newacronym{Non-RT RIC}{Non-RT RIC}{non-real-time RIC}
\newacronym{Near-RT RIC}{Near-RT RIC}{near-real-time RIC}
\newacronym{CN}{CN}{core network}
\newacronym{CP}{CP}{control plane}
\newacronym{ISG}{ISG}{Industry Specification Group}
\newacronym{UP}{UP}{user plane}
\newacronym{MP}{MP}{management plane}
\newacronym{RRM}{RRM}{radio resource management}
\newacronym{PM}{PM}{performance metric}
\newacronym{SMO}{SMO}{service management and orchestration}
\newacronym{O-Cloud}{O-Cloud}{O-RAN cloud platform}
\newacronym{MAC}{MAC}{medium access control}
\newacronym{VNF}{VNF}{virtualized network function}
\newacronym{PNF}{PNF}{physical network function}
\newacronym{OOF}{OOF}{ONAP optimization framework}
\newacronym{SO}{SO}{service orchestrator}
\newacronym{UUI}{UUI}{use case user interface}
\newacronym{AAI}{AAI}{active and available inventory}
\newacronym{SP}{SP}{slice profile}
\newacronym{FH}{FH}{fronthaul}
\newacronym{MH}{MH}{midhaul}
\newacronym{BH}{BH}{backhaul}
\newacronym{TN}{TN}{transport network}
\newacronym{MD}{MD}{management domain}
\newacronym{NSSAI}{NSSAI}{network slice selection assistance information}
\newacronym{S-NSSAI}{S-NSSAI}{single NSSAI}
\newacronym{SST}{SST}{slice/service type}
\newacronym{SD}{SD}{slice differentiator}
\newacronym{UE}{UE}{user equipment}
\newacronym{RAB}{RAB}{radio access bearer}
\newacronym{SLS}{SLS}{service level specification}
\newacronym{O-NSSI}{O-NSSI}{O-RAN NSSI}
\newacronym{MEC}{MEC}{multi-access edge computing}
\newacronym{COTS}{COTS}{commercial off-the-shelf}
\newacronym{OPEX}{OPEX}{operational expenditure}
\newacronym{CAPEX}{CAPEX}{capital expenditure}
\newacronym{OSC}{OSC}{O-RAN software community}
\newacronym{LF}{LF}{Linux Foundation}
\newacronym{RRC}{RRC}{radio resource control}
\newacronym{SDAP}{SDAP}{service data adaptation protocol}
\newacronym{PDCP}{PDCP}{packet data convergence protocol}
\newacronym{PHY}{PHY}{physical layer}
\newacronym{RLC}{RLC}{radio link control}
\newacronym{xApp}{xApp}{Near-RT RIC application}
\newacronym{rApp}{rApp}{Non-RT RIC application}
\newacronym{RF}{RF}{radio frequency}
\newacronym{O-FH}{O-FH}{open fronthaul}
\newacronym{TCO}{TCO}{total cost of ownership}
\newacronym{TSG}{TSG}{Technical Specification Group}
\newacronym{WG}{WG}{working group}
\newacronym{LTE}{LTE}{long term evolution}
\newacronym{TIP}{TIP}{Telecom Infra Project}
\newacronym{ONF}{ONF}{open networking foundation}
\newacronym{SD-RAN}{SD-RAN}{software-defined RAN}
\newacronym{D-RAN}{D-RAN}{distributed RAN}
\newacronym{NSaaS}{NSaaS}{network slice as a service}
\newacronym{OSS}{OSS}{operations support system}
\newacronym{BSS}{BSS}{business support system}
\newacronym{VM}{VM}{virtual machine}
\newacronym{ETSI}{ETSI}{European Telecommunications Standards Institute}
\newacronym{OS}{OS}{operating system}
\newacronym{SDO}{SDO}{standards development organization}
\newacronym{OSM}{OSM}{open source MANO}
\newacronym{VPN}{VPN}{virtual private network}
\newacronym{NRP}{NRP}{network resource partition}
\newacronym{TNE}{TNE}{transport network equipment}
\newacronym{O-gNB}{O-gNB}{O-RAN gNodeB}
\newacronym{gNB}{gNB}{next generation NodeB}
\newacronym{eNB}{eNB}{evolved NodeB}
\newacronym{O-eNB}{O-eNB}{O-RAN eNB}
\newacronym{ng-eNB}{ng-eNB}{next generation eNB}
\newacronym{UMTS}{UMTS}{universal mobile telecommunications system}
\newacronym{eCPRI}{eCPRI}{evolved CPRI}
\newacronym{CPRI}{CPRI}{common public radio interface}
\newacronym{5GC}{5GC}{5G core}
\newacronym{EN-DC}{EN-DC}{E-UTRAN NR dual connectivity}
\newacronym{PLMN}{PLMN}{public land mobile network}
\newacronym{AMF}{AMF}{access and mobility management function}
\newacronym{CCSDK}{CCSDK}{common controller software development kit}
\newacronym{CPS}{CPS}{configuration persistence service}
\newacronym{DCAE}{DCAE}{data collection, analysis and event}
\newacronym{CDS}{CDS}{controller design studio}
\newacronym{SA5}{SA5}{service and system aspects 5}
\newacronym{SDC}{SDC}{service design and creation}
\newacronym{SDN-C}{SDN-C}{SDN controller}
\newacronym{SDN-R}{SDN-R}{SDN radio}
\newacronym{CaaS}{CaaS}{container-as-a-service}
\newacronym{IT}{IT}{information technology}
\newacronym{ZSM}{ZSM}{zero touch network and service management}
\newacronym{CIS}{CIS}{container infrastructure service}
\newacronym{CISM}{CISM}{CIS management}
\newacronym{VNFC}{VNFC}{virtual network function component}
\newacronym{CCM}{CCM}{CIS cluster management}
\newacronym{CIR}{CIR}{container image registry}
\newacronym{EM}{EM}{element management}
\newacronym{IMS}{IMS}{infrastructure management service}
\newacronym{DMS}{DMS}{deployment management service}
\newacronym{WIM}{WIM}{WAN infrastructure manager}
\newacronym{CLAMP}{CLAMP}{closed loop automation management platform}
\newacronym{GUI}{GUI}{graphical user interface}
\newacronym{PoP}{PoP}{point-of-presence}
\newacronym{MSCS}{MSCS}{multi-site connectivity service}
\newacronym{SLO}{SLO}{service level objective}
\newacronym{SDP}{SDP}{service demarcation point}
\newacronym{NSC}{NSC}{network slice controller}
\newacronym{NBI}{NBI}{north bound interface}
\newacronym{SBA}{SBA}{services based architecture}
\newacronym{SBMA}{SBMA}{services based management architecture}
\newacronym{ONAP}{ONAP}{open network automation platform}
\newacronym{FCAPS}{FCAPS}{fault, configuration, accounting, performance, security}
\newacronym{FDD}{FDD}{frequency division duplexing}
\newacronym{TDD}{TDD}{time division duplexing}
\newacronym{BBU}{BBU}{baseband unit}
\newacronym{MP-BGP}{MP-BGP}{multi-protocol border gateway protocol}
\newacronym{MPLS}{MPLS}{multi-protocol label switching}
\newacronym{L3VPN}{L3VPN}{layer 3 VPN}
\newacronym{EVPN}{EVPN}{ethernet VPN}
\newacronym{IP}{IP}{internet protocol}
\newacronym{WDM}{WDM}{wavelength division multiplexing}
\newacronym{SRv6}{SRv6}{SR over IPv6}
\newacronym{SR}{SR}{segment routing}
\newacronym{IGP}{IGP}{interior gateway protocol}
\newacronym{BGP}{BGP}{border gateway protocol}
\newacronym{SID}{SID}{segment ID}
\newacronym{DC}{DC}{data center}
\newacronym{PE}{PE}{provider edge}
\newacronym{RoE}{RoE}{radio over ethernet}
\newacronym{VPWS}{VPWS}{virtual private wire service}
\newacronym{VLAN}{VLAN}{virtual LAN}
\newacronym{NB-IoT}{NB-IoT}{narrowband IoT}
\newacronym{TE}{TE}{traffic engineering}
\newacronym{5QI}{5QI}{5G QoS identifier}
\newacronym{DSCP}{DSCP}{differentiated services code point}
\newacronym{GTP-U}{GTP-U}{GPRS tunneling protocol user plane}
\newacronym{FIM}{FIM}{federated information model}
\newacronym{NRM}{NRM}{network resource model}
\newacronym{IETF}{IETF}{internet engineering task force}
\newacronym{MnS}{MnS}{management service}
\newacronym[
  longplural={information object classes}]
  {IOC}{IOC}{information object class}
\newacronym{CE}{CE}{customer edge}
\newacronym{SLE}{SLE}{service level expectation}
\newacronym{PDU}{PDU}{packet data unit}
\newacronym{SR-IOV}{SR-IOV}{single root I/O virtualization}
\newacronym{TNO}{TNO}{transport network orchestrator}
\newacronym{ETN}{ETN}{edge TN node}
\newacronym{AC}{AC}{attachment circuit}
\newacronym{CRUD}{CRUD}{create, read, update, and delete}
\newacronym{C-RAN}{C-RAN}{centralized RAN}
\newacronym{vRAN}{vRAN}{virtualized RAN}
\newacronym{CU}{CU}{centralized unit}
\newacronym{RU}{RU}{radio unit}
\newacronym{SCF}{SCF}{small cell forum}
\newacronym{DARTs}{DARTs}{disaggregated RAN transport study}
\newacronym{FAPI}{FAPI}{functional API}
\newacronym{nFAPI}{nFAPI}{network FAPI}
\newacronym{NG-RAN}{NG-RAN}{next-generation RAN}
\newacronym[
  longplural={radio access technologies}]
  {RAT}{RAT}{radio access technology}
\newacronym{RRH}{RRH}{remote radio head}
\newacronym{DL}{DL}{deep learning}
\newacronym{HMTC}{HMTC}{high-performance machine type communication}
\newacronym{HLS}{HLS}{high layer split}
\newacronym{LLS}{LLS}{low layer split} 
\newacronym{OAM}{OAM}{operations and maintenance}
\newacronym{OTIC}{OTIC}{open testing and integration center}
\newacronym{TSC}{TSC}{technical steering committee}
\newacronym{FG}{FG}{focus group}
\newacronym{MVP-C}{MVP-C}{minimum viable plan - committee}
\newacronym{RIA}{RIA}{radio intelligence automation}
\newacronym{ROMA}{ROMA}{OpenRAN orchestration and management automation}
\newacronym{PG}{PG}{project group}
\newacronym{S-RU}{S-RU}{small cell RU}
\newacronym{OAI}{OAI}{OpenAirInterface}
\newacronym{M5G}{M5G}{MOSAIC 5G}
\newacronym{OSA}{OSA}{OpenAirInterface software alliance}
\newacronym{SRS}{SRS}{software radio system}
\newacronym{KPM}{KPM}{key performance measurement}
\newacronym{SDR}{SDR}{software-defined radio}
\newacronym{MF}{MF}{management function}
\newacronym{OOM}{OOM}{ONAP operations manager}
\newacronym{CPU}{CPU}{central processing unit}
\newacronym{RAM}{RAM}{random access memory}
\newacronym{NIC}{NIC}{network interface card}
\newacronym{BIOS}{BIOS}{basic input and output system}
\newacronym{BMC}{BMC}{baseboard management controller}
\newacronym{DES}{DES}{data exposure service}
\newacronym{DMaaP}{DMaaP}{data movement as a platform}
\newacronym{NFV-MANO}{NFV-MANO}{network functions virtualization management and orchestrierung}
\newacronym{NSSMS-P}{NSSMS\_P}{NSSMS provider}
\newacronym{NSSMS-C}{NSSMS\_C}{NSSMS cunsumer}
\newacronym{NFMS-P}{NFMS\_P}{NFMS provider}
\newacronym{MO}{M\&O}{management \& orchestration}
\newacronym{DWDM}{DWDM}{dense WDM}
\newacronym{FFT}{FFT}{fast fourier transform}
\newacronym{L1}{L1}{layer 1}
\newacronym{L2}{L2}{layer 2}
\newacronym{L3}{L3}{layer 3}
\newacronym{EU}{EU}{european union}
\newacronym{ARC}{ARC}{aerial RAN CoLab}
\newacronym{GPU}{GPU}{graphics processing unit}
\newacronym{EI}{EI}{enrichment information}
\newacronym{FPGA}{FPGA}{field programmable gate array}
\newacronym{FIR}{FIR}{finite impulse response}
\newacronym{POWDER}{POWDER}{platform for open wireless data-driven experimental research}
\newacronym{MIMO}{MIMO}{multiple-input multiple-output}
\newacronym{5G-PPP}{5G-PPP}{5G Public-Private Partnership}
\newacronym{FOCOM}{FOCOM}{federated O-Cloud orchestration and management}
\newacronym{RL}{RL}{reinforcement learning}
\newacronym{MARL}{MARL}{multi-agent RL}
\newacronym{LSTM}{LSTM}{long short-term memory}
\newacronym{FL}{FL}{federated learning}
\newacronym{CNN}{CNN}{convolutional neural network}
\newacronym{UI}{UI}{user interface}
\newacronym{RSRP}{RSRP}{reference signal received power}
\newacronym{SINR}{SINR}{signal-to-interference-plus-noise ratio}%
\renewcommand{\glossarysection}[2][]{}
\newglossarystyle{mytable}
{
    \setglossarystyle{super}%
}%
\renewcommand{\glossarypreamble}{%
 \glsfindwidesttoplevelname[\currentglossary]
 \begin{list}{}%
 {\setlength{\leftmargin}{-0.50em}\rightmargin\leftmargin}
 \item\relax}
\renewcommand{\glossarypostamble}{\end{list}}

\usepackage{orcidlink}
\usepackage{comment}
\usepackage{svg}
\usepackage{soul}

\usepackage{pifont}
\usepackage[most]{tcolorbox}

\hypersetup{hidelinks}
\newif\iftrackrivision
\trackrivisiontrue
\iftrackrivision

\else

\fi

\providecommand{\DenKrCommentsUsage}{0}%
\expandafter\ifstrequal\expandafter{\DenKrCommentsUsage}{0}{%
    \newcommand{\disablewr}[1]{}%
}{%
	\newcommand{\disablewr}[1]{#1}%
}%
\newcommand{\newcommanddisw}[3]{\newcommand{#1}[1]{\disablewr{{\color{#2}#3}}}}%
\definecolor{todocol}{named}{red}
\definecolor{notecol}{named}{gray}
\definecolor{denkrComColDenKr}{RGB}{145,20,145}%
\newcommanddisw{\dktodo}{todocol}{ToDo: #1}%
\newcommanddisw{\notice}{notecol}{Notice: #1}%
\newcommanddisw{\dekr}{denkrComColDenKr}{{\#}DenKr: #1}
\newcommanddisw{\khal}{blue}{{\#}KhAl: #1}
\newcommanddisw{\AsHa}{orange}{{\#}AsHa: #1}

\newcommand*\rot{\rotatebox{90}}

\IEEEoverridecommandlockouts

\makeatother
\usepackage{eso-pic}
\newcommand\AtPageUpperMyright[1]{\AtPageUpperLeft{
\put(\LenToUnit{0.08\paperwidth},\LenToUnit{-1cm}){
    \parbox{1\textwidth}{\fontsize{8}{11}\selectfont #1}}
}}
\newcommand{\conf}[1]{
\AddToShipoutPictureBG*{
\AtPageUpperMyright{#1}
}
}

\begin{document}

\title{%
A Comprehensive Tutorial and Survey of O-RAN: Exploring Slicing-aware Architecture, Deployment Options, Use Cases, and Challenges

\conf{This is the author's version of an article that has been accepted for publication in IEEE Communications Surveys \& Tutorials.\\The final version of record is available at: https://doi.org/10.1109/COMST.2025.3598406}

\thanks{This manuscript was received on 04 November $\mathrm{2024}$, revised on 15 July $\mathrm{2025}$, and accepted on 05 August $\mathrm{2025}$. It was published on DD MM $\mathrm{2025}$, and the current version was last updated on DD MM $\mathrm{2025}$.}
\thanks{Khurshid Alam\orcidlink{0009-0003-3344-8629}, Dennis Krummacker\orcidlink{0000-0001-9799-4870}, and Hans D. Schotten\orcidlink{0000-0001-5005-3635} are with the Department of Intelligent Networks (IN), German Research Center for Artificial Intelligence (DFKI), $\mathrm{67663}$ Kaiserslautern, Germany. 
Mohammad Asif Habibi\orcidlink{0000-0001-9874-0047}, Matthias Tammen\orcidlink{0009-0005-4599-8876}, and Hans D. Schotten\orcidlink{0000-0001-5005-3635} are with the Division of Wireless Communications and Radio Navigation (WiCoN), Department of Electrical and Computer Engineering (EIT), University of Kaiserslautern (RPTU), $\mathrm{67663}$ Kaiserslautern, Germany. 
Walid Saad\orcidlink{0000-0003-2247-2458} is affiliated with the Department of Electrical and Computer Engineering, Virginia Tech, 22203 Virginia, United States of America. 
Marco Di Renzo\orcidlink{0000-0003-0772-8793} is associated with Universit\'e Paris-Saclay, CNRS, CentraleSup\'elec, Laboratoire des Signaux et Syst\`emes, 3 Rue Joliot-Curie, 91192 Gif-sur-Yvette, France. 
Tommaso Melodia\orcidlink{0000-0002-2719-1789} is affiliated with the Institute for the Wireless Internet of Things, Northeastern University, Boston, 02115 Massachusetts, United States of America. 
Xavier Costa-P\'erez\orcidlink{0000-0002-9654-6109} is with the 6G Networks R\&D Department, NEC Laboratories Europe, $\mathrm{69115}$ Heidelberg, Germany. He is also associated with the Department of AI-Driven Systems of the i2CAT Research Center and the Department of Engineering Sciences of the Catalan Institution for Research and Advanced Studies (ICREA), $\mathrm{08034}$ Barcelona, Spain. 
M\'erouane Debbah\orcidlink{0000-0001-8941-8080} is with Khalifa University of Science and Technology, P. O. Box 127788, Abu Dhabi, United Arab Emirates. 
Ashutosh Dutta\orcidlink{0000-0002-6182-9395} is affiliated with the Applied Physics Laboratory, Johns Hopkins University, 20723 Maryland, United States of America.
The corresponding authors are Khurshid Alam\orcidlink{0009-0003-3344-8629} (khurshid.alam@dfki.de) and Mohammad Asif Habibi\orcidlink{0000-0001-9874-0047} (asif@eit.uni-kl.de).}
}

\author{Khurshid Alam\orcidlink{0009-0003-3344-8629}, Mohammad Asif Habibi\orcidlink{0000-0001-9874-0047}, Matthias Tammen\orcidlink{0009-0005-4599-8876}, Dennis Krummacker\orcidlink{0000-0001-9799-4870}, \\
Walid Saad\orcidlink{0000-0003-2247-2458}~\IEEEmembership{(Fellow,~IEEE)},
Marco Di Renzo\orcidlink{0000-0003-0772-8793}~\IEEEmembership{(Fellow,~IEEE)}, 
Tommaso Melodia\orcidlink{0000-0002-2719-1789}~\IEEEmembership{(Fellow,~IEEE)}, 
Xavier Costa-P\'erez\orcidlink{0000-0002-9654-6109}~\IEEEmembership{(Senior Member,~IEEE)}, M\'erouane Debbah\orcidlink{0000-0001-8941-8080}~\IEEEmembership{(Fellow,~IEEE)}, 
Ashutosh Dutta\orcidlink{0000-0002-6182-9395}~\IEEEmembership{(Fellow,~IEEE)},  
and Hans D. Schotten\orcidlink{0000-0001-5005-3635}~\IEEEmembership{(Member,~IEEE)}
\vspace{-8.5mm}
}

\maketitle

\vspace{-2.5mm}
\begin{abstract}
Open-radio access network (O-RAN) seeks to establish the principles of openness, programmability, automation, intelligence, and hardware-software disaggregation with interoperable and standard-compliant interfaces. It advocates for multi-vendorism and multi-stakeholderism within a cloudified and virtualized wireless infrastructure, aimed at enhancing the deployment, operation, and management of RAN architecture. These enhancements promise increased flexibility, performance optimization, service innovation, energy efficiency, and cost effectiveness across fifth-generation (5G), sixth-generation (6G), and beyond networks. A silent feature of O-RAN architecture is its support for network slicing, which entails interaction with other domains of the cellular network, notably the transport network (TN) and the core network (CN), to realize end-to-end (E2E) network slicing. The study of this feature requires exploring the stances and contributions of diverse standards development organizations (SDOs). In this context, we note that despite the ongoing industrial deployments and standardization efforts, the research and standardization communities have yet to comprehensively address network slicing in O-RAN. To address this gap, this paper provides a comprehensive exploration of network slicing in O-RAN through an in-depth review of specification documents from O-RAN Alliance and research papers from leading industry and academic institutions. The paper commences with an overview of the relevant standardization and open source contributions, subsequently delving into the latest O-RAN architecture with an emphasis on its slicing aspects. Furthermore, the paper explores O-RAN deployment scenarios, examining options for the deployment and orchestration of RAN and TN slice subnets. It also discusses the slicing of the underlying infrastructure and provides an overview of various use cases related to O-RAN slicing. Finally, it summarizes the potential research challenges identified throughout the study. 
\end{abstract}

\begin{keywords}
3GPP, 5G, 6G, Disaggregation, ETSI, Management and Orchestration, Intelligence, Near-RT RIC, Network Slicing, NFV-MANO, Non-RT RIC, O-Cloud Site, O-RAN, O-RAN Alliance, Open Interfaces, Openness, RAN Architecture, RAN Slicing, RIC, SMO Framework, Standards, TN Slicing
\vspace{-2.5mm}
\end{keywords}

\vspace{-2.5mm}
\section{Introductory Remarks}\label{sec:Introduction}
\vspace{-1.5mm}
\IEEEPARstart{T}{he} \gls{RAN} {is a critical domain of} cellular network, providing wireless connectivity between \gls{UE} and base stations across a specified geographical footprint. It employs various \glspl{RAT} to ensure efficient {bidirectional} data transmission \cite{pana_5g_2022}. Its architecture {has} evolved with increasing density of \glspl{UE}, diverse access technologies, performance demands (e.g., higher data rates, lower latency), and emerging trends such as virtualization and cloudification \cite{8723481,9124820,chaccour2022data,s24031038}. This evolution---from \gls{4G} to \gls{5G} and now towards \gls{6G}---has enabled a wide range of advanced services, applications, and use cases \cite{9349624,9923927,10286980,10525242}.

Introduced in \gls{4G}, the \gls{D-RAN} architecture separated radio and baseband functions into distinct {components}---the \gls{RRH} and \gls{BBU}---though both components remained co-located at the same site  \cite{10464310,husabelCran2Oran,8723481}. {To enhance the scalability and efficiency}, \gls{C-RAN} decoupled the \gls{BBU} {from the} \gls{RRH} and relocating it to a centralized \gls{DC} \cite{pana_5g_2022}. {This architectural shift} enabled centralized control and optimization of multiple \glspl{RRH} via high-speed \gls{FH} interface, commonly using \gls{CPRI} \cite{10464310,husabelCran2Oran,thiruvasagam2023Oran}. The {transition from} \gls{D-RAN} to \gls{C-RAN} marked a key milestone in the evolution of modern \gls{RAN} architectures, paving the way for innovations in \gls{4G}, \gls{5G}, and beyond \cite{pana_5g_2022,frauendorf2023}.

The advent of \gls{5G} heralded a paradigm shift in \gls{RAN} to address the diverse demands of {emerging} industrial applications and vertical markets~\cite{8723481,10835138,9627832}. To support this evolution, \gls{3GPP} introduced the \gls{NG-RAN}, {where} base stations---\glspl{gNB}---are functionally split into the \gls{CU} for higher layer functions and the \gls{DU} for lower layer processing \cite{3GPP-TS-38.401,rouwet2022open,bonati2020open5G,9750106}.

{A defining} feature of \gls{NG-RAN} is its cloud-native and virtualized {design}, allowing the deployment of \gls{CU} and \gls{DU} on \gls{COTS} hardware within virtualized infrastructures \cite{9750106,O-RAN.WG1.OAD}. {However,} \gls{NG-RAN} remains largely closed and proprietary, limiting opportunities for innovation~\cite{10024837}. Its components often built with tightly coupled interfaces optimized for performance, {tailored} to specific manufacturers, {thus restricting} multi-vendor interoperability \cite{9839628}.

To overcome the limitations of proprietary \gls{NG-RAN} systems and to promote openness and interoperability, the \gls{O-RAN} Alliance introduced the \gls{O-RAN} architecture---a transformative initiative aimed to redefine traditional \gls{RAN} design principles \cite{9933014,eBook-Keysight,maxenti2025autoran}. Building upon the \gls{3GPP}-defined \gls{gNB}, the \gls{O-RAN} introduces further functional {disaggregation} by separating the \gls{DU} into two distinct entities: the \gls{O-DU} and the \gls{O-RU}. {As a result}, the \gls{O-gNB} consists of three interoperable components---\gls{O-CU}, \gls{O-DU}, and \gls{O-RU}---supporting a scalable and flexible architecture where a single \gls{O-CU} can control multiple \glspl{O-DU}, and each \gls{O-DU} can connect to several \glspl{O-RU}. The \gls{O-RU} handles the transmission and reception of radio signals to and from \glspl{UE} \cite{O-RAN.WG1.OAD,ericssonWPDORAN,maxenti2025autoran}.

This architectural shift replaces closed and proprietary {\gls{RAN} solutions} with open, cloud-native, interoperable, and intelligent systems. {By} standardizing open interfaces and specifications, {\gls{O-RAN} fosters} multi-vendor interoperability, innovation, and flexible deployments across diverse network environments~\cite{10178010}. {A core enabler of this vision is} the integration of \gls{AI} and \gls{ML} via \glspl{RIC}, which enhances automation, real-time optimization, {and service assurance} \cite{O-RAN.WG1.OAD, 10008676,thomas2024causal,8755300}. {Achieving an} open and intelligent \gls{NG-RAN} necessitates the adoption of \gls{SDN} and \gls{NFV} to decouple control and user planes, virtualize \gls{RAN} components, and implement standardized open interfaces between them \cite{9754560,Rehman-NFV}. {In addition, the \gls{O-RAN} architecture underpins advanced} use cases such as network slicing, dynamic spectrum sharing, and \gls{RAN} resource orchestration---{key for delivering next-generation services with greater flexibility, efficiency, and scalability}~\cite{9839628,9204600,ericsson5Gran,8385093}.

Network slicing enables the partitioning of a physical network into multiple virtual slices, each operating independently and configured for specific application requirements, {thereby supporting differentiated \gls{QoS} across verticals} \cite{10178010,O-RAN.WG1.StdSA,9204600,10225846}. In contrast to the rigid, \textit{one size fits all} architectural solutions, slicing introduces a flexible, and dynamic approach for resource allocation and network optimization, {aligning with the needs of emerging \gls{5G}, \gls{6G}, and beyond services}~\cite{3GPP-TR-28801MgmtOrch,sallentRanSlicingRrmp,9204709,9277891}. In addition, network slicing empowers \glspl{MNO} to fulfill \glspl{SLA} with tenants \cite{8320765} by {addressing diverse} performance {and functional} requirements using standardized service types as defined by \gls{3GPP}. These service types, as of this writing, encompass \gls{eMBB}, \gls{URLLC}, \gls{mMTC}, \gls{HMTC}, and \gls{V2X}~\cite{9750106}. {With the evolution of} \gls{5G}, \gls{6G} and beyond, this set is expected to grow significantly to accommodate the expanding communication and non-communication services~\cite{9349624}.

{While} \gls{3GPP} {provides foundational support for} \gls{RAN} slicing, the \gls{O-RAN} Alliance plays a {complementary and critical} role in {enabling} intelligent slicing through open interfaces and components \cite{9844089,9933014}. 
To unlock the full potential of slicing within \gls{O-RAN} {framework, a strong unification} between \gls{O-RAN} and \gls{3GPP} standards is essential. 
Beyond traditional \gls{RAN} slicing, \gls{O-RAN} introduces several key enhancements that improve customization, interoperability and automation: \textbf{(a)} Through its disaggregated architecture, \gls{O-RAN} enables fine-grained control over slice specific resources. Operators can select specialized hardware and software from different vendors to optimize the performance per slice. \textbf{(b)} \gls{O-RAN}'s open interfaces eliminates vendor lock-in, making it possible to mix and match interoperable components. This allows operators to choose the best-in-breed solutions for each slice, leading to cost savings and innovation. \textbf{(c)} The integration of \gls{Non-RT RIC} and \gls{Near-RT RIC} introduces intelligent control via \glspl{rApp} and \glspl{xApp}, which enable close-loop automation and optimization of slice-specific resource management \cite{O-RAN.WG1.SA,openRANGymToolbox,wu2020dynamic-RAN-Slicing}. \textbf{(d)} \gls{O-RAN} further supports automation of slice lifecycle management--encompassing slice creation, configuration, and orchestration--through its interfaces and protocols, significantly reducing operational complexity and improving reliability.

Note that, while both \gls{O-RAN} and \gls{3GPP} address key aspects of \gls{RAN} slicing, neither framework explicitly focuses on the virtualization aspects. Thus, a broader harmonization---encompassing \gls{O-RAN}, \gls{3GPP}, and \gls{ETSI}---is required to support both the physical and virtual infrastructure. Such alignment is anticipated to deliver unified and scalable solutions for the realization of flexible, service-specific \gls{O-RAN} slices \cite{BuildingSMOFramework}.

\begin{table*}[!ht]
\centering
\caption{Comparing the contributions of our paper to the most recent state-of-the-art overview and survey papers related to O-RAN} 
\renewcommand{\arraystretch}{1.4}
\small
\begin{tabular}{|l|l|l|l|l|l|l|l|} 
\hline \hline
\rowcolor{lightgray} 
\multicolumn{1}{|p{1.4cm}|}{\textbf{Ref.}} &
\multicolumn{1}{p{1.0cm}|}{\textbf{Year}} &
\multicolumn{1}{p{2.0cm}|}{\textbf{Network Architecture}} &
\multicolumn{1}{p{2.0cm}|}{\textbf{Deployment Scenarios}} &
\multicolumn{1}{p{2.0cm}|}{\textbf{Open Source Initiatives}} &
\multicolumn{1}{p{2.0cm}|}{\textbf{Network Slicing}} &
\multicolumn{1}{p{2.2cm}|}{\textbf{Management \& Orchestration}} &
\multicolumn{1}{p{2.0cm}|}{\textbf{Use Cases \& Examples}} \\ \hline \hline

\cite{bonati2020open5G} & 2020 & \checkmark &  & \checkmark & & &
\\ \hline

\cite{waypior2022OpenRan} & 2022 & \checkmark & & & & &
\\ \hline

\cite{9754560} & 2022 & \checkmark & & & & &
\\ \hline

\cite{9798822} & 2022 & \checkmark & & & & \checkmark & \checkmark
\\ \hline

\cite{9695955} & 2022 & \checkmark & & & & &
\\ \hline

\cite{9839628} & 2022 & \checkmark & & & & &
\\ \hline

\cite{10024837} & 2023 & \checkmark & \checkmark & \checkmark & & & \checkmark
\\ \hline

\cite{s23218792} & 2023 & \checkmark & & & & &
\\ \hline

\cite{103389} & 2023 & \checkmark & & & & &
\\ \hline

\cite{LIYANAGE2023103621} & 2023 & \checkmark & & & & &
\\ \hline

\cite{10178010} & 2023 & \checkmark & & & & &
\\ \hline

\cite{thiruvasagam2023Oran} & 2023 & \checkmark & \checkmark & & & &
\\ \hline

\cite{s24031038} & 2024 & \checkmark &  & \checkmark & \checkmark &  & 
\\ \hline

\cite{10330580} & 2024 & \checkmark & \checkmark & \checkmark & & &
\\ \hline

\cite{10329947} & 2024 & \checkmark & & & & & \checkmark
\\ \hline

\cite{10439167} & 2024 & \checkmark & & & & \checkmark & \checkmark
\\ \hline

\cite{10601697} & 2024 & \checkmark & & & & & \checkmark
\\ \hline

This paper & 2025 & \checkmark & \checkmark & \checkmark & \checkmark & \checkmark & \checkmark
\\ \hline 

\multicolumn{8}{|l|}{\begin{tabular}[c]{@{}l@{}}\textbf{Note:} For each column, a \checkmark indicates that the aspect is discussed in detail, while a blank space signifies that the aspect is absent. \end{tabular}} \\ \hline

\end{tabular}
\label{tab:ComparisonTable}
\vspace{-6mm}
\end{table*}

\vspace{-2.5mm}
\subsection{Literature Review and Research Gap Analysis}\label{Review&ResearchGap}
\vspace{-1.5mm}
The research community has made significant contributions to diverse research challenges associated with \gls{O-RAN}. These efforts span multiple domains, encompassing the \gls{MO} of \gls{O-RAN} slicing, the design {of intelligent applications such as} \glspl{xApp} and \glspl{rApp}, and {the resolution of} numerous optimization problems. In addition, several comprehensive survey and overview papers have been published by prominent academic institutions. Table \ref{tab:ComparisonTable} presents a curated and up-to-date list of these works, highlighting their major contributions to the broader \gls{O-RAN} landscape, and where applicable, to the specific domain of \gls{O-RAN} slicing.

While several of the works referenced {in Table~\ref{tab:ComparisonTable}} offer valuable insights into the architectural components of \gls{O-RAN} and {explore aspects of} network slicing, \textbf{they fall short in delivering a comprehensive analysis of slicing mechanisms tailored to the \gls{O-RAN} framework}. In particular, \textbf{these surveys lack a detailed examination of the current landscape of \gls{O-RAN} slicing}, including the exploration of various deployment strategies proposed in the literature and the complexities involved in tight integration of \gls{O-RAN} components necessary to support intelligent and flexible slicing.

Moreover, \textbf{there is a significant gap in the existing literature regarding the seamless integration and interoperability of \gls{O-RAN}’s disaggregated components and open interfaces}, a crucial factor for the successful development and deployment of diverse \gls{O-RAN} slice types. The absence of thorough discussion on these vital aspects hampers a complete understanding of how \gls{O-RAN} can unlock the full potential of slicing across \gls{5G}, \gls{6G}, and beyond. This is particularly important when considering the challenges of achieving \textbf{interoperability and scalability in multi-vendor environments}, where cohesive integration across different components is essential for ensuring efficient network performance.

Therefore, \textbf{the lack of a detailed and cohesive analysis of the entire \gls{O-RAN} and underlying infrastructure slicing}, from deployment scenarios to seamless integration {of disaggregated} components, reveals a critical gap in the literature. This underscores the need for further research to bridge these gaps and advance the development of optimized, interoperable network slices within the \gls{O-RAN} framework.

\vspace{-2.5mm}
\subsection{Goals and Contributions} \label{Subsec:Goals&Contributions}
\vspace{-1.5mm}
To address the identified gap in the literature, this paper provides a comprehensive {exploration of slicing-aware \gls{O-RAN} architecture. It offers a detailed tutorial} and {critical} overview of network slicing within the context of the \gls{O-RAN} architecture. The key contributions of this paper are as follows:

\begin{itemize}[noitemsep, topsep=0pt, left=0pt]
    \item An exploration {of \textbf{open source initiatives, standardization efforts, and the design of experimental platforms}} {supporting the development and validation} of \gls{O-RAN}.
    \item A focused analysis of {\textbf{the latest \gls{O-RAN} architecture}} as defined by the \gls{O-RAN} Alliance, with particular emphasis on both {\textbf{theoretical advancements from academic research and practical deployment by industry leaders}}.
    \item A {comprehensive and \textbf{holistic view of the network slicing paradigm within \gls{O-RAN}}}, including {\textbf{the functional components and open interfaces essential for implementing slicing capabilities}} across \gls{O-RAN}.
    \item An {exploration of key \textbf{deployment scenarios for \gls{O-RAN} slicing and the \gls{SMO} framework}}, alongside {\textbf{detailed insight of several high-level use cases}} that \gls{O-RAN} is expected to support.
    \item An in-depth discussion of {\textbf{the \glspl{NF} and \gls{TN} elements, along with their \gls{MO}}---}particularly \gls{FH} and \gls{MH}---that collectively constitute the \gls{O-RAN} slice subnet. Additionally, this paper examines {\textbf{the aspects of the underlying infrastructure associated with the \gls{TN}}} and the slicing of its resources. 
\end{itemize}
Finally, Table \ref{tab:ComparisonTable}, presents a comparison between the contributions of this article and those of other survey papers.

\begin{figure}[!ht]
    \centering
    \includegraphics[width=\columnwidth]{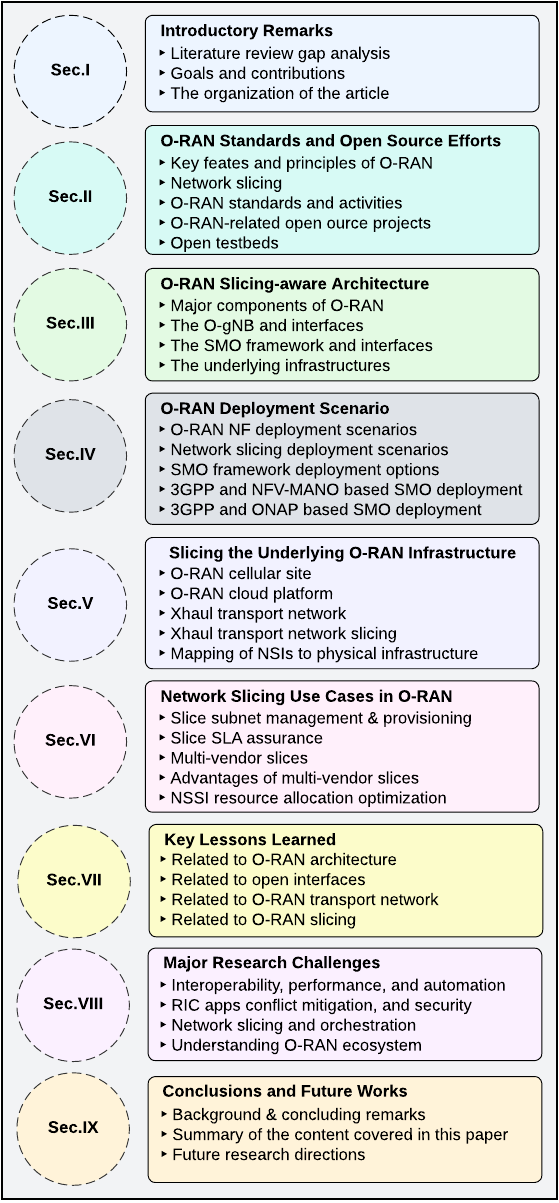}
    \caption{Overview of the organization and structure of this survey paper. Each box {in the figure} represents one of the chapters of the paper, encapsulating their respective contributions and themes.}
    \label{fig:paperstrcuture}
\end{figure}

\vspace{-2.5mm}
\subsection{The Organization of the Article} \label{Subsec:Structure}
\vspace{-1.5mm}
The remainder of this paper is structured as follows: Section~\ref{Sec:SecondChapter} highlights the key features of \gls{O-RAN} and provides an overview of open source projects, activities, and contributions, along with the standardization efforts associated with \gls{O-RAN}. In Section~\ref{sec:Architecture}, we introduce the architectural components and open interfaces of the \gls{O-RAN} slicing-aware architecture, highlighting its features and interactions with \gls{3GPP}-defined network components, service management, and service orchestration. Section~\ref{sec:DeploymentOptions} covers various deployment scenarios for \gls{O-RAN}, slicing, and the \gls{SMO} proposed by the \gls{O-RAN} Alliance tailored to different use cases. Section~\ref{sec:UnderlyingInfrastructure} discusses slicing the underlying infrastructure, elaborating \gls{O-Cloud} slicing, \gls{TN} slicing, and \gls{TN} slice orchestration. In Section~\ref{sec:UseCases}, we outline high-level use cases expected to be prioritized by the \gls{O-RAN} community, particularly regarding \gls{RAN} slicing. Section~\ref{Sec:LessonsLearned} provides an overview of the lessons learned through this survey, while Section~\ref{Sec:ResearchChallenges} identifies key research challenges that require further research and investigation. Finally, Section~\ref{sec:conclusion} summarizes our work, draws conclusions, and suggests potential directions for future research. An overview of the organization of this survey is illustrated in Figure~\ref{fig:paperstrcuture}.

\vspace{-2.5mm}
\section{Ongoing Standardization Efforts and Open Source Contributions to O-RAN}\label{Sec:SecondChapter}
\vspace{-1.5mm}
This section provides a detailed overview of the key features of the \gls{O-RAN} architecture. It also examines ongoing open source initiatives that are actively shaping \gls{O-RAN}, as well as publicly available experimental platforms that support the development and validation of \gls{O-RAN} components and interfaces. In addition, we provide an overview of state-of-the-art contributions across various \textit{de facto and de jure} \glspl{SDO} involved in the evolution of the \gls{O-RAN}. The goal is to provide a comprehensive understanding of the current advancements and collaborative efforts driving innovation within the \gls{O-RAN} ecosystem.

\vspace{-2.5mm}
\subsection{Key Features and Principles of O-RAN}
\vspace{-1.5mm}
The primary objective behind \gls{O-RAN} for service providers and network operators is to diversify vendor partnerships and avoid vendor lock-in by enabling the use of non-proprietary software and hardware components source from multiple vendors \cite{6GChapterHaibi}. Traditionally, \gls{RAN} has been proprietary and vertically integrated, wherein both hardware and software were tightly coupled and delivered by a single vendor \cite{10178010}. \gls{O-RAN} aims to disrupt this conventional model by promoting open interfaces and fostering interoperability among \gls{RAN} components. The \gls{O-RAN} architecture emphasizes modularity and flexibility through open standards, enabling operators to integrate hardware and software solutions from different suppliers \cite{ag_easy_2020,s23218792}. This flexibility facilitates enhanced customization of network configuration and encourages broader participation from second- and third-tier equipment manufacturers.

Beyond open interfaces, \gls{O-RAN} enables full access to \gls{NG-RAN} through \gls{AI}-based control mechanisms that support real-time monitoring, proactive resource allocation, and adaptive responsiveness to dynamic radio conditions \cite{9023918,10056646}. It further promotes a disaggregated, virtualized, cloud-native, and interoperable \gls{RAN}~\cite{9750106}, empowering service providers to deploy a fully programmable, intelligent, autonomous, and multi-vendor \gls{RAN} suited for \gls{5G}, \gls{6G}, and beyond networks \cite{10508126,vmwareOran}.

In the following, we summarize the core principles and defining characteristics of \gls{O-RAN}, highlighting aspects particularly relevant to service providers, network operators, and other stakeholders in the telecommunications ecosystem.

\paragraph{Intelligent and Programmable Network} The \gls{O-RAN} architecture is inherently intelligent and programmable. This enables dynamic optimization of network \gls{OAM} \cite{TowardsAIMLORAN}. This programmability allows cellular networks to efficiently adapt to diverse traffic demands and deployment scenarios, supporting the evolving requirements of next-generation wireless communication systems \cite{5GamericasWp}.

\paragraph{Data Center Economics in the RAN} \gls{O-RAN} brings the economic principles of \gls{DC} into the \gls{RAN}. By leveraging virtualization, \gls{COTS} hardware, and centralized resource management, it enhances the scalability and economic sustainability of \gls{RAN} infrastructure by aligning it with the cost-efficient practices of modern \gls{DC} environments \cite{gavrilovska_cloud_2020}. Through optimized resource utilization, simplified deployment, and improved maintenance processes, \gls{O-RAN} reduces \gls{OPEX} and significantly lowers the \gls{TCO} \cite{ag_deutsche_2023}.
    
\paragraph{Automation and Manageability} \gls{O-RAN} places strong emphasis on automation and centralized manageability, aiming to reduce manual intervention in network \gls{OAM} \cite{5GamericasWp}. It improves operational efficiency while enhancing the reliability, consistency, and scalability of network management processes.

\paragraph{Faster Time to Market and Innovation Agility} \gls{O-RAN} accelerates the deployment of network solutions by enabling service providers and operators to introduce new features and services more rapidly \cite{Pw-OpenRan2020}. The modular and open architecture of \gls{O-RAN} fosters innovation agility by enabling the rapid development, testing, and deployment of new technologies and features. Through standardized interfaces and software-defined components, it facilitates seamless integration of emerging solutions—such as \gls{AI}/\gls{ML}-driven \glspl{xApp} and \glspl{rApp}—into the \gls{RAN} \cite{Pw-OpenRan2020}. This agility is essential for maintaining competitiveness in the rapidly evolving telecoms landscape, ensuring that the network keeps pace with technological advancements and shifting industry requirements.
    
\paragraph{Vendor Diversity} \gls{O-RAN} promotes a competitive, multi-vendor ecosystem by decoupling hardware and software components through standardized interfaces. This architectural openness allows network operators and service providers to select equipment and solutions from a broad range of vendors, promoting flexibility in network design and deployment. Such diversity not only encourages innovation and accelerates technology evolution but also enables operators to tailor their networks and services to specific technical and business requirements \cite{ViaviWP}. Furthermore, it reduces \gls{CAPEX} by mitigating vendor lock-in and supporting cost-effective integration of interoperable components.

\paragraph{Open Source Software} \gls{O-RAN} leverages open source software to support the development of reference implementations aligned with its open and intelligent \gls{RAN} architecture \cite{gavrilovska_cloud_2020}. {The \gls{OSC}---a collaboration between the \gls{O-RAN} Alliance and the \gls{LF}---plays a central role in implementing \gls{O-RAN} specifications through open source projects.} This approach fosters interoperability, accelerates innovation, and promotes collaboration among diverse industry stakeholders. As a result, it significantly facilitates the evolution, validation, and deployment of disaggregated, vendor-neutral \gls{RAN} solutions within the telecommunications ecosystem.

\paragraph{AI/ML for O-RAN Optimization and Resource Management}
One primary enabler of \gls{O-RAN} is \gls{AI}/\gls{ML}. Some \gls{AI}/\gls{ML} algorithms are already proposed to optimize performance, enhance autonomous management, and support real-time, data-driven intelligence across \gls{O-RAN} \cite{10103771, 9812489,9367572}. A key aspect of this integration is the evolution of \gls{AI}/\gls{ML}-driven \glspl{RIC} \cite{9367572}. This empowers service providers, vendors, and third party developers to deploy intelligent applications (e.g., xApps and rApps) for automated, closed-loop network optimization \cite{9367572,10243548,10604823,TowardsAIMLORAN}. In addition, \gls{O-RAN} enables embedding intelligent control across multiple functional domains, including \gls{RRM} and service orchestration \cite{9812489}. In this context, \gls{AI}/\gls{ML} facilitates predictive real-time decision-making to address a broad spectrum of use cases, such as interference mitigation, energy efficiency, dynamic load balancing, {QoS}-aware scheduling, and self-healing capabilities \cite{10103771}. Furthermore, \gls{AI}/\gls{ML} is integral to \gls{O-RAN} management in order to offer advanced predictive maintenance, traffic forecasting, and continuous optimization of performance in response to dynamic operating conditions.
 
Moreover, \gls{AI}/\gls{ML} serves as critical enablers for dynamic and intelligent slicing in \gls{O-RAN}. It can facilitate the creation, adaptation, and real-time management of multiple virtualized, service-specific slices \cite{10935636}. It empowers \gls{O-RAN} to intelligently monitor, predict, and respond to fluctuating network demands, thereby optimizing \gls{E2E} \gls{QoS} and resource utilization. Various types of \gls{AI}/\gls{ML} algorithms can be beneficial for addressing different aspects of \gls{O-RAN} slicing. For example, \gls{RL} and \gls{MARL} enable adaptive policy learning and slice orchestration, while \gls{LSTM} networks support proactive traffic forecasting and timely reconfiguration of slice parameters \cite{10476972}. Additionally, \gls{FL} has emerged as a promising approach for distributed slice management, preserving data privacy while ensuring coordinated learning across decentralized nodes \cite{10935636}. Lastly, \gls{DL} models, including \glspl{CNN} and Transformers, are increasingly applied for tasks such as anomaly detection and intelligent beamforming.

The \gls{O-RAN} Alliance plays a crucial role in the development and promotion of \gls{O-RAN} standards. By embracing the principles set forth by the alliance, the telecommunication industry aims to accelerate innovation, reduce deployment costs, and cultivate a more dynamic and competitive marketplace for \gls{RAN} solutions towards \gls{5G}, \gls{6G}, and beyond networks \cite{10178010}.

\vspace{-2.5mm}
\subsection{Network Slicing}
\vspace{-1.5mm}
Network slicing is a transformative architectural paradigm that enables multiple, logically isolated virtual networks---referred to as ``slices"---{to coexist} over a shared physical infrastructure~\cite{9933014}. {As a foundational element of} \gls{5G}, \gls{6G}, and future systems, it addresses {heterogeneous} service demands and unlocks the full potential of next-generation networks. At its core, it partitions the physical network into discrete virtualized slices, each {independently} configured to specific service requirements and performance criteria---such as bandwidth, latency, reliability, and security \cite{9410215,10467183}. These slices operate autonomously, {allowing concurrent support for diverse use cases---ranging} from latency-sensitive industrial automation to bandwidth-intensive mobile broadband{---within the same physical network infrastructure} \cite{8985329,9079548}.

An \gls{E2E} network slice spans all network domains, encompassing the \gls{RAN}, \gls{TN}, and \gls{CN} segments \cite{GSMA-E2E-nsr,101145}. Each network slice is {carefully engineered} to meet the specific requirements of different services, ensuring logical isolation among slices. This isolation preserves the integrity of individual network slices by preventing faults or malfunctions in one from affecting others, thereby fostering the autonomy and reliability across virtualized networks \cite{10375939}.

To guarantee performance and service quality, operators allocate dedicated resources to each slice---such as computing capacity, bandwidth, \gls{QoS} provisions, and other critical elements \cite{9003208}. This resource assurance underscores the commitment to supporting the diverse service types, while ensuring efficiency, scalability and robustness across a cellular network.

Despite the substantial progress in \gls{E2E} slicing, several challenges {persist in the realization of} \gls{NG-RAN} slicing \cite{8385093,8985329}. The complexity arises primarily from the need to balance varying degrees of isolation and resource sharing, while tailoring the \gls{UP} and \gls{CP} {functionalities to meet the specific requirements of} individual slices \cite{sallentRanSlicingRrmp,schmidtRanSlicingSystem,9927252}. The key challenges include managing the trade-off between resource utilization efficiency and isolation, harmonizing inter-\gls{RAN} and intra-\gls{RAN} resource allocation algorithms, and prioritizing slices effectively across different layers of the \gls{RAN} \cite{9750106}. In addition, the limited availability of radio resources demands highly efficient resource management {strategies to sustain} optimal network performance. The introduction of advanced \gls{5G} \gls{NR} features--such as \gls{BWP} and physical numerology--further amplifies these {challenges by increasing configuration complexity and the need for dynamic adaptation} \cite{karim5GNetworkSlicing,10494372}. 

To address the above challenges, the \gls{3GPP} provided guidelines in Release 17 for realizing slicing in \gls{NG-RAN}. These guidelines encompass various aspects, including support for diverse \gls{QoS} types, resource segregation, \gls{SLA} enforcement, among others \cite{3GPP-TR-38832-NrSlicing}. The \gls{3GPP} specifications further enhance architectural flexibility by presenting multiple implementation options for \gls{RAN} slicing, such as \gls{L1}, \gls{L2}, or the \gls{MAC}-based approaches \cite{3GPP-TR-38832-NrSlicing, kuklinskiOranMecSonIntegration}. In addition, they specify a \gls{MO} framework to support the efficient lifecycle management of \gls{RAN} slices and their associated resources across the \gls{NG-RAN}. {This framework also ensures interoperability} with other standardized architectures for the realization of \gls{E2E} slicing \cite{9750106}.

\vspace{-2.5mm}
\subsection{O-RAN Standards and Activities}
\vspace{-1.5mm}
As of this writing, numerous de facto and de jure organizations are actively involved---both directly and indirectly---in developing standards for software and hardware components that align with the principles of \gls{O-RAN}. These efforts are coordinated through collaborative initiative under various \glspl{SDO} across the globe. In this subsection, we dive into an exploration of these \glspl{SDO} and their respective contributions towards the advancement and realization of \gls{O-RAN}.

\subsubsection{3GPP}
\gls{3GPP} does not directly define standards specific to the \gls{O-RAN}. However, many architectural components of the \gls{3GPP}-defined \gls{RAN} architecture---including \glspl{NF}, \gls{MO} frameworks, functional split options, and interface specifications---have been adopted and further extended by other \glspl{SDO}, most notably the \gls{O-RAN} Alliance, to establish comprehensive standards for the \gls{O-RAN} architecture. The \gls{3GPP} specifications provide a foundational and system-level definition of \gls{RAN} architecture, distributed across its various \glspl{TSG}.

{In the course of} \gls{5G} evolution, \gls{3GPP} evaluated eight functional split options and ultimately standardized two \gls{NG-RAN} split architectures. The first is the \gls{HLS}, corresponding to option 2 from the \gls{3GPP} study. It involves dividing the \gls{BBU} into \gls{CU} and \gls{DU}. The second split involves \gls{CP} and \gls{UP} separation within the \gls{CU}, introducing a logical division of signaling and data-handling responsibilities \cite{s23218792}. To support this architecture, \gls{3GPP} introduced the F1 interface, which connects \glspl{CU} to \glspl{DU} and the E1 interface, which facilitates coordination between \gls{CP} and \gls{UP} \cite{sasha5GRan,2018NGMNOO}. The functional split options are analyzed in our earlier work \cite{9750106}.

The introduction of functional split in \gls{3GPP} represents a critical step towards disaggregating the standard protocol stack. This involves separating the processing of a specific layer within the protocol stack from the computing entity, thereby promoting architectural openness, intelligent cellular interface, and the feasibility of network slicing \cite{103389}. This functional split has served as a pivotal catalyst for the development of subsequent \gls{O-RAN}-related specifications.

\subsubsection{O-RAN Alliance}
The \gls{O-RAN} Alliance, established in 2018, is {a global industry consortium} committed to the ambitious task of modernizing traditional \gls{RAN} architecture. Its central mission revolves around steering the wireless communication industry towards a future defined by openness, intelligence, automation, cloudification, virtualization, and interoperability within the \gls{RAN} \cite{OranIntro,lacava2022programmable}. This transformative journey is underpinned by a shift towards virtualized and cloud-native network components, the adoption of white-box hardware, and the implementation of open, {standardized} interfaces that facilitates {seamless} communication between various software and hardware components of the \gls{O-RAN} architecture \cite{OranIntro}.

To achieve this vision, the \gls{O-RAN} Alliance follows a systematically organized technical specification governed by its \gls{TSC}. The \gls{TSC} plays a pivotal role in decision-making and provides essential guidance on \gls{O-RAN} technical matters. It assumes the crucial responsibility for approving specifications prior to their submission for board approval and eventual publication. The current structure of the \gls{TSC} encompasses eleven technical \glspl{WG}, five \glspl{FG}, a dedicated research group, an open source software community, and a \gls{MVP-C}. These specialized divisions collaborate to focus on specific aspects of \gls{O-RAN}, contributing collectively to its development, deployment, and evolution of its technical standards. An overview of the specific objectives and focus areas of each division is provided in Table \ref{tab:OranWorkgroups}. 

\begin{table*}[!htbp]
\centering
\caption{Summary of contributions and focus areas across multiple WGs supervised by the TSC within the O-RAN Alliance}
\renewcommand{\arraystretch}{1.4}
\small
\begin{tabular}{|p{1cm}|p{3.0cm}|p{12.85cm}|}
\hline \hline
\rowcolor{lightgray} 
\multicolumn{1}{|c|}{\textbf{{Group}}} & \multicolumn{1}{c|}{\textbf{{Title}}} &  \multicolumn{1}{c|}{\textbf{{Principal Areas of Focus and Notable Contributions}}} \\ \hline \hline

\gls{WG}1 & Use Cases and Overall Architecture \gls{WG} & 
\begin{minipage}[t]{\linewidth}
\begin{itemize} [topsep=0ex,partopsep=0ex]
\item Exploring {a number of} use cases, {defining} system-level requirements, {introducing numerous} deployment scenarios, and {proposing} a comprehensive architecture for \gls{O-RAN}
\item Investigation into network slicing within \gls{O-RAN}, including defining {several} use cases, {key} requirements, and introducing slicing-aware architecture with interface extensions
\item Coordination of proof of concepts to demonstrate \gls{O-RAN} products to the market
\end{itemize}
\end{minipage}
\\ \hline

\gls{WG}2 & \gls{Non-RT RIC} and A1 Interface \gls{WG} & 
\begin{minipage}[t]{\linewidth}
\begin{itemize} [topsep=0ex,partopsep=0ex]
\item Defining an architecture for \gls{Non-RT RIC} {and its functionalities}, {and} incorporating the R1 interface to connect the \gls{Non-RT RIC} framework with \glspl{rApp}
\item Expanding R1 services within {the functionalities of the} \gls{Non-RT RIC}, {and enabling interoperability among the} interfaces {of the various} management components of the \gls{SMO} framework
\item {Discussing} the A1 interface, {the interface} between the \gls{Non-RT RIC} and the \gls{Near-RT RIC}, including associated use cases{, deployment scenarios,} and applications
\end{itemize}
\end{minipage}
\\ \hline

\gls{WG}3 & \gls{Near-RT RIC} and E2 Interface \gls{WG} &
\begin{minipage}[t]{\linewidth}
\begin{itemize} [topsep=0ex,partopsep=0ex]
\item Specifying E2 interface {-- an interface} between the \gls{Near-RT RIC} and the E2 nodes
\item Defining the \gls{Near-RT RIC} architecture and introducing \glspl{API} {to connect} the \gls{Near-RT RIC} platform and the \glspl{xApp}
\item Defining {several} use cases, requirements, and management specifications for the \gls{Near-RT RIC}, and contributing to service models for E2 interface and E2 nodes 
\end{itemize}
\end{minipage}
\\ \hline

\gls{WG}4 & Open Fronthaul Interfaces \gls{WG} & 
\begin{minipage}[t]{\linewidth}
\begin{itemize} [topsep=0ex,partopsep=0ex]
\item Establishing specifications for an \gls{O-FH} interface between {the} \gls{O-DU} and {the} \gls{O-RU} {within an \gls{O-gNB}}
\item Setting standards for Control, User, Synchronization, and Management Plane protocols with {their} corresponding YANG models for {the} \gls{O-FH} link
\item Developing specifications for transport interfaces and conducting \gls{O-FH} interoperability tests
\end{itemize}
\end{minipage}
\\ \hline

\gls{WG}5 & Open F1/W1/E1/X2/Xn Interface \gls{WG} & 
\begin{minipage}[t]{\linewidth}
\begin{itemize} [topsep=0ex,partopsep=0ex]
\item Providing interoperable multi-vendor specifications aligned with \gls{3GPP}{-defined} standards for F1, W1, E1, X2, and Xn interfaces, enhancing {the overall} \gls{O-RAN} architecture
\item Defining specifications for O1 interface, covering {interaction between the} \gls{O-CU} and \gls{O-DU} {with \gls{SMO} and discussing the} \gls{OAM} functions
\item Developing open \gls{MH} and \gls{BH} interoperability test specifications
\end{itemize}
\end{minipage}
\\ \hline

\gls{WG}6 & Cloudification and Orchestration \gls{WG} &  
\begin{minipage}[t]{\linewidth}
\begin{itemize} [topsep=0ex,partopsep=0ex]
\item Specifying cloud-native and virtualized infrastructure for hosting {the} \gls{O-CU} and \gls{O-DU} {of an \gls{O-gNB}}, focusing on hardware-software decoupling {within the underlying infrastructure}
\item Providing technology and reference designs for leveraging commodity hardware platforms
\item Identifying use cases, deployment scenarios, and requirements for cloud resource hosting, and defining high-level orchestration architecture for \gls{SMO} framework and \gls{O-Cloud} interaction
\end{itemize}
\end{minipage}
\\ \hline

\gls{WG}7 & White-box Hardware \gls{WG} & 
\begin{minipage}[t]{\linewidth}
\begin{itemize} [topsep=0ex,partopsep=0ex]
\item Specifying standards for comprehensive reference design of high-performance, spectral-efficient, and energy-efficient white box base stations within the \gls{O-RAN} architecture
\item Promoting decoupled software and hardware platform for \gls{O-RAN} components {and interfaces}
\item Addressing outdoor and indoor cells with various split options, along with \gls{O-FH} interface
\end{itemize}
\end{minipage}
\\ \hline

\gls{WG}8 & Stack Reference Design \gls{WG} & 
\begin{minipage}[t]{\linewidth}
\begin{itemize} [topsep=0ex,partopsep=0ex]
\item Developing a software architecture {as well as define a comprehensive} design and release plan for \gls{O-CU} and \gls{O-DU} {of an \gls{O-gNB}}, tailored for \gls{NR} protocol stack
\item Providing specifications for interoperability testing of {various} \gls{O-CU} and \gls{O-DU} deployment scenarios with other \gls{O-RAN} components and interfaces
\end{itemize}
\end{minipage}
\\ \hline

\gls{WG}9 & Open X-haul Transport \gls{WG} & 
\begin{minipage}[t]{\linewidth}
\begin{itemize} [topsep=0ex,partopsep=0ex]
\item Designing an open \gls{TN} within \gls{O-RAN}, meeting \gls{FH}, \gls{MH}, and \gls{BH} service requirements
\item Concentrating on open transport domain, including transport equipment, physical media, and associated control {and/or} management protocols within the {open} \gls{TN}
\end{itemize}
\end{minipage}
\\ \hline

\gls{WG}10 & OAM for O-RAN & 
\begin{minipage}[t]{\linewidth}
\begin{itemize} [topsep=0ex,partopsep=0ex]
\item Specifying \gls{OAM} architecture for \gls{O-RAN} and management services for O1 interface, {such as proposing a set of} unified operation and notification mechanisms
\item Developing information models and data models for \gls{OAM} architecture in \gls{O-RAN}
\end{itemize}
\end{minipage}
\\ \hline

\gls{WG}11 & Security \gls{WG} & 
\begin{minipage}[t]{\linewidth}
\begin{itemize} [topsep=0ex,partopsep=0ex]
\item Establishing specifications for \gls{O-RAN}'s security, including its \glspl{NF}, interfaces, and (r/x)Apps
\item Defining requirements, use cases, architectures, and protocols to ensure security and privacy of {various types of data and} stakeholders within the \gls{O-RAN} {architecture}
\end{itemize}
\end{minipage}
\\ \hline

SDFG & Standard Development Focus Group & 
\begin{minipage}[t]{\linewidth}
\begin{itemize} [topsep=0ex,partopsep=0ex]
\item Leading in formulating standardization strategies for the \gls{O-RAN} Alliance and serving as the primary interface {between the \gls{O-RAN} Alliance and} other relevant \glspl{SDO}
\item Managing coordination of both incoming and outgoing liaison statements
\end{itemize}
\end{minipage}
\\ \hline

IEFG & Industry Engagement Focus Group & 
\begin{minipage}[t]{\linewidth}
\begin{itemize} [topsep=0ex,partopsep=0ex]
\item Engaging with leading industry players and members of the \gls{O-RAN} Alliance to drive adoption, spread, and ongoing innovation of O-RAN-based technologies and solutions
\end{itemize}
\end{minipage}
\\ \hline

OSFG & Open Source Focus Group & 
\begin{minipage}[t]{\linewidth}
\begin{itemize} [topsep=0ex,partopsep=0ex]
\item Managing \gls{O-RAN} Alliance's open source activities, including establishing the \gls{OSC} and developing open source related strategies
\item Collaborating with other open source communities to drive innovation and adoption of \gls{O-RAN}
\end{itemize}
\end{minipage}
\\ \hline

TIFG & Test \& Integration Focus Group & 
\begin{minipage}[t]{\linewidth}
\begin{itemize} [topsep=0ex,partopsep=0ex]
\item Defining testing and integration approaches, coordinating specifications across {various} \glspl{WG}, including \gls{E2E} test specifications and productization profiles
\item Planning PlugFests and offering guidelines for third-party \glspl{OTIC}, facilitating integration and verification processes
\end{itemize}
\end{minipage}
\\ \hline

\multicolumn{3}{r}{\footnotesize\textit{Continued on the next page}}
\end{tabular}
\label{tab:OranWorkgroups}
\end{table*}

\begin{table*}[!htbp]
\ContinuedFloat
\centering
\caption{{\normalfont\textit{Continued from previous page}}}
\renewcommand{\arraystretch}{1.4}
\small
\begin{tabular}{|p{1cm}|p{3.0cm}|p{12.85cm}|}
\hline \hline
\rowcolor{lightgray} 
\multicolumn{1}{|c|}{\textbf{{Group}}} & \multicolumn{1}{c|}{\textbf{{Title}}} &  \multicolumn{1}{c|}{\textbf{{Principal Areas of Focus and Notable Contributions}}} \\ \hline \hline

SuFG & Sustainability Focus Group & 
\begin{minipage}[t]{\linewidth}
\begin{itemize} [topsep=0ex,partopsep=0ex]
\item Focusing on enhancing energy efficiency and reducing environmental impact in \gls{O-RAN}
\item Collaborating with \gls{MVP-C} to align initiatives across all \glspl{WG} and \glspl{FG} {within \gls{O-RAN} Alliance}
\end{itemize}
\end{minipage}
\\ \hline

nGRG & next Generation Research Group & 
\begin{minipage}[t]{\linewidth}
\begin{itemize} [topsep=0ex,partopsep=0ex]
\item Researching intelligent O-RAN principles for {standardizing \gls{6G} and} beyond {systems}
\item Driving network evolution towards greater intelligence and performance using new technologies
\end{itemize}
\end{minipage}
\\ \hline

MVP-C & Minimum Viable Plan -- Committee & 
\begin{minipage}[t]{\linewidth}
\begin{itemize} [topsep=0ex,partopsep=0ex]
\item Providing roadmap for implementing comprehensive \gls{O-RAN} solutions in commercial networks
\item Managing \gls{O-RAN} features, including creation, prioritization, and tracking documents
\item Approving feature creation and inclusion in relevant releases; collaborating and coordinating with all \glspl{WG} and \glspl{FG} {within the \gls{O-RAN} Alliance}
\vspace{0.5mm}
\end{itemize}
\end{minipage}
\\ \hline

OSC & O-RAN Software Community & 
\begin{minipage}[t]{\linewidth}
\begin{itemize} [topsep=0ex,partopsep=0ex]
\item Leading the development {and possible deployment} of open source software for \gls{O-RAN} architecture in collaboration with the \gls{LF}
\item Focusing on aligning with the open architecture and criteria of the \gls{O-RAN} Alliance to deliver a solution suitable for commercial deployment {and \gls{O-RAN} components and interfaces}
\end{itemize}
\end{minipage}
\\ \hline

\end{tabular}
\vspace{-6mm}
\end{table*}

\subsubsection{Telecom Infra Project}
The \gls{TIP} is a global {consortium} of over 500 companies and organizations working to accelerate the development and {adoption} of open, disaggregated, intelligent, and standards-based technologies for telecommunications infrastructure. It has contributed significantly to open standards and specifications, with its solutions adopted by major operators worldwide to address specific commercial {and operational requirements}. \gls{TIP}'s work is organized into multiple \glspl{PG} focused on products, solutions, and software across domains such as \gls{RAN}, \gls{TN}, and \gls{CN}, {along with associated \gls{MO}} layers.

Within \gls{TIP}'s various \glspl{PG}, the OpenRAN \gls{PG} {plays a central role in} enabling an open ecosystem that redefines the \gls{3GPP}-based \gls{NG-RAN} architecture using open components and standardized interfaces. It supports the evolution of \gls{4G}, \gls{5G}, and future networks by promoting interoperability {and vendor diversity}. The goals of the OpenRAN \gls{PG} includes developing and validating interoperable OpenRAN solutions, advancing innovative platforms for network management, and fostering collaboration among operators, vendors, integrators, and global stakeholders \cite{TipIntro}. The \gls{PG} focuses to align requirements for key OpenRAN components, including \glspl{O-RU}, \glspl{O-DU}, and \glspl{O-CU}, which are extensively tested, and validated in \gls{TIP} Community Labs and PlugFests \cite{adrian_kliks_towards_2023}.

The OpenRAN \gls{PG} is organized into two main workstreams:
\paragraph{Component Subgroups} They are dedicated on enhancing the performance of individual OpenRAN software and hardware components such as \gls{RU}, \gls{DU}, \gls{CU}, \gls{RIA}, and \gls{ROMA}.
\paragraph{Segment Subgroups} They concentrate on developing integrated \gls{RAN} solutions tailored to {diverse} deployment scenarios across indoor and outdoor use cases. 

In June 2024, the OpenRAN \gls{PG} released its Release 4 Technical Priorities Document, presenting critical deployment requirements \cite{TipOpenRanMouGroup}. It addresses radio configurations, hardware/software requirements for each OpenRAN building block, and evolving requirements for the \gls{SMO}, \gls{RIC}, and the cloud infrastructure hosting \gls{O-RAN} components. It emphasizes security, energy efficiency, consolidating related requirements and identifying new priorities across workstreams. {These efforts aim} to accelerate the development {and global adoption} of competitive OpenRAN solutions \cite{TipOpenRanMouGroup,TipIntro}.

\subsubsection{Small Cell Forum}
The \gls{SCF} is a global organization focused on developing technical specifications and tools to accelerate the adoption of flexible, cost-effective, and scalable cellular network infrastructure. The \gls{SCF} has played a key role in standardizing essential elements of network technology, including the \gls{FAPI}, \gls{nFAPI}, and enhancement to the X2 interface. These specifications enable an open, multi-vendor platform, thereby reducing the barriers to the densification of stakeholders in the wireless communications industry \cite{SCF239.10.01}.

The \gls{SCF} has established its own Open \gls{RAN} ecosystem with a particular emphasis on small cell {deployments}. {A notable contribution is} the introduction of the \gls{nFAPI} protocol, which pioneered {\gls{3GPP}'s} split option 6 \cite{5GamericasWp}, dividing the \gls{MAC} and \gls{PHY} layers, with the \gls{PHY} hosted in the \gls{S-RU}. The \gls{nFAPI} is pivotal in empowering {multi-vendor} interoperability, allowing a small cell \gls{CU}/\gls{DU} to connect seamlessly {with independently deployed} \glspl{S-RU} \cite{SCF225.3.0}. It also provides tools and integration support, including the \gls{SCF} \gls{DARTs} suite. 

{In collaboration} with \gls{O-RAN} Alliance and \gls{TIP}, \gls{SCF} {has contributed significantly to the advancement of} standardized testing processes across the industry, which includes active participation in plugfests \cite{SCF239.10.01}. {Beyond this,} \gls{SCF} {engages with other prominent organizations such as} \gls{3GPP}, \gls{OAI}, and a wide range of stakeholders across technical, commercial, and regulatory domains. The forum's goal is to accelerate open \gls{RAN} adoption across all domains, driving the widespread deployment of virtualized open \gls{RAN} infrastructure. {By promoting the convergence of} open systems, open source code, and shared spectrum, \gls{SCF} aims to enable a broader range of network deployers. This is particularly impactful in areas like enterprise and smart city environments, where small cells are indispensable \cite{SCF239.10.01}.

\vspace{-2.5mm}
\subsection{O-RAN-Related Open Source Projects}
\vspace{-1.5mm}
The software community {plays an essential role} in ensuring that software reference implementations {are closely} aligned with \gls{O-RAN} and its technical specifications. In this context, the \gls{OSC} undertakes various responsibilities, including the development and maintenance of open source software, fostering collaboration with other open source initiatives, and promoting related projects and activities {that contribute to the advancement of \gls{O-RAN}}. As of this writing, multiple open source platforms compliant with \gls{O-RAN} principles are {publicly available and actively utilized} by researchers and academic institutions. {These platforms provide a foundational environment for experimentation, prototyping and validation of \gls{O-RAN} functionalities. The subsequent section examines} key contributors and collaborators involved in the implementation and evolution of open source solutions in \gls{O-RAN}.

\subsubsection{ONAP}
The \gls{ONAP}, launched by \gls{LF} in 2017, is an open source platform for orchestrating, managing, and automating network and edge computing services. It addresses the needs of network operators, cloud providers, and enterprises {through} real-time, policy-driven orchestration and {lifecycle} automation of physical and virtual \glspl{NF}~\cite{adrian_kliks_towards_2023}. {\gls{ONAP} leverages} \gls{SDN} and \gls{NFV} technologies and {implements a} complete \gls{MANO} layer aligned with \gls{ETSI} \gls{NFV} architecture. {Beyond supporting} \gls{FCAPS} functionalities, \gls{ONAP} offers a {robust} framework for network service design.

\gls{ONAP} {collaborates with the} \gls{OSC}, particularly on deploying the \gls{SMO} and integrating the \gls{Non-RT RIC} functionalities \cite{9367572}. This partnership enhances coordination, minimizes duplication of efforts, and streamlines development. Shared priorities are outlined in \cite{OnapOranCommonArea}, including a joint study on the \enquote{ONAP/3GPP \& \gls{O-RAN} Alignment-Standards Defined Notifications over VES} use case, which seeks to align \gls{ONAP} with \gls{O-RAN} Alliance and \gls{3GPP}, fostering interoperability and broader adoption~\cite{Onap3gppOranAlignment}.
The Kohn release of \gls{ONAP} further advances this integration by enhancing cloud-native \gls{NF} orchestration, supporting intent-driven closed-loop automation, and \gls{E2E} network slicing across \gls{5G} \gls{RAN}, \gls{CN}, and \gls{TN} domains~\cite{OnapKuhn,OnapE2ENetworkSlicing}.

\subsubsection{OpenAirInterface}
The \gls{OSA} is a nonprofit organization founded by the French research institute EURECOM. It supports a global community of researchers and industry contributors in the development of open source software for the \gls{CN} and \gls{RAN} domains of a \gls{3GPP} cellular networks. The Alliance supports the advancement of the \gls{3GPP} \gls{5G} cellular stack, which is {maintained within} the \gls{OAI} software packages and are designed to operate on \gls{COTS} hardware. The \gls{OSA} {is responsible for} roadmap development, quality assurance, and {community engagement, including the promotion} of \gls{OAI} software packages. These packages are widely used by academic institutions and business communities for a broad range of use-cases. The goal of the \gls{OSA} is to accelerate the adoption of the \gls{OAI}.

In the context of \gls{5G}, the \gls{OAI} community and software assets {have been} expanding rapidly. Current active projects include: \gls{OAI} \gls{5G} \gls{RAN}, \gls{OAI} \gls{5G} CORE, \gls{M5G}, and \gls{CI}/\gls{CD}. The newly created \gls{M5G} \gls{PG} aims to transform both the \gls{RAN} and \gls{CN} into agile and open platform for network service delivery. The \gls{M5G} \gls{PG} {focuses on developing} software implementations of the \gls{O-RAN} E2 protocol, as well as FlexRIC (a flexible \gls{RIC}), FlexCN (a flexible core controller), and intelligent orchestration tools for \gls{RAN} and \gls{CN} domains \cite{mosaic-intro}. {Additionally, researchers at} Northeastern University have successfully integrated \gls{OAI} with the \gls{OSC} \gls{RIC}, enabling interoperability between open source development and standardized \gls{O-RAN} implementations. Further details are provided in \cite{KALTENBERGER2025,polese2024ColDT}.

\subsubsection{Open Networking Foundation}
\Gls{ONF} is a consortium led by several major network operators, playing a pivotal role in driving the transformation of network infrastructure {through open and disaggregated solutions. Among its key initiatives,} the \gls{SD-RAN} project contributes open source components to the open \gls{RAN} by developing and testing \gls{O-RAN} compliant network elements. {The project promotes} multi-vendor \gls{RAN} solutions and {demonstrates the potential of modular component integration} to foster further innovation in \gls{RAN} \cite{SD-RAN-whitepaper}.

In close collaboration with the \gls{O-RAN} Alliance and \gls{OSC}, {\gls{SD-RAN} aims} to develop open source components for the \gls{O-CU} \gls{CP}, \gls{O-CU} \gls{UP}, and \gls{O-DU} \cite{s23218792}. {A cornerstone of the platform is the} cloud-native uONOS-RIC (micro-ONOS-RIC), a fully functional \gls{Near-RT RIC}, which includes an \gls{xApp} development environment, and a set of reference \glspl{xApp} for managing open \gls{RAN} elements \cite{SD-RAN-whitepaper}.

Notably, Deutsche Telekom has deployed a fully disaggregated \gls{5G} field trial using the \gls{SD-RAN} platform, integrating components from more than eight vendors via uONOS-RIC. This deployment represents the first comprehensive realization of \gls{O-RAN}, encompassing \gls{O-RU}, \gls{O-DU}, \gls{O-CU}, \gls{RIC}, and multiple \glspl{xApp} sourced from various providers---marking a significant milestone in the evolution of open \gls{RAN}.

\subsubsection{srsRAN} 
The srsRAN project, developed by \gls{SRS}, is an open source \gls{RAN} solution that supports both \gls{4G} and \gls{5G} technologies. It features an \gls{O-RAN}-native \gls{gNB}, providing a comprehensive implementation of the {\gls{L1}/\gls{L2}/\gls{L3}} protocol stack with minimal {external} dependencies \cite{Upadhyaya2022PrototypingNO}. The solution adheres to standards defined by both \gls{3GPP} and the \gls{O-RAN} Alliance. It adopts the \gls{3GPP} \gls{5G} architecture, implementing functional splits between the \gls{DU} and \gls{CU}, with further separation into \gls{CU-CP} and \gls{CU-UP}. The srsRAN platform also supports integration with third-party \glspl{Near-RT RIC} and \glspl{xApp} through FlexRIC, with the ultimate goal of achieving full compliance with the E2 interface \cite{srsRAN-intro}.

The srsRAN offers deployment flexibility, allowing users to operate a monolithic \gls{gNB} on a single machine or distribute \gls{RAN} functions across multiple machines and geographic locations. It supports seamless integration with third-party \glspl{RIC}, \gls{PHY} solutions, and other \gls{O-RAN} compliant hardware and software components, making it well-suited for a wide range of use cases, and experimental scenarios.

\subsubsection{OpenRAN Gym}
OpenRAN Gym, led by Northeastern University, is a collaborative open source initiative designed for large-scale, data-driven experimental research within the open \gls{RAN} ecosystem \cite{bonati2023openrangympawr,openRANGymToolbox}. Its goal is to unite researchers from academia and industry in a cooperative environment to accelerate the development of intelligent and \gls{AI}-driven solutions for open \gls{RAN}. 

OpenRAN Gym builds on frameworks for data collection and \gls{RAN} control, which enables \gls{E2E} design and testing of data-driven \glspl{xApp} by providing an \gls{O-RAN} compliant \gls{Near-RT RIC} and E2 termination interface \cite{openRANGymToolbox}. This allows users to collect runtime data, prototype new strategies, and evaluate them in diverse wireless environments before transitioning them into production networks.

The architecture of OpenRAN Gym, as detailed in \cite{openRANGymToolbox,bonati2023openrangympawr}, consists of the following key components:
\begin{itemize}[noitemsep, topsep=0pt, left=0pt]
    \item Remotely accessible wireless testbeds, like Colosseum, Arena, and PAWR platforms, which support large-scale data collection and validation in real-world scenarios.
    \item A softwarized \gls{RAN} using open protocol stacks like srsRAN and \gls{OAI} to emulate cellular networks.
    \item A data collection and control framework, such as SCOPE, that extracts \glspl{KPI} and \glspl{KPM} and supports runtime \gls{RAN} control \cite{openRANGymToolbox}.
    \item An \gls{O-RAN} compliant control architecture, such as ColO-RAN, which connects to the \gls{RAN} through standardized E2 interface, receives runtime \glspl{KPM}, and coordinates intelligent control through \gls{AI}/\gls{ML}-based xApps and rApps.
\end{itemize}

\vspace{-2.5mm}
\subsection{Open Access Testbeds}
\vspace{-1.5mm}
In addition to the aforementioned open source projects, several experimental testbeds have been developed to support the implementation of softwarized \gls{5G} networks by leveraging a variety of open source components. The following subsections provide an overview of selected {open-access} testbeds.

\subsubsection{Colosseum} Colosseum is a publicly accessible, large-scale wireless testbed designed to support advanced experimental research through the use of virtualized and softwarized waveforms and protocol stacks, deployed on a fully programmable white-box platform. Equipped with 256 state-of-the-art \glspl{SDR} and a {powerful} channel emulator, Colosseum has the capability to simulate almost any scenario. This allows the comprehensive design, development, and {validation of} solutions at scale across various deployment topologies and channel conditions. It achieves high-fidelity reproduction of radio frequency scenarios through \gls{FPGA}-based emulation employing \gls{FIR} filters. These filters accurately model channel taps and apply them to \gls{SDR}-generated signals, enabling realistic replication of real-world radio environments---crucial for evaluating performance under controlled and repeatable conditions \cite{bonati2021colosseum,OranTestBedProfile}.

{Colosseum also serves as the foundational infrastructure} for OpenRAN Gym, which is tightly integrated within the testbed. This integration enables experimentation with \gls{E2E} \gls{O-RAN} compliant networks and services, facilitates data collection, and supports the development and validation of \gls{AI}/\gls{ML}-driven models, among other essential research activities \cite{openRANGymToolbox}.

\subsubsection{POWDER} \Gls{POWDER} is a city-scale testbed tailored to support a wide range of innovative research experiments. Its infrastructure includes an outdoor area incorporating multiple \glspl{SDR}-equipped nodes, an indoor laboratory for over-the-air experiments, and a wired attenuator matrix \cite{OranTestBedProfile}, {providing an environment for filed tests and controlled lab evaluations.}

The primary goal of \gls{POWDER} is to enable experimental research across heterogeneous wireless technologies, with a particular emphasis on \gls{5G} \gls{RAN} architecture, network orchestration, massive \gls{MIMO}, and many more. The platform includes integrated support for the rapid deployment of the \gls{O-RAN} architecture, facilitating the configuration and testing of advanced \gls{RAN} functionalities. {Researchers can efficiently prototype and evaluate} components such as the \gls{Near-RT RIC}, xApps, the \gls{O-CU} subsystem, and open source \gls{SMO} within \gls{POWDER} \cite{powder-intro,johnson22NexRAN}.

\subsubsection{COSMOS} The cloud enhanced open software-defined mobile wireless (COSMOS) is a city-Scale testbed, deployed as a component of the \gls{POWDER} initiative. The COSMOS is aims to create, develop, and operate an advanced, city-scale wireless testbed that facilitates real-world experimentation with next-generation wireless technologies and applications. It has been certified by the \gls{O-RAN} Alliance as an \gls{OTIC} \cite{OranTestBedProfile,cosmos3380891}. The COSMOS architecture prioritizes ultra-high bandwidth and low-latency wireless communication, tightly integrated with edge cloud computing. The testbed consists of approximately 40-50 advanced \gls{SDR} nodes, interconnected via fiber-optic \gls{FH} and \gls{BH} \glspl{TN}, alongside dedicated edge and core cloud computing infrastructure. Researchers can access COSMOS remotely through a web-based portal, which provides comprehensive tools for experiment orchestration, real-time measurements, and data collection \cite{cosmos3380891}.

\subsubsection{Arena} Arena stands as an innovative open-access wireless testing platform designed to research in sub-6 GHz \gls{5G} and beyond spectrum. Located in an indoor office environment, the testbed is anchored by a grid of ceiling-mounted antennas, each connected to programmable \glspl{SDR}, enabling real-time, scalable, and reproducible experimentation \cite{bertizzolo2020arena}. The platform integrates 12 high-performance computational servers, 24 symbol-level synchronized \glspl{SDR}, and a total of 64 antennas, providing a unique combination of processing capabilities and spatial diversity. This architecture makes Arena particularly well-suited for exploring technologies in dense spectrum environment \cite{bertizzolo2020arena}. Arena operates on a three-tier physical design: servers and \glspl{SDR} are housed in a dedicated room, while antennas are strategically distributed across the office ceiling and connected via 100-foot long \gls{RF} cables. This layout ensures minimal interference, precise control over experimental parameters, and a realistic representation of indoor wireless propagation characteristics \cite{bertizzolo2020arena,OranTestBedProfile}.

\subsubsection{X5G}
X5G is a pioneering private \gls{5G} network testbed at Northeastern University, Boston. It integrates open source and programmable components across the entire network stack---from the \gls{PHY} to the \gls{CN}. Notably, it stands as the first fully programmable multi-vendor and \gls{O-RAN} compliant testbed of its kind. Developed through a collaborative endeavor involving Northeastern University, NVIDIA, and \gls{OAI} \cite{villa2024x5g}. The testbed leverage NVIDIA \glspl{GPU} to accelerate \gls{L1} (\gls{PHY} operations), while \gls{L2} and \gls{L3} are implemented using the \gls{OAI} software stack. This integration is based on the \gls{SCF} \gls{FAPI} for seamless interaction between the \gls{MAC} and \gls{PHY} layers. This integration enables the inline hardware acceleration of computationally demanding \gls{PHY} tasks on the \gls{GPU}, fostering scalability and facilitating the integration of \gls{AI}/\gls{ML} within the \gls{RAN}. The NVIDIA \gls{ARC} platform operates on a specialized multi-vendor infrastructure comprising eight servers for \gls{CU} and \gls{DU}, along with four \glspl{RU} designed for lab installations. It also incorporates \gls{O-RAN} 7.2x \gls{FH}, precise timing hardware, and a dedicated \gls{5G} \gls{CN}. {By combining the performance advantages of} NVIDIA \gls{ARC} with \gls{OAI}, X5G provides flexible, high-performance environment for exploring open, programmable, and intelligent wireless systems---paving the way for future advancements in \gls{5G}, \gls{6G}, and beyond \cite{villa2024x5g}.

\subsubsection{5GENESIS} The 5GENESIS initiative, funded by the European Union (EU), aims to validate \gls{5G} \glspl{KPI} across a broad spectrum of applications, ranging from controlled laboratory environments to large-scale public events. The project builds upon the collective outcomes of multiple EU projects along with internal research and development contributions from its consortium partners, to establish a unified, \gls{E2E} \gls{5G} infrastructure spanning five test platforms across Europe. Each platform within 5GENESIS is characterized by distinct capabilities and specialized features. However, they are engineered for interoperability within a cohesive architecture, forming a flexible and distributed testing facility. This infrastructure enables comprehensive \glspl{KPI} validation, supports targeted demonstrations, and facilitates the evaluation of critical \gls{5G} and beyond use cases, including diverse deployment scenarios. The trials conducted within the 5GENESIS are primarily focused on assessing and confirming the \glspl{KPI} defined by \gls{5G-PPP}. These evaluations inherently serve to benchmark the performance, scalability and readiness of each individual platform and the collective infrastructure as a whole \cite{koumaras5genesis}.

\subsubsection{Insights from Open-Source O-RAN Deployments}
Building on the open-source platforms and testbeds introduced earlier, this section synthesizes representative findings from recent evaluations of open {5G} deployments.

Our study \cite{TechRxiv25} compares the deployment of commercial and open-source \glspl{O-gNB} in an industrial environment. The findings indicate that the commercial \glspl{gNB} offer better coverage, while the open-source \glspl{gNB} achieve lower and more consistent latency. Despite being less energy-efficient, the open-source \glspl{gNB} provide greater flexibility, easier configuration, and better maintainability, supported by a vibrant developer community. In contrast, the commercial \glspl{gNB} operate within a closed system, offering limited scope for customization.

The work in \cite{10816630} emphasizes how component configurations and interdependency significantly influence \gls{E2E} performance. Through over-the-air measurements of packet loss and one-way delay under concurrent uplink and downlink transmissions, the authors show that open-source \gls{O-RAN}-based systems can, under certain conditions, match or surpass conventional solutions in performance, while offering industrial-grade connectivity at lower cost. Interoperability is evaluated in \cite{10571340}, where various open-source \gls{RAN} and \gls{CN} combinations are tested, providing critical insights into integration and performance trade-offs.

A performance evaluation presented in \cite{10439170} highlights the suitability of the srsRAN platform for private \gls{5G} network deployments. The study also identifies key limitations, emphasizing that the full potential of \gls{5G} remains unrealized—primarily due to constraints associated with general-purpose hardware and the current maturity level of open-source software components. The study \cite{10827366} employs an \gls{OAI}-based testbed to evaluate \gls{O-RAN} performance, with a focus on throughput measurements. The findings expose limitations of the setup and suggest potential underlying causes.

The authors in \cite{11038873} present a standalone testbed to evaluate downlink performance across two \gls{O-RAN} implementations: srsRAN and \gls{OAI}. The results indicate that functional split architectures enhance flexibility and resource efficiency by offloading processing to the \gls{O-DU}. In \cite{10826795}, the authors propose a cloud-native \gls{O-RAN} testbed that supports dynamic deployment and \gls{O-DU} scaling. When \gls{CPU} load exceeds 80\%, an additional \gls{O-DU} is instantiated, improving throughput and latency under heavy traffic and demonstrating efficient network resource management.

The demonstration in \cite{10741499} presents a modular, cloud-native \gls{5G} \gls{O-RAN}, deployed across multiple locations. It enables flexible resource allocation and supports on-demand monitoring applications, facilitating autonomous, and feedback-driven orchestration. The study in \cite{11028861} evaluates resource scheduling strategies within an \gls{O-RAN}-compliant \gls{5G} network using the ns-3 simulator. By comparing different scheduling mechanisms, the authors demonstrate that the integration of \gls{RIC}-driven control enhances flexibility and adaptability in scheduling policies.

Further advancing practical implementation, the work in \cite{11028899} provides a comprehensive guide for developing and deploying \gls{O-RAN} applications in both simulated and real-world environments. It equips developers with methodologies for architectural evaluation, \gls{xApp} migration to testbeds, and deployment of key components such as the \gls{SMO}, \gls{Non-RT RIC}, and \glspl{rApp} to enable full \gls{E2E} integration.

\vspace{-2.5mm}
\section{In-depth Analysis of Slicing-aware O-RAN Architecture}\label{sec:Architecture}
\vspace{-1.5mm}
The \gls{O-RAN} architecture is built on a foundation of open and standardized interfaces, protocols, transport links, \glspl{NF}, and \glspl{MF}. By employing open interfaces and open source software, \gls{O-RAN} separates the control and user plane, enabling a modular and flexible software stack \cite{tripathi2025fundamentals}. Furthermore, the \gls{O-RAN} Alliance {proposes the integration of the} \gls{AI}/\gls{ML} {capabilities} to enhance automation and operational efficiency \cite{cst-gov-sa-oran}. In this section, we explore the latest \gls{O-RAN} architecture in a detailed manner, with a particular emphasis on the features that support network slicing. We describe its key components and interfaces, and review network slicing related \glspl{MF} as defined by \gls{3GPP}, \gls{ETSI}, and \gls{ONAP}. Moreover, we provide an overview of the underlying infrastructure, including the \gls{O-Cloud} sites, the open \gls{TN}, and open cellular network site.

\vspace{-2.5mm}
\subsection{Major Components of the O-RAN Architecture}
\begin{figure*}[!htbp]
    \centering
    \includegraphics[width=\textwidth]{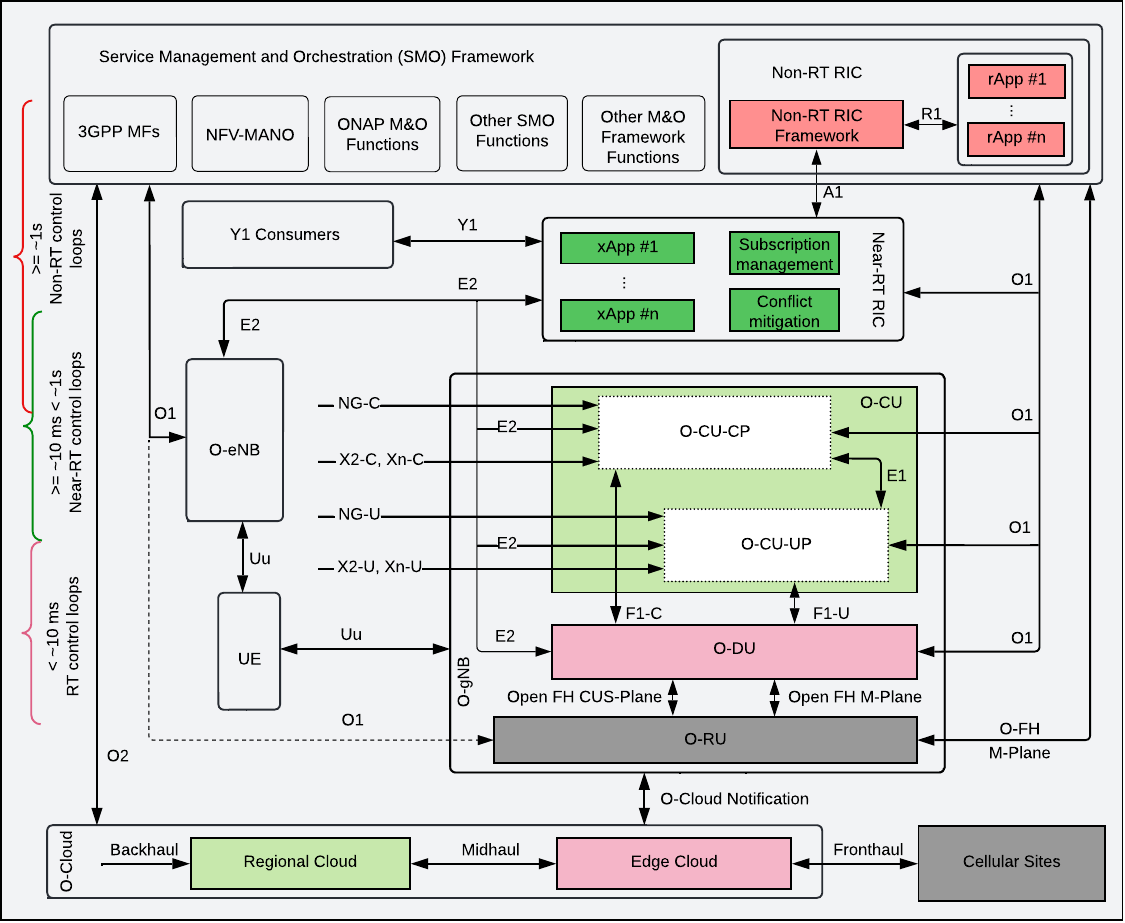}
    \caption{{The latest} O-RAN slicing-aware architecture}
    \label{fig:RefArchitecture}
    \vspace{-6mm}
\end{figure*}

\vspace{-1.5mm}
The \gls{O-RAN} architecture {follows a} disaggregated {design} paradigm, {dividing the} cellular base station into multiple logical and physical units, {each} responsible for {specific layers} and interfaces of the radio network protocol stack \cite{tripathi2025fundamentals}. The \gls{O-RAN} architecture, as illustrated in Figure~\ref{fig:RefArchitecture}, consists of four major components: the \gls{O-gNB}, the \glspl{RIC}, the \gls{SMO} framework, and the underlying infrastructure.

The \gls{O-gNB} encompasses the radio functionalities, including tasks such as modulation, coding, resource scheduling, and many others in both uplink and downlink directions. A \gls{RIC} is a software-defined component within the \gls{O-RAN} architecture responsible for the control and optimization of \gls{O-RAN} functions \cite{9627832,NGO2024680}. The \gls{SMO} framework serves as an automation platform dedicated for the \gls{MO} of \gls{O-RAN} \glspl{NF}, radio resources and network slices, supporting lifecycle management of \gls{O-RAN} at scale {in an intelligent and autonomous manner} \cite{TowardsAIMLORAN,ericssonSmo}. The underlying infrastructure is responsible for hosting the \gls{O-RAN} components and includes \gls{O-Cloud} sites, cellular network sites, and transport links.

\vspace{-2.5mm}
\subsection{O-gNB (E2 Nodes) and its Corresponding Interfaces}
\vspace{-1.5mm}
{In the \gls{O-RAN} architecture, the traditional \gls{gNB} is systematically disaggregated into distinct logical entities. As} illustrated in Figure~\ref{fig:RefArchitecture}, these nodes include the \gls{O-CU}, the \gls{O-DU}, and the \gls{O-RU}, or a combined \gls{O-eNB}. Out of these nodes, the \gls{O-CU} and \gls{O-DU} are collectively referred to as E2 nodes in \gls{O-RAN} Alliance terminology \cite{Marcin21248173}. {Each E2 node is associated with specific functionalities and standardized open interfaces. This modular decomposition underpins the architectural principles of \gls{O-RAN} and enables an interoperable next-generation \gls{RAN} ecosystem.} Below, we provide a detailed analysis of each node, together with an in-depth description of its corresponding open interface specifications.

\subsubsection{O-RAN Centralized Unit}
The \gls{O-CU} is a logical {network} node responsible for implementing the higher layer protocols of the \gls{RAN} stack. {These} include the \gls{RRC} layer, which controls the life cycle of radio connections; the \gls{SDAP} layer, which manages the \gls{QoS} of traffic flows {within individual} bearers; and the \gls{PDCP} layer, which handles {essential functions such as} packet reordering, duplication, and encryption {over} the air interface \cite{9750106,Motalleb2019JointPA}. As shown in Figure~\ref{fig:RefArchitecture}, the \gls{O-CU} terminates the E2 interface towards the \gls{Near-RT RIC} and the O1 interface towards the \gls{SMO} framework \cite{O-RAN.WG1.OAD}. {Architecturally, it comprises a single} \gls{O-CU-CP} and potentially multiple \glspl{O-CU-UP}, which communicate {over the standardized} E1 interface \cite{10329947}. {The \gls{O-CU-CP} and \gls{O-CU-UP} interface with the \gls{O-DU} through the F1-C and F1-U components of the standardized F1 interface, respectively.}

According to \gls{3GPP} specifications, the \gls{O-CU} {is required to} support functionalities associated with network slicing. The \gls{O-CU-UP} may be deployed either as a {dedicated entity} for each network slice or shared among multiple slices, depending on the specific requirements and design of each slice \cite{9750106,O-RAN.WG1.SA}. 
The \gls{O-RAN} architecture {further enhances these capabilities by} extending network slicing functionalities beyond those defined in \gls{3GPP}. {This is achieved through} dynamic slice optimization {mechanisms facilitated} by the \gls{Near-RT RIC} via the E2 interface \cite{10279730}. 
Additionally, the O1 interface supports the configuration of {extended slice related} parameters to further improve the capabilities of \gls{O-RAN} architecture.
The \gls{O-CU} is also expected to perform slice-aware resource allocation and {enforce} isolation mechanisms {to ensure compliance with slice specific \glspl{SLA}}. 
The \gls{O-CU} is initially configured through the O1 interface based on the requirements of individual slices and is subsequently reconfigured dynamically by the \gls{Near-RT RIC} through the E2 interface {to support evolving} slicing use cases. 
The \gls{O-CU} may be required to generate and send certain \glspl{PM} over both the O1 and E2 interfaces in response to requests from the \gls{SMO} framework and the \gls{Near-RT RIC}. These \glspl{PM} {serve as critical inputs} for slice performance monitoring and \gls{SLA} assurance \cite{O-RAN.WG1.SA,10329597}.

\paragraph{E1 Interface}
The E1 interface functions as a control interface connecting the \gls{O-CU-CP} and \gls{O-CU-UP} entities within an \gls{O-gNB} \cite{LARSEN2024110292}. It is standardized by the \gls{3GPP} and plays a pivotal role in \gls{O-RAN} {by enabling} efficient {coordination between control and user plane components of \gls{O-CU}}. The {adoption} of this standardized interface not only ensures the efficiency of \gls{O-RAN} but also offers flexibility and scalability for future {network evolution and innovation} \cite{3GPP-TS-38460E1Interface}.

\subsubsection{O-RAN Distributed Unit}
The \gls{O-DU} is a logical network node that hosts the lower layer protocols of the \gls{RAN} stack and serves as a baseband processing unit that handles the high \gls{PHY}, \gls{MAC}, and \gls{RLC} layers \cite{10200260}. {It is typically deployed as} a \gls{VNF} that can be hosted within a virtual machine or container at the edge cloud \cite{9750106}. {The \gls{O-DU}} terminates {multiple critical interfaces, including} the E2 {interface (towards the \gls{Near-RT RIC})}, F1 {interface (towards the \gls{O-CU})}, and the \gls{O-FH} interface {(towards the \gls{O-RU})}. Additionally, it terminates the O1 interface towards the \gls{SMO} framework to enable the management and {orchestration functionalities} \cite{O-RAN.WG1.OAD,O-RAN.WG1.SA}. {It also serves as the aggregation point} for multiple \glspl{O-RU}, terminating the \gls{O-FH} M-Plane interface to facilitate hierarchical or hybrid management of \gls{O-RU} within the \gls{O-RAN} architecture.

{In the context of network slicing,} the \gls{O-DU} {plays a key role} to enable the slice-aware resource management. The \gls{MAC} layer {is responsible} to allocate and isolate \glspl{PRB} per network slice according to the configuration {received through the O1 interface}, along with \gls{O-CU} directives over the F1 interface, and dynamic guidance received from the \gls{Near-RT RIC} through the E2 interface \cite{10335921,10225994}. Similar to the \glspl{O-CU}, the \glspl{O-DU} must also generate and report \glspl{PM} through both the O1 and E2 interfaces, {in response} to requests from the \gls{SMO} framework and the \gls{Near-RT RIC}, respectively. These \gls{PM} can be used for network slice performance monitoring and \gls{SLA} assurance \cite{O-RAN.WG1.SA}.

\subsubsection{O-RAN Radio Unit}
The \gls{O-RU} is a physical node that implements the lower \gls{PHY} and \gls{RF} processing functions within an \gls{O-gNB}, based on the lower layer functional split---specifically, the 7.2x split option \cite{9627832,oranJie22}. It serves as the termination point for the \gls{O-FH} interface, {as well as} the lower \gls{PHY} functionalities interfacing with \glspl{UE}. Additionally, the \gls{O-RU} terminates the \gls{O-FH} M-Plane interface, which connects to the \gls{O-DU} and/or the \gls{SMO} framework, depending on the specific deployment scenario. In \gls{O-RAN}, a single \gls{O-RU} is expected to support multiple slice instances {to enable radio-level resource sharing} \cite{O-RAN.WG1.Use-Cases}.

\subsubsection{O-eNB}
The \gls{O-RAN} architecture also supports the integration of \gls{LTE} base stations, referred to as \glspl{O-eNB} in \gls{O-RAN} Alliance terminology. An \gls{O-eNB} may take the form of either a {legacy} \gls{eNB} or a \gls{ng-eNB}. 
To ensure compatibility within the \gls{O-RAN} ecosystem, the associated interfaces and protocols required by these base stations---particulary the E2 and O1 interfaces---must be fully supported \cite{O-RAN.WG1.OAD}.

\subsubsection{E2 Interface}
The E2 interface {establishes the logical} connection between the \gls{Near-RT RIC} and E2 nodes. It supports two categories of functions: \gls{RIC} services and E2 support services. \gls{RIC} services---namely, Report, Insert, Control, Policy, and Query---are enabled through functional procedures, including subscription management, control operations, and information queries. E2 support services facilitate interface and \gls{RAN} function management through global procedures such as E2 setup, E2 reset, E2 removal, E2 node configuration updates, \gls{RIC} service updates, and reporting of general error situations \cite{oranJie22,O-RAN.WG3.E2GAP,O-RAN.WG3.E2SM}.

An E2 node comprises a logical E2 agent that terminates the E2 interface and facilitates communication between \gls{RIC} services and \gls{RAN} functions. The \glspl{xApp} in \gls{Near-RT RIC} deliver value-added services leveraging \gls{RIC} functional procedures over the E2 interface \cite{O-RAN.WG3.RICARCH,O-RAN.WG3.E2GAP}. Furthermore, the E2 interface facilitates the collection of measurements from the \gls{RAN} to the \gls{Near-RT RIC}, either periodically or {based on predefined} trigger events. {It supports control and data collection across multiple network dimensions,} including individual cells, slices, \gls{QoS} classes, and specific \glspl{UE} \cite{O-RAN.WG3.E2GAP}.

Slice-aware \glspl{xApp} utilize the E2 interface to influence the behavior of E2 nodes in a slice {specific manner} \cite{tripathi2025fundamentals}. This includes the configuration of \gls{RRM}, \gls{MAC} scheduling policies, and other control parameters {across the} \gls{O-RAN} protocol stacks \cite{10335921}. Moreover, the \gls{Near-RT RIC} employs the E2 interface to configure and collect slice-specific performance indicators and reports from E2 nodes, supporting real-time monitoring and closed-loop optimization of network slice performance \cite{MarcinOranSlicing,O-RAN.WG1.SA}.

\subsubsection{F1 Interface}
The F1 interface connects the \gls{O-CU} and \gls{O-DU} within an \gls{O-gNB} \cite{LARSEN2024110292}. {In} \gls{O-RAN}, it follows the protocol {architecture and specifications} defined by the \gls{3GPP}, complemented by \gls{O-RAN}-defined interoperability profile specification. Mirroring the \gls{O-CU}'s division into control and user planes, the F1 interface consists of two components: F1-C and F1-U \cite{O-RAN.WG1.OAD}.

\paragraph{F1-C} The F1-C interface handles control and signaling functions between the \gls{O-CU-CP} and the \gls{O-DU}. It facilitates the exchange of control plane information such as connection setup and release, handover management, and radio resource coordination.

\paragraph{F1-U} The F1-U interface is responsible for the actual user data transmission between the \gls{O-CU-UP} and the \gls{O-DU}. {It ensures} efficient and reliable transport of user plane data {across the} \gls{O-RAN} {architecture}.

\subsubsection{X2 Interface}
The \gls{O-RAN} adopts the X2 interface from \gls{3GPP} standards {to support} interoperability profile specifications. It connects the \gls{O-CU} with other \glspl{eNB} in an \gls{EN-DC} configuration. {The interface is divided into} X2-C and X2-U, which handle control plane and user plane information, respectively \cite{O-RAN.WG1.OAD}.

\subsubsection{Xn Interface}
The \gls{O-RAN} adopts the principles and protocol stack of the Xn interface as defined in \gls{3GPP} standards, to support interoperability profile specifications. The Xn interface contains two components Xn-C and Xn-U, which connect the \gls{O-CU-CP} and \gls{O-CU-UP}, respectively, to other \glspl{O-gNB} within O-RAN architecture \cite{O-RAN.WG1.OAD}.

\subsubsection{NG Interface}
The NG interface, adopted from \gls{3GPP}, connects the \gls{O-CU} to the \gls{5GC}. It comprises two components: NG-C for the control plane and NG-U for the user planes. NG-C connects the \gls{O-CU-CP} with the \gls{AMF}, while NG-U connects the \gls{O-CU-UP} to the \gls{UPF} \cite{O-RAN.WG1.OAD}.

\subsubsection{Uu Interface}
The \gls{3GPP} defines the interface between the \gls{UE} and the e/\gls{gNB} as the Uu interface. The Uu interface encompasses a comprehensive protocol stack spanning from layer 1 to layer 3 and terminates within the \gls{NG-RAN} architecture. {In the decomposed} \gls{NG-RAN}, {protocol terminations occur} at distinct reference points, none explicitly defined by the \gls{O-RAN} Alliance. As Uu messages continue to traverse from the \gls{UE} to the {appropriate} e/\gls{gNB} managed function, the \gls{O-RAN} architecture does not define it as a distinct interface towards a specific managed function \cite{3GPP-TS-38.401,O-RAN.WG1.OAD}.

\subsubsection{Open Fronthaul Interface}
The \gls{O-RAN} \gls{FH} Specification defines the disaggregation and virtualization of the traditional cellular network site to enhance the efficiency of \gls{FH} \gls{TN} in next-generation telecommunication networks. {It adopts the \gls{3GPP} 7.2x functional split option in the physical layer, partitioning} it into high-PHY and low-PHY \cite{O-RAM-WP-2020}. The low-PHY functions reside in \gls{O-RU}, while high-PHY processing is implemented at the \gls{O-DU} \cite{ericssonWPDORAN,ag_bundled_2023}. 
{The intra-PHY lower layer \gls{FH} split imposes stringent bandwidth and latency requirements, necessitating a dedicated \gls{FH} service profile for the \gls{FH} \gls{TN} \cite{O-RAM-WP-2020,10711198}. The service profile must be adaptable to diverse deployment scenarios, network topologies, and specific use case requirements. Its framework and latency model align with the reference points outlined in the \gls{eCPRI} specification.}
The Open \gls{FH} interface connects the \gls{O-DU} and \gls{O-RU} via the CUS-Plane{---comprising control, user, and synchronization---}and M-Plane, {which supports \gls{MO}} operations \cite{O-RAN.WG4.M-Plane,O-RAN.WG4.CUS-Plane}. {Furthermore, it supports centralized control of the \gls{O-RU} by the \gls{O-DU} and allows, in certain configurations, a single \gls{O-DU} to manage multiple \glspl{O-RU}, including those operating across different carrier networks \cite{O-RAN.FH.Spec}.}

\paragraph{C-Plane}
The control plane {(C-Plane) in the \gls{O-RAN} architecture} refers to the real-time control {signaling exchanged} between the \gls{O-DU} and \gls{O-RU}. C-Plane messages convey data-associated control information essential for user data processing, including scheduling and beamforming \cite{O-RAN.FH.Spec} {instructions---particularly when} such information is not provided via the M-Plane. These messages are transmitted independently for both downlink and uplink {directions}. To enhance flexibility, C-Plane messages can be {transmitted either} collectively or individually, depending on the associated channel {and specific transmission requirements} \cite{O-RAN.WG4.CUS-Plane}.

\paragraph{U-Plane}
The user plane {(U-Plane)} handles the transmission of \gls{IQ} sample data between the \gls{O-DU} and \gls{O-RU} \cite{10711198}. To ensure {proper} coordination with the C-Plane, the \gls{FH} interface mandates that C-Plane messages arrive at the \gls{O-RU} prior to the latest permissible time for the corresponding U-Plane messages. U-Plane data is encapsulated in two-layer header structure: the first layer contains an \gls{eCPRI} or IEEE 1914.3 header indicating the message type, while the second layer constitutes an application specific header with fields necessary for control and synchronization \cite{O-RAN.WG4.CUS-Plane,ag_bundled_2023}.

\paragraph{S-Plane}
The synchronization plane {(S-Plane) ensures coordination} between the \gls{O-RU}, \gls{O-DU} and a synchronization controller, typically an IEEE 1588 Grand Master, {which} may be integrated into the \gls{O-DU}. The \gls{O-RAN} {supports \gls{E2E}} synchronization of frequency, phase, and time across all relevant network elements---including \glspl{O-DU}, intermediate switches, and \glspl{O-RU}{---to meet the requirements of} both the \gls{TDD} and \gls{FDD} operations \cite{O-RAN.WG4.CUS-Plane}.

\paragraph{M-Plane}
The management plane {(M-Plane) in \gls{O-RAN}} handles non-real-time management operations between the \gls{O-DU} and the \gls{O-RU}. Depending on the \gls{TN} topology, various connectivity models may exist between the \gls{O-RU}, \gls{O-DU}, and the \gls{SMO} framework \cite{O-RAN.WG4.M-Plane,O-RAN.FH.Spec}. The primary requirement of the M-Plane is to ensure \gls{E2E} connectivity between the \gls{O-RU} and entities responsible for its management, which may include the \gls{O-DU}, the \gls{SMO}, or {designated} \gls{O-RU} controllers \cite{O-RAN.WG4.M-Plane}. 

The Open \gls{FH} M-Plane utilizes a NETCONF/YANG-based {framework} to manage the \gls{O-RU}, supporting functions such as installation, configuration, software update, performance monitoring, fault, and file management. Two architectural models are defined: First, the hierarchical model, where one or more \glspl{O-DU} manage the \gls{O-RU} via a NETCONF interface. Second, the hybrid model, which allows additional direct logical interfaces between the \gls{SMO} framework and the \gls{O-RU}, alongside the existing \gls{O-DU}--\gls{O-RU} link \cite{O-RAN.WG4.M-Plane,O-RAN.FH.Spec}. In the hybrid model, the \gls{O-RU} can establish \gls{E2E} connectivity with the \gls{SMO} either directly or through the \gls{O-DU}. Notably, there is no explicit signaling {mechanism to indicate whether a} hierarchical or hybrid {model is in use.} All NETCONF servers compliant with the M-Plane specification must support multiple {concurrent} sessions, and all the \glspl{O-RU} {expected to be compatible with} both architectural models \cite{O-RAN.WG4.M-Plane}.

\vspace{-2.5mm}
\subsection{RAN Intelligent Controller}
\vspace{-1.5mm}
The \gls{RIC} {represents a significant advancement within} the \gls{O-RAN} architecture, {introducing a centralized abstraction layer that facilitates greater control and flexibility in \gls{RAN} operations \cite{10271688,10617647,10872859}. As an integral architectural innovation, the \gls{RIC} empowers \glspl{MNO} to design and deploy custom control plane functionalities, thereby enhancing the agility, efficiency and programmability of the \gls{RAN} \cite{10617647}}. It manifests as a software-defined \gls{NF} designed to manage specific control functionalities, such as mobility management and \gls{RRM}, which have traditionally been confined to the base stations. 

By offering real-time visibility and {centralized} control over \gls{RAN} resources, the \gls{RIC} plays a pivotal role in advancing the \gls{O-RAN} disaggregation {paradigm \cite{maxenti2025autoran}. It enables critical capabilities including} multi-vendor interoperability, intelligent {decision making, and dynamic resource allocation---}that collectively redefine the operational landscape of \gls{O-RAN} \cite{10271688,10617647,10872859,3700838,brown_ric_2020,NGO2024680}. Its integration into the \gls{O-RAN} architecture facilitates intelligent \gls{MO}, particularly in the implementation of key concepts such as network slicing \cite{aws_open_2022}.

Moreover, the \gls{RIC} is responsible for configuring network slices, orchestrating \glspl{NF}, monitoring network performance, and conducting real-time optimizations of \gls{RAN} resources via open interfaces \cite{brown_ric_2020,oran_risk_analysis}. As illustrated in Figure~\ref{fig:RefArchitecture}, the \gls{RIC} {is implemented} in two distinct forms---\gls{Non-RT RIC} and \gls{Near-RT RIC}---each designed to address specific control loop dynamics and latency requirements \cite{ag_bundled_2023}. The following sections provide a detailed examination of both {variants, elaborating on} their functionalities, applications, and {strategic importance} in the broader context of \gls{RAN} optimization and {automation}.

\subsubsection{Near-Real-Time RAN Intelligent Controller}
The \gls{Near-RT RIC} serves as a logical entity {in \gls{O-RAN}}, facilitating precise and close-to-real-time control and optimization of E2 nodes and {their} resources \cite{10877797}. {It is positioned in} close proximity to the E2 nodes and interacts directly with them to enhance performance by leveraging continuous data collection and executing real-time control actions via the E2 interface \cite{brik2023survey,aws_open_2022}. Operating within a control loop {bounded by latencies between} 10 milliseconds and 1 second, {it ensures timely responsiveness to network dynamics} \cite{O-RAN.WG3.RICARCH}.

The \gls{Near-RT RIC} serves as a software platform for hosting xApps{---modular,} microservice-based applications that are intelligent, autonomous, {and tailored for specific control functions} \cite{10355063,elyasixapp}. These \glspl{xApp} are deployed to the \gls{Near-RT RIC} as needed to offer targeted functionalities, such as intelligent \gls{RRM} \cite{9277891,10741499,gavrilovska_cloud_2020}. During the onboarding process, xApps {define their data dependencies, including the types of} data they collect, process, consume, and expose \cite{O-RAN.WG1.OAD,10355063}. The integration of xApps within the \gls{Near-RT RIC} enables dynamic management and optimization of \gls{RAN} resources, effectively addressing the diverse service requirements of modern cellular networks \cite{10178010,10741499}. By leveraging \gls{UE} and cell-specific metrics collected through the E2 interface, xApps facilitate real-time optimization of \gls{RAN} resources and functionalities \cite{elyasixapp}. This capability ensures efficient resource utilization and contributes to improved user experience in high-demand network environments \cite{10329915,Marcin21248173}.

Furthermore, the \gls{Near-RT RIC} gains direct control over E2 nodes and their resources by policies and information delivered through A1 interface from the \gls{Non-RT RIC} \cite{O-RAN.WG3.RICARCH}. In specific scenarios, it is authorized to monitor, suspend, override, or control E2 nodes and their resources on rules defined within the E2 service model \cite{O-RAN.WG3.E2SM,O-RAN.WG1.OAD}. {Additionally, the \gls{Near-RT RIC} exposes E2-related \glspl{API} that support access to E2 functions, xApp subscription management, and conflict mitigation mechanisms} \cite{O-RAN.WG3.RICARCH,O-RAN.WG3.E2GAP}.

A critical function of the \gls{Near-RT RIC} lies in network slicing, where it supports near-real-time optimization of \gls{RAN} slice subnet by coordinating with \gls{O-CU} and \gls{O-DU} components through the E2 interface. To accomplish this, xApps must be slice-aware, allowing them to implement algorithms to meet each slice \gls{SLA} \cite{O-RAN.WG1.SA}. Addressing this challenge, a growing body of literature presents various \glspl{xApp} designed for slice management and optimization tasks in \gls{O-RAN} \cite{10329915,9814869}.

The xApps employ \gls{AI}/\gls{ML}-based models, guided by A1 policies generated by the \gls{Non-RT RIC} \cite{10103771,9754560}, to make intelligent context-aware decisions. Once a slice is active slice-specific \glspl{PM} are collected from E2 nodes and integrated with slice configuration data at the \gls{Near-RT RIC}. It facilitates dynamic and automated slice optimization. The collaborative intelligence between the \gls{Non-RT RIC} and \gls{Near-RT RIC}, powered by \gls{AI}/\gls{ML} framework, plays a pivotal role in maximizing the overall efficiency and scalability of network slicing within the \gls{O-RAN} architecture \cite{s24031038,O-RAN.WG3.RICARCH}.

\paragraph{Y1 Interface}
The \gls{Near-RT RIC} provides \gls{RAN} analytics information services to an authorized third party, known as Y1 consumer, via the Y1 service interface \cite{O-RAN.WG3.Y1GAP}. Access to these services is granted following mutual authentication and authorization. Within a \gls{PLMN} trusted domain, Y1 consumers can subscribe to or request {analytics data through this} interface. Entities outside the \gls{PLMN} trust domain may also access Y1 services securely via a standardized exposure function. Notably, Y1 consumers are external entities and are not represented as logical \gls{O-RAN} functions within the architecture, as shown in Figure~\ref{fig:RefArchitecture} \cite{O-RAN.WG1.OAD}.

\subsubsection{Non-Real-Time RAN Intelligent Controller}
The \gls{Non-RT RIC} is a core component of \gls{O-RAN} architecture, responsible for non-real-time management and optimization of the \gls{RAN} components and underlying resources \cite{Marcin21248173}. As an integral part of the \gls{SMO} framework, it communicates with the \gls{Near-RT RIC} over the A1 interface and orchestrates \gls{AI}/\gls{ML} workflows \cite{9681936}. As illustrated in Figure~\ref{fig:RefArchitecture}, it also facilitates the execution of third-party applications known as \glspl{rApp}. These modular applications {leverage the} R1 interface of the \gls{Non-RT RIC} framework to deliver value-added services such as managing policy-driven suggestions via the A1 interface and enabling control actions for potential implementations through the O1 and O2 interfaces \cite{O-RAN.WG1.OAD,ag_deutsche_2023}.

The \gls{Non-RT RIC} facilitates the autonomous configuration of \gls{O-RAN} components, minimizing the need for manual operator intervention. It provides \glspl{MNO} with insights into network behavior and supports high-level optimization strategies \cite{brik2023survey}. It performs data analytics and \gls{AI}/\gls{ML} model training leveraging \gls{SMO} provided services such as data collection and provisioning from \gls{RAN} nodes \cite{MarcinOranSlicing,9681936}. Once trained, these models are distributed to the \gls{Near-RT RIC} for real-time inference and execution. 

The \gls{Non-RT RIC} plays a critical role in network slicing, providing advanced orchestration capabilities. It collects slice-specific \gls{PM}, configuration parameters, and optional internal metrics from \glspl{DC} to support \gls{AI}/\gls{ML} driven optimization \cite{10353004}. Trained \gls{AI}/\gls{ML} models enable non-real-time optimization of slice-specific parameters over the O1 interface addressing complex tasks such as \gls{RRM} policy enforcement \cite{Marcin21248173,9277891}.

The gathered data is also forwarded to the \gls{Near-RT RIC}, which utilizes it for dynamic slice optimization to mitigate potential \gls{SLA} violations across network slice instances \cite{O-RAN.WG1.StdSA}. 
While the \gls{Near-RT RIC} controls the E2 resources via the E2 interface, the \gls{Non-RT RIC} manages the cloud resources through the O2 interface, with decisions informed by the collected analytics and trained \gls{AI}/\gls{ML} models \cite{O-RAN.WG2.Non-RT-RIC-ARCH}.

\paragraph{R1 Interface}
The R1 interface located within the internal structure of the \gls{Non-RT RIC}, provides access to the framework services that empower rApps retrieve data for initiating intelligent policy generation and \gls{RAN} optimization \cite{OranMvpWhitepaper}. It also supports authorized enrichment data exchange with the \gls{Near-RT RIC} and allows rApps to share services and analytics within the \gls{Non-RT RIC} framework \cite{5GamericasUpdate,O-RAN.WG2.Non-RT-RIC-ARCH}.

\subsubsection{A1 Interface}
The A1 interface establishes communication between the \gls{Non-RT RIC} and the \gls{Near-RT RIC} \cite{OranMvpWhitepaper,O-RAN.WG1.OAD}. It enables the \gls{Non-RT RIC} to provide policy guidance, known as A1 policies, to the \gls{Near-RT RIC} \cite{10439167}. These policies support functions such as provisioning directives for an individual or groups of \glspl{UE}, monitoring policy states, providing feedback, and exchanging enrichment information required for \gls{RIC} operation \cite{O-RAN.WG2.A1GAP,10439167}. The A1 interface also facilitates coordination for \gls{AI}/\gls{ML} workflows, including model training, distribution, and inference.

{In network slicing scenarios, A1 services are essential for} \gls{SLA} assurance. For example, the \gls{Non-RT RIC} can use A1-based policy management to transmit slice-specific policies that guide the \gls{Near-RT RIC} in resource allocation and slice-specific control actions, while also receiving feedback on policy compliance and performance \cite{O-RAN.WG1.SA}.

\vspace{-2.5mm}
\subsection{The SMO Framework and its Corresponding Interfaces}
\vspace{-1.5mm}
\gls{O-RAN} is designed to deliver flexibility, reliability, scalability, and interoperability across multi-vendor environments. It operates on \gls{COTS} hardware within a cloud-native, virtualized infrastructure, {and relies on} automation, and \gls{AI}/\gls{ML} {to support intelligent and adaptive network management} \cite{10506767,9562627}. The \gls{SMO} framework {forms the foundation of autonomous and intelligent} \gls{MO} within \gls{O-RAN} \cite{10617647}. As shown in Figure~\ref{fig:RefArchitecture}, it integrates {a suit of} \glspl{MF} and services tailored to \gls{O-RAN}. The \gls{SMO} incorporates management {capabilities that are defined by multiple \glspl{SDO} and ensures} interoperability among their \glspl{MF} through standardized service-based management interfaces \cite{BuildingSMOFramework}. Operating on a \gls{SBA}, the \gls{SMO} framework {enables seamless} provisioning and consumption of key services, including authentication, authorization, service registration and discovery, data management, and trained model distribution sharing \cite{O-RAN.WG1.OAD,O-RAN.WG1.D-SMO}.

The \gls{SMO} manages \gls{FCAPS} operations through the O1 interface, facilitates intelligent \gls{RRM} via the \gls{Non-RT RIC}, and orchestrates \gls{O-Cloud} infrastructure. {It should support \gls{O-Cloud} orchestration by integrating \gls{VNF} orchestration and \gls{FOCOM} via the O2 interface \cite{O-RAN.WG1.D-SMO}. It also handles} workload management and resource provisioning. The \gls{SMO} communicates with the \gls{O-RU} for \gls{FCAPS}-related functions via the \gls{O-FH} M-Plane interface \cite{O-RAN.WG1.SA}. The A1 interface connects the \gls{Non-RT RIC} and \gls{Near-RT RIC}, {enabling the former} to collect data, train or select \gls{ML} models, and transmit them to the {latter for real-time execution}.

The architecture of \gls{SMO}---particularly the \gls{Non-RT RIC}---offers flexible implementation {options, enabling the} operators to selectively adopt the desired features. The \gls{SMO} can integrate with an \gls{E2E} multi-domain service orchestrator, coordinating domain-specific modules across the \gls{RAN}, \gls{TN}, and \gls{CN}. It facilitates on-demand creation and management of \glspl{NSI} over distributed \gls{5G} infrastructure \cite{O-RAN.WG1.D-SMO}.

{To support network slicing,} the \gls{SMO} {must adhere to} the architectural requirements {defined by} \gls{3GPP}, \gls{ETSI}, and \gls{ONAP}, {utilizing standardized \glspl{MF} that align with} their respective specifications. These \glspl{MF} of perform tasks such as slice creation, operation, modification, termination as well as scaling the underlying resources. The \gls{O-RAN} Alliance maintains alignment with network slicing concepts and architectural principles established by the aforementioned \glspl{SDO} \cite{10604823} while extending them with general guidelines tailored to \gls{O-RAN}'s modular and open architecture. For example, \gls{O-RAN} mandates compatibility with \gls{3GPP} interface specifications \cite{10872859}, standardized management service interfaces for slicing operations, support for multi-vendor interoperability, {flexibility} in deployment options, and support for multi-operator slice subnet management \cite{OranMvpWhitepaper,O-RAN.WG1.SA}.

\subsubsection{O1 Interface}
The \gls{O-RAN} managed elements and the management entities within the \gls{SMO} framework are logically connected through the O1 interface, as shown in Figure~\ref{fig:RefArchitecture}. The O1 interface {facilitates essential} operations and management tasks for \gls{O-RAN} components, including \gls{FCAPS}, software and file management, among many others. The key components such as the \gls{O-CU}, \gls{O-DU}, and \gls{Near-RT RIC} are managed through this interface, enabling the \gls{SMO} to access {and control relevant} \gls{O-RAN} \glspl{NF} \cite{O-RAN.WG1.O1-Interface}.

In the context of \gls{O-RAN} slicing, the O1 interface supports the configuration of \gls{O-RAN} nodes with slice-specific parameters tailored to the service requirements of individual network slices. The \gls{3GPP} defines a slice-specific information model, which includes \gls{RRM} policy attributes{---for example, the distribution} of \glspl{PRB} across slices \cite{9204709}. These models can be extended to include slice profiles and additional configuration parameters to support \gls{O-RAN} slicing use cases over the O1 interface. {Moreover, the interface enables} the collection of slice-specific performance metrics and fault reports from \gls{O-RAN} nodes {, supporting \gls{SLA} assurance and closed-loop automation within the slicing framework} \cite{O-RAN.WG1.SA}.

\subsubsection{O2 Interface}
The O2 interface is an open, logical interface that facilitates secure communication between the \gls{SMO} framework and distributed \gls{O-Cloud} sites \cite{O-RAN-WG6-CADS}. It supports the lifecycle management of \glspl{VNF} that operates on the \gls{O-Cloud} infrastructure. Within the \gls{O-RAN}, the \gls{O-Cloud} hosts essential \glspl{NF}, while the O2 interface allows the \gls{SMO} to coordinate \gls{O-Cloud} infrastructure management and deployment activities.

In addition to lifecycle management of \gls{O-Cloud} infrastructure, the O2 interface {supports} the orchestration of resource management such as inventory, monitoring, provisioning, and software management. It provides logical services that govern the deployment of \gls{O-RAN} \glspl{NF} on cloud resources. 
The O2 interface is inherently extensible, allowing new features to be incorporated without altering the underlying protocol or management processes. It is vendor-neutral and is unaffected by particular \gls{SMO} framework and \gls{O-Cloud} implementations. Through the O1 and/or O2 interfaces, operators can dynamically manage, reconfigure, and upgrade network components hosted within the \gls{O-Cloud} environment \cite{O-RAN.WG6.O2-GAnP}.

\subsubsection{3GPP Network Slicing MFs within the SMO Framework} 
\begin{figure}[!tp]
    \centering
    \includegraphics[width=\columnwidth]{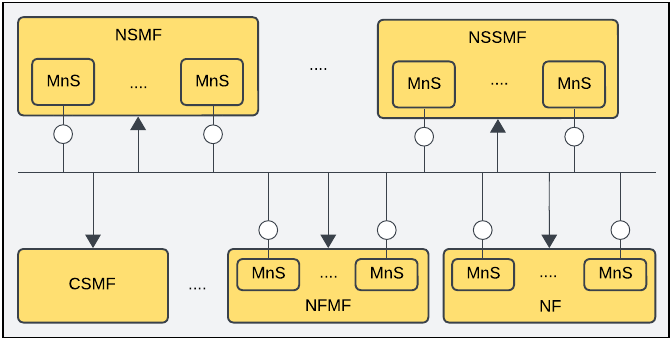}
    \caption{3GPP management architecture within the SMO framework}
    \label{fig:Ref3GPPManagementFunctions}
\end{figure}

The \gls{3GPP}-defined \glspl{MF} for the \gls{MO} of network slicing {include} the \gls{CSMF}, \gls{NSMF}, \gls{NSSMF}, and \gls{NFMF} \cite{3GPP-TR-28801MgmtOrch}. These \glspl{MF} can be {integrated} within the \gls{SMO} framework in accordance with the requirements outlined in the \gls{3GPP} specifications. Additionally, the \gls{O-RAN} Alliance has defined both functional and non-functional requirements for network slicing architecture, as detailed in \cite{O-RAN.WG1.SA}. The requirements specified by both \gls{3GPP} and the \gls{O-RAN} Alliance are critical for the successful realization of network slicing within the \gls{O-RAN} architecture and for ensuring the effective operation of the \glspl{MF}. The provision of \glspl{MnS} for mobile networks---including network slicing---can be achieved through a set of functional blocks, as illustrated in Figure~\ref{fig:Ref3GPPManagementFunctions}.

\paragraph{\glsreset{CSMF}\Gls{CSMF}} It is responsible for translating communication service requirements---received from third parties such as \gls{OSS}/\gls{BSS}, \gls{NSaaS} tenants, and other {external entities}---into network slicing requirements for an \gls{E2E} network slice \cite{3GPP-TS-28.533,Juniper-WP-NS}.

\paragraph{\glsreset{NSMF}\Gls{NSMF}} The \gls{NSMF} manages an \gls{E2E} network slice absed on the requirements determined by the \gls{CSMF}. It is responsible for coordinating and managing the necessary resources to support the associated communication services and interfaces with the \glspl{NSSMF} responsible to manage individual network slice subnets within that domain (e.g., RAN NSSMF)\cite{3GPP-TS-28.533,Juniper-WP-NS}.

\paragraph{\glsreset{NSSMF}\Gls{NSSMF}} The \gls{NSSMF} operates within individual network domains (i.e., \gls{RAN}, \gls{TN}, and \gls{CN}) and is responsible to instantiate the required resources based on the instructions provided by the \gls{NSMF}. Each \gls{NSSMF} orchestrates the domain-specific resources within its respective subnet to fulfill the service requirements assigned to that domain. For example, the \gls{RAN} \gls{NSSMF} orchestrates the \gls{RAN} segment of a network slice by performing life-cycle management, configuration management, performance monitoring, and fault management. Additionally, the \gls{RAN} \gls{NSSMF} interfaces with the \gls{RIC} to execute control plane operations related to \gls{O-RAN} slices within the \gls{O-RAN} architecture \cite{3GPP-TS-28.533,Juniper-WP-NS}.

\paragraph{\glsreset{NFMF}\Gls{NFMF}} The \gls{NFMF} offers \gls{NF} management services within the \gls{NF} management framework. It is responsible for managing multiple \glspl{NF}, including application-level management of both \gls{VNF} and \gls{PNF} \cite{BuildingSMOFramework}. Additionally, the \gls{NFMF} produces \gls{NF} provisioning services, which includes configuration, fault, and performance management. {At the same time}, it also {acts as a} consumer of the \gls{NF} provisioning service exposed by individual \glspl{VNF} and \glspl{PNF} \cite{3GPP-TS-28.533}.

To provide comprehensive \gls{MO} solutions, the \gls{SMO} framework can be customized to include either all or a subset of the aforementioned \gls{3GPP}-defined \glspl{MF}. The {selection and integration} of these \glspl{MF} depend on deployment-specific considerations, which are further discuss in Section \ref{sec:DeploymentOptions}.

In addition, the \gls{SMO} framework can {be extended to} incorporate management functions from both the \gls{NFV-MANO} and \gls{ONAP} \cite{Rehman-NFV}. Within this context, the \gls{NFV-MANO} is responsible for the management and orchestration of \glspl{VNF} as well as the virtualized resources associated with an \gls{O-RAN} network slice. Readers seeking detailed architectural information are encouraged to refer to the \cite{ETSI-GR-NFV-IFA-046,ETSI-GS-NFV-006-R4}.

Furthermore, \gls{ONAP} is currently being integrated by the \gls{OSC}, with selected \gls{SMO} functionalities leveraging and extending components of \gls{ONAP}’s existing infrastructure \cite{NGO2024680}. This integration enables faster automation of new services and comprehensive lifecycle management{---capabilities that} are essential for \gls{5G} and next-generation network---through real-time, policy-driven orchestration and automation of both \glspl{PNF} and \glspl{VNF} \cite{OnapArchitectureWhitepaper}. Further details regarding the roles of \gls{NFV-MANO} and \gls{ONAP}, along with their deployment scenarios within the \gls{SMO} framework, are provided in Section~\ref{sec:DeploymentOptions}.

\vspace{-2.5mm}
\subsection{The Underlying O-Cloud and O-Transport Infrastructure}
\vspace{-1.5mm}
The underlying infrastructure of \gls{O-RAN} comprises the \gls{O-Cloud} sites (which include the regional cloud and edge cloud sites), the cellular network sites, and the \glspl{TN}. The regional and edge cloud sites along with cellular network sites, {provide necessary hosting environment for} essential \gls{O-RAN} \glspl{NF}. The \gls{TN} ensures connectivity between various virtual or physical \gls{NF} deployed across cellular and/or cloud sites. In this section, we provide a concise overview of these key infrastructure components within the \gls{O-RAN} architecture.

\subsubsection{Cellular Network Site}
A cellular network site refers to the physical locations where \glspl{O-RU} are deployed, which may be mounted on the same structure as the antenna or situated at the base of the installation. Typically, a cellular site is engineered to support multiple sectors, enabling the deployment of several \glspl{O-RU} {within a single} site. They facilitate the exchange of user data, control plane signaling, and synchronization information between the \gls{O-RU} and the \gls{O-DU}. The distribution of cellular sites may follow uniform or non-uniform pattern, {depending on factors} such as user density, geographical terrain, and network topology. {Based on their coverage area and transmit power,} cellular sites are typically categorized as Macro, Micro, Pico, and Nano types. 

\subsubsection{Cloud Site}
A cloud site refers to a physical location equipped with cloud infrastructure resources, suitable for \glspl{O-Cloud}, and possibly accommodating other non-\gls{O-Cloud} resources. Within the \gls{O-RAN} architecture, \gls{O-Cloud}s are strategically deployed at both regional and edge cloud sites. These sites serve as centralized platforms for hosting \glspl{VNF}, \gls{SDN} controllers, and other cloud-native applications. The regional cloud sites typically provide broader geographical coverage and greater computational capacity, while the edge cloud sites bring compute and storage resources closer to the network edge. It enables low-latency and high-bandwidth support for a wide range of services and applications.

\paragraph{Edge Cloud}
Edge cloud refers to a site that hosts virtualized \gls{RAN} functions, {particularly those required to support multiple nearby} cellular network sites. {It provides} centralized processing capabilities for these sites, {enabling efficient coordination and management}. The physical coverage of an edge cloud can vary depending on the operator’s deployment strategy and use case---it may serve either broad region or a more localized area. Regardless of scale, the edge cloud must maintain sufficient proximity to the associated \glspl{O-RU} to satisfy the stringent latency requirements of \gls{O-DU} functions. This proximity ensures low-latency communication between \glspl{O-RU} and \glspl{O-DU}, thereby enabling efficient network operations and {timely} service delivery.

\paragraph{Regional Cloud}
A regional cloud refers to a cloud site that supports virtualized \gls{RAN} functions for a large number of cellular network sites, typically spanning multiple edge clouds. It enables a higher degree of centralization by hosting the key functions, including \gls{O-CU} and the \gls{Near-RT RIC}. To fulfill the latency requirements of hosted functions, the regional cloud site must be located sufficiently close to the associated \glspl{O-DU}. This proximity ensures that communication between the \gls{O-CU}, \gls{O-DU}, and \gls{Near-RT RIC} remains within acceptable latency bounds. It enables effective coordination and optimization of \gls{RAN} resources across a wide geographical area, while maintaining {high level of performance, scalability,} and responsiveness.

\subsubsection{O-RAN Cloud Platform}
The \glsreset{O-Cloud} \gls{O-Cloud} refers to a cloud computing environment composed of physical network infrastructures. It supports the deployment of critical \gls{O-RAN} components such as the \gls{SMO}, \gls{Near-RT RIC}, and E2 nodes, along with the associated software and required \gls{MO} services. Each \gls{O-Cloud} is composed of a group of \glspl{CPU}, \gls{RAM}, storage, \glspl{NIC}, \gls{BIOS}, \glspl{BMC}, and hardware accelerators. {These elements work together to} handle computationally intensive tasks across the platform \cite{O-RAN.WG1.OAD}.

Depending on the deployment scenario, the \gls{O-Cloud} can virtualize a variety of \glspl{NF} and take {on the execution of} \gls{RAN} functionalities within the overall \gls{O-RAN} architecture. A more detailed discussion on these aspects can be found in Section~\ref{sec:UnderlyingInfrastructure}.

\subsubsection{O-Cloud Notification API}
The \gls{O-Cloud} notification \gls{API} facilitates event subscription for consumers such as the \gls{O-DU}, {which operates} within the \gls{O-Cloud} environment. This {\gls{API} allows} connsumers to subscribe to receive event notifications and status updates from the \gls{O-Cloud}. Additionally, the \gls{O-Cloud} exposes event producers, making it possible for cloud workloads to access relevant notifications and statuses that would otherwise {remain internal to the platform} \cite{5GamericasUpdate,O-RAN.WG1.OAD}.

\subsubsection{Transport Network}
In disaggregated \gls{O-RAN} deployments, components such as the \gls{O-CU} and \gls{O-DU} may be deployed on {separate, geopgraphically} distributed \gls{O-Cloud} sites. To {support seamless} communication among \gls{O-CU}, \gls{O-DU} and \gls{O-RU}, a robust networking infrastructure must {interconnect these elements} across the cellular network site and distributed \gls{O-Cloud} sites through open and highly reliable \glspl{TN} \cite{Han2022ORAN}.

The \gls{TN} encompasses multiple segments, including \gls{FH}, \gls{MH}, and \gls{BH}. It supports both the \gls{NR} and legacy technologies such as \gls{LTE} and \gls{UMTS}. The transport services span the C-Plane, U-Plane, S-Plane, and M-Plane, and are designed to support the operational requirements of diverse operators and support various \gls{E2E} services, including \gls{URLLC} and \gls{eMBB}.

The \gls{TN} must be highly flexible to support various use cases, applications, and heterogeneous \gls{RAN} architecture. Each segment of the physical \gls{TN} may need to concurrently support multiple network slice instances, distinct \gls{5G} services, and various \gls{3GPP} interfaces, tailored to specific {deployment scenarios and performance} requirements.

\paragraph{Fronthaul}
In \gls{O-RAN}, \gls{FH} {refers to the communication link} between the \gls{O-DU} and \gls{O-RU} within the \gls{RAN} infrastructure. It encompasses control, user, synchronization, and management planes. To meet the stringent latency requirements associated with \gls{FH}, \glspl{O-RU} and their corresponding \glspl{O-DU} are deployed in close physical proximity \cite{keysight-OranFH-TN}.

\paragraph{Midhaul}
The \gls{MH} network is a logical segment of the \gls{TN} that facilitates communication between the \gls{O-DU} and \gls{O-CU}, enabling the transport of \gls{3GPP} F1/W1/E1 interfaces \cite{O-RAN.WG1.OAD}. When the \gls{O-DU} and \gls{O-CU} {are deployed} as a unified entity, these interfaces {are internal and not exposed, effectively eliminating the need for a distinct} \gls{MH} segment. It also provides inter-\gls{O-CU} communication, specifically supporting the transport of the \gls{3GPP} Xn interface. In deployments where \glspl{MNO} have nott adopted a split architecture between the \gls{O-DU} and \gls{O-CU}, these interface functions are instead handled within the \gls{BH} network.

\paragraph{Backhaul}
In \gls{O-RAN} architecture, the \gls{BH} network connects the \gls{O-CU} to the \gls{5G} \gls{CN}. It {comprises both} \gls{CP} and \gls{UP} components to ensure a clear separation between user data and control {signaling defined by} \gls{3GPP}. The \gls{CP} includes multipoint interfaces such as N1, N2, N4 and Xn-c, which facilitates communication between the \gls{O-CU-CP}, the \gls{UPF} and other \gls{5G} \gls{CN} components. The \gls{UP} includes interfaces such as N3 (between \gls{O-CU-UP} and \gls{UPF}), N9 (between \gls{UPF} instances), and Xn-u (between \gls{O-CU-UP} nodes), supporting efficient transmission of user data.

\vspace{-2.5mm}
\section{O-RAN NFs, Network Slicing, and SMO Deployment Options} \label{sec:DeploymentOptions}
\vspace{-1.5mm}
The \gls{O-RAN} \glspl{NF} can be implemented as \glspl{VNF} or \glspl{PNF}. {They} can be hosted by the underlying \glspl{O-Cloud} {infrastructure} and cellular sites, {respectively}. Regardless of whether the \glspl{NF} are virtualized or physical, they must be {strategically allocated to suitable} hosts within the \gls{O-RAN} infrastructure. This mapping {process plays a key role} in implementing the logical network functionalities of an \gls{O-RAN} slice, particularly in cloud computing environments, {where resource optimization and performance considerations are critical}.

Deployments can {vary} from fully distributed to {highly} centralized configurations depending on {the placement of \glspl{NF} on} edge and regional \gls{O-Cloud} sites---referred to as \gls{PoP} in \gls{ETSI} terminology. The decision entails determining the optimal execution location for each logical function, with potential impacts on performance, scalability, cost, and other crucial factors \cite{Cisco5GOranDeployment}. In this regard, the \gls{O-RAN} Alliance has introduced the \gls{O-Cloud} architecture and outlined several deployment scenarios for \gls{O-RAN} \glspl{NF} within the cloud-native framework \cite{O-RAN-WG6-CADS}. In addition, the document highlights numerous considerations essential for the effective deployment of logical \glspl{NF} across multiple \glspl{O-Cloud} environment. The diverse slicing and \gls{NF} deployment options within \gls{O-RAN} require a range of \gls{MO} solutions, leading to the multiple deployment options for the \gls{SMO} framework.

In the following subsections, we {explore various} deployment alternatives for \gls{O-RAN} \glspl{NF}, aligning them with the underlying infrastructure. Additionally, we analyzed different network slicing deployment strategies within \gls{O-RAN}. Furthermore, we {highlight multiple deployment options for} the \gls{SMO} framework, emphasizing the significance of network slice \glspl{MF}.

\vspace{-2.5mm}
\subsection{O-RAN NF Deployment Scenario}
\vspace{-1.5mm}
The \gls{O-RAN} Alliance has {explored various approaches} for virtualizing the \gls{O-RAN} \glspl{NF} in regional and edge clouds proposing different deployment scenarios {aligned with} \gls{O-RAN} specifications. These scenarios {are characterized} by {the specific} grouping of functionalities {across} key locations such as cellular sites, edge clouds, and regional clouds \cite{AndresOran}. {Additionally, they are distinguished} by whether the functionality at a particular location {is implemented} by an \gls{O-RAN} \gls{PNF}-based solution---where software and hardware are tightly integrated and share a {unified} identity---or through cloud-based services.

\begin{figure}[!htp]
    \centering
    \includegraphics[width=\columnwidth]{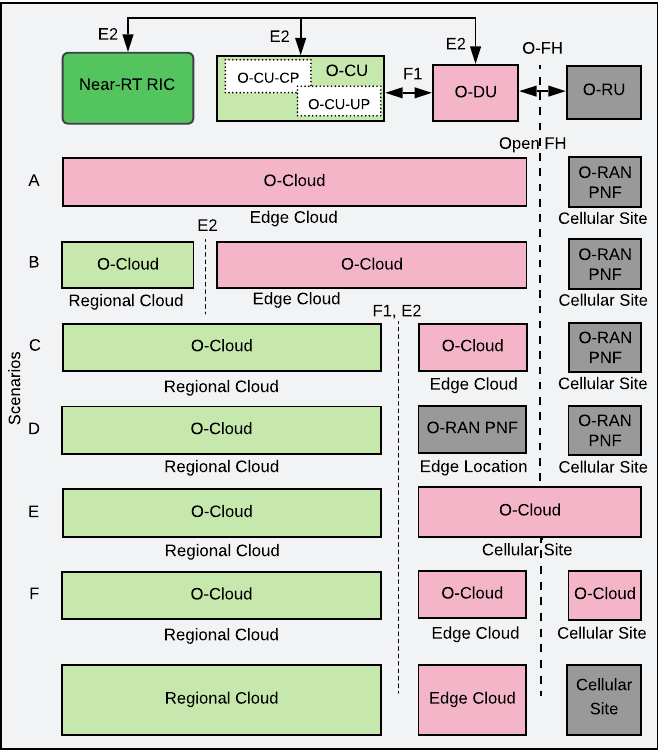}
    \caption{O-RAN NFs deployment scenarios onto the underlying O-Cloud sites and cellular network sites}
    \label{fig:RefDeploymentScenarios}
\end{figure}

Figure~\ref{fig:RefDeploymentScenarios} illustrates several \gls{NF} deployment scenarios as presented in \cite{O-RAN-WG6-CADS,dryjanski_o-ran_2023}. At the top, it provides an overview of the \glspl{NF}, while each scenario exhibits how these \glspl{NF} are deployed---either as cloudified \glspl{NF} within \gls{O-Cloud} or as \glspl{PNF} at the cellular site. {A detailed explanation of each deployment scenario is provided in the following}.

\paragraph{Scenario A} In this scenario, the \gls{Near-RT RIC}, \gls{O-CU}, and \gls{O-DU} are deployed at the edge cloud as \glspl{VNF}, whereas the \glspl{O-RU} are deployed on cellular network sites. This scenario is ideal for dense urban deployments with ample \gls{FH} capacity, enabling the pooling of \gls{BBU} functionalities at a central location. It reduces latency but comes with potentially higher deployment costs compared to other scenarios.

\paragraph{Scenario B} In this deployment scenario, the \gls{O-CU} and \gls{O-DU} are deployed at the edge cloud site in order to reduce latency, while the \gls{Near-RT RIC} is deployed at the regional cloud site in order to gain a wider network perspective for performance optimization {and improvement}.

\paragraph{Scenario C} In this deployment scenario, the \gls{O-CU} is co-located with the \gls{Near-RT RIC} in the regional cloud site, and the \gls{O-DU} is positioned at the edge cloud site. This scenario is tailored to support deployments in areas with limited remote Open \gls{FH} capacity, imposing restrictions on the number of \glspl{O-RU}. Two additional variations, C.1 and C.2, have been introduced to address the specific requirements of certain network slice instances \cite{O-RAN-WG6-CADS,AndresOran}.

\paragraph{Scenario D} This deployment scenario is akin to Scenario C (see above). However, the \gls{O-DU} is deployed as \gls{PNF} at the edge \gls{O-Cloud} site in this scenario. 

\paragraph{Scenario E} This deployment scenario mirrors Scenario D, with the key distinction that all components, including both \gls{O-DU} and \gls{O-RU}, are fully virtualized within the same edge cloud site. This approach is being considered for future use, acknowledging that the virtualized versions of the low-PHY layer and other \gls{O-RU} aspects are not currently available.

\paragraph{Scenario F} This deployment scenario involves the virtualization of both \gls{O-DU} and \gls{O-RU}, but they are hosted on separate \gls{O-Cloud} sites. Like Scenario E, this scenario is also considered for future use for the similar reason.

Within the context of \gls{O-RAN} deployment, {as discussed in the above scenarios,} the Open \gls{FH} plays a pivotal role {in the definition of the interface} between \glspl{VNF} deployed within the \gls{O-Cloud} and the cellular sites. \glspl{O-RU} are always located at the cellular site, while \gls{O-DU} can {reside} at the edge cloud site. {To meet strict latency requirements, \gls{O-DU} placement can be adjusted} closer to the cellular site; however, extending it farther from the cellular sites may violate \gls{RAN} internal or \gls{RAN} service-specific timing constraints \cite{dryjanski_o-ran_2023}.

A common deployment scenario involves moving \gls{O-DU} instances toward or even {directly} to the cellulari site {alongside} \gls{O-RU}, particularly when the edge cloud site must be closer to the cellular site due to fiber availability or other constraints. However, such adjustments may compromise the benefits of centralization and {resource} pooling \cite{ag_bundled_2023}.

The placement of \gls{O-CU} and its associated \gls{UPF} is determined by the latency requirements of the F1 interface or specific service constraints. For example, \gls{O-CU-UP} and \gls{UPF} for \gls{URLLC} services must remain at the edge cloud site. whereas for \gls{eMBB}, deployment at regional cloud site is feasible. Additionally, for services without specific latency requirements, \gls{O-CU-UP} and \gls{UPF} can be placed in the core cloud site \cite{O-RAN-WG6-CADS}. Centralizing \gls{O-DU} is particularly beneficial in densely populated networks where multiple cellular sites remain within the latency limits between \gls{O-RU} and \gls{O-DU}. Conversely, in sparsely populated areas, centralizing only the \gls{O-CU} is often more practical {in O-RAN deployment}.

\vspace{-2.5mm}
\subsection{Network Slicing Deployment Options in O-RAN}
\vspace{-1.5mm}
{Network slicing is a foundational concept in next-generation network architectures, centered around the creation of logical} \gls{E2E} virtual connections between end users or vertical customers and their {target} applications and services \cite{9079548,101145}. This is achieved {through the strategic allocation of} network resources to ensure that {each} service or application {receives the necessary support to satisfy its} specific \gls{QoS} requirements and meet predefined \gls{SLA}, thereby enabling reliable and differentiated service delivery \cite{10375939,3GPP-TS-23.501,3GPP-TS-28.530,park_technology_2023}.

The architecture {of network slicing} is structured into three distinct layers: the \gls{IL}, \gls{NFL}, and \gls{SL} \cite{IdrisE2eNetworkSlicing,8985329}.
\subsubsection{Infrastructure Layer} The \gls{IL} encompasses the entire physical network infrastructure, comprising \gls{RAN}, \gls{CN}, and \gls{TN} components. It is responsible for the deployment, control, and management of network infrastructure, as well as allocation of computing, storage, network, and radio resources to network slices. Additionally, it manages how these resources are {exposed to upper architectural} layers.
\subsubsection{Network Function Layer} The \gls{NFL} manages the lifecycle of \glspl{NF} including both physical and virtual. These functions are {deployed over} virtualized infrastructure and interconnected to deliver \gls{E2E} service that adheres to specific constraints defined in the service design of a network slice.
\subsubsection{Service Layer} The \gls{SL} focuses on service definition and orchestration. It maps service descriptions onto the underlying network infrastructure and encompasses the functional design of slice management and orchestration entities. The \gls{SL} plays a pivotal role in {translating service demands into actionable configurations for lower layers, to ensure} efficient and scalable network slicing operations \cite{thiruvasagam2023Oran}.

The deployment of slicing within \gls{O-RAN} is {facilitated} by decoupling software and hardware components integrated with \gls{NFV} \cite{park_technology_2023}. Determining the allocation of specific logical functions to {appropriate} \gls{O-Cloud} platforms, and identifying which functions should be co-located, is essential for the implementation of slicing in \gls{O-RAN} architecture \cite{MarcinOranSlicing}. Certain \gls{O-RAN} components such as the \gls{Near-RT RIC}, \gls{O-CU-CP}, \gls{O-DU}, and \gls{O-RU}, {are designed to }be shared across multiple network slices. {In contrast} the \gls{O-CU-UP} is {typically} dedicated to individual slices {to ensure isolation and performance}.

\begin{figure}[!htp]
    \centering
    \includegraphics[width=\columnwidth]{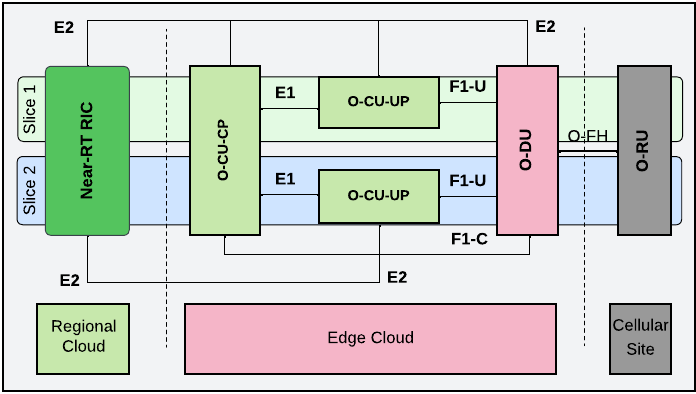}
    \caption{O-RAN reference slicing deployment option}
    \label{fig:RefO-RANSlicingDeployment}
\end{figure}

Moreover, the strategy for mapping \gls{NF} to {either shared or separate} cloud platforms must {align with the specific service requirements and deployment constraints of each use case} \cite{O-RAN.WG1.StdSA}. 
One of the potential deployment models proposed by the \gls{O-RAN} Alliance for slicing is illustrated in Figure~\ref{fig:RefO-RANSlicingDeployment}, where the \gls{O-RU} is deployed  as a \gls{PNF} at a cellular site. The \gls{Near-RT RIC} {is virtualized at a regional cloud site, while both} the \gls{O-CU} and \gls{O-DU} are virtualized on a location-independent edge cloud platform. The \gls{O-CU}/\gls{O-DU} are connected to the \gls{Near-RT RIC} with the E2 interface, and the \gls{O-CU} communicates with the \gls{O-DU} through the F1 interface \cite{O-RAN.WG1.SA}. 

The deployment scenario depicted in Figure~\ref{fig:RefO-RANSlicingDeployment} {can be realized through multiple configurations} by virtualizing {the shared components} such as \gls{Near-RT RIC}, \gls{O-CU}, and \gls{O-DU} across the regional and edge cloud platform as shown in Figure~\ref{fig:RefDeploymentScenarios}. For example, instead of sharing a common \gls{O-DU} for all slice instances, a dedicated \gls{O-DU} may be created for {each} slice instance in the scenario illustrated in Figure~\ref{fig:RefO-RANSlicingDeployment}.

It is important to {recognize that while the} requirements for \glspl{PNF}, cloudified network services, and \gls{O-Cloud} platforms may vary. However, the logical network function {requirements} always remain the same \cite{O-RAN.WG1.SA,O-RAN-WG6-CADS}. For Example, in the scenario illustrated, a single \gls{O-CU-CP} instance {manages} the control {plane operations for} both network slices, whereas each slice {is assigned a dedicated} \gls{O-CU-UP} instance. If the \gls{UE} is connected to both slices, only one \gls{RRC} connection {is established,} handling handover and cell assignments through the shared \gls{O-CU-CP}. However, each service {associated with} a different \gls{NSI} can {benefits} personalized \gls{QoS} management and independent flow control through an individual SDAP/PDCP stacks within {their respective} \gls{O-CU-UP} \cite{O-RAN.WG1.SA}.

{For the management and orchestration of slicing,} the \gls{O-RAN} slicing-aware architecture leverages the \gls{SMO} framework, {which incorporates a dedicated} slice \gls{MF} block. {This block} contains \gls{3GPP}-defined \gls{NSMF}, \gls{NSSMF}, and \gls{NFMF}. {Additionally, it may integrate} additional \glspl{MF} specified by the \gls{ETSI} \gls{ISG} \gls{NFV} or {derived} from the \gls{ONAP}. The following section {provides a comprehensive analysis of} different deployment options for the \gls{SMO} framework in the context of network slicing management.

\vspace{-2.5mm}
\subsection{SMO Framework Deployment Options}
\vspace{-1.5mm}
As discussed in Section~\ref{sec:Architecture}, the \gls{SMO} framework is responsible for the \gls{MO} of \gls{O-RAN} components and resources. The \gls{SMO} framework {may comprise} management components and systems {developed by} various \glspl{SDO}. To date, several \gls{SMO} {solutions emerged} in the market, claiming {compliance} with the latest specifications of the \gls{O-RAN} Alliance. However, {these frameworks often} lack transparency, as their internal architectural and operational mechanisms are not publicly {disclosed}, thereby constraining comprehensive insight into their functional capabilities \cite{10024837}. 

To promote openness and interoperability, the \gls{O-RAN} Alliance has proposed two open source solutions: the \gls{ONAP} and the \gls{NFV-MANO} framework. \gls{ONAP} serves as a comprehensive platform for the \gls{MO} of virtualized and software-defined elements within \gls{O-RAN} architecture. Its affiliation with the \gls{LF} enables integration with {other key} projects such as Kubernetes, Akraino, Acumos, and OpenDaylight \cite{OnapArchitectureWhitepaper}. \gls{ONAP} is also the preferred \gls{SMO} platform used by \gls{OSC} in their open source \gls{O-RAN} code releases. 

{Alternatively}, the \gls{OSM}, {developed under} the \gls{NFV-MANO} framework, by the \gls{ETSI} offers comparable \gls{SMO} {functionalities} in a more lightweight {design compared to} \gls{ONAP}. Notably, in May 2021, \gls{ETSI} {entered into} a cooperation agreement with the \gls{O-RAN} Alliance, {marking the beginning of collaborative} efforts to integrate the \gls{OSM} framework within the \gls{O-RAN} architecture \cite{osm_nodate}.

In the {remainder} of this subsection, we examine the deployment scenarios of \gls{NFV-MANO} and \gls{ONAP}, {along with their potential impacts} on the network slicing architecture, as elaborated in \cite{O-RAN.WG1.SA}.

\subsubsection{3GPP and NFV-MANO-based SMO Deployment}
The deployment options of the \gls{SMO} framework{---aligned} with both the \gls{3GPP} management system and the \gls{NFV-MANO} framework---emphasize the {core principles and essential requirements} of network slicing. {These include} the virtualization and softwarization of \gls{RAN} resources and components, as well as the seamless integration of \gls{AI}/\gls{ML} capabilities and programmable {control} within the \gls{SMO} framework \cite{10056710,10353004}.

\begin{figure}[!htbp]
    \centering
    \includegraphics[width=\columnwidth]{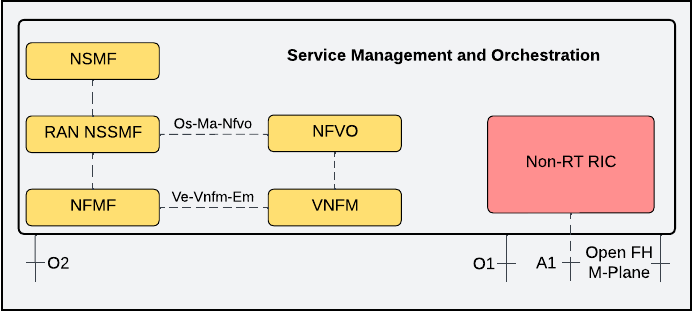}
    \caption{The 3GPP and NFV-MANO-based SMO deployment option with a particular emphasis on O-RAN slicing}
    \label{fig:Ref3GPP-MANOBasedSMODeployment}
\end{figure}

This deployment option integrates the slice \glspl{MF} and network \glspl{MF} defined by \gls{3GPP} with the functional blocks defined by the \gls{ETSI} \gls{ISG} \gls{NFV}. The \gls{NFV-MANO} framework is responsible for the \gls{MO} of \glspl{VNF}, including processes such as the automation, {monitoring}, and operation of virtualized functions {deployed over} a multi-tenant and virtualized infrastructure. Figure \ref{fig:Ref3GPP-MANOBasedSMODeployment} illustrates this deployment scenario, where \gls{3GPP}-defined slice \glspl{MF} (such as the \gls{NSMF}, \gls{NSSMF}, and \gls{NFMF}) {are integrated with} the functional blocks defined within the \gls{NFV-MANO}.  The detailed functionalities of the \gls{3GPP}-defined \glspl{MF} are discussed in Section~\ref{sec:Architecture}.

Moreover, the study group under \gls{O-RAN} \gls{WG} 1 has identified four different options for the deployment of \gls{SMO} framework. These options vary exclusively in the placement of the \gls{3GPP}-defined slice \glspl{MF} and are designed to support different network slice management topologies. Each deployment option explore the potential implications for the \gls{O-RAN} slicing-aware architecture. The four possible deployment options for the \gls{SMO} framework are explained as follows:

\begin{itemize}[noitemsep, topsep=0pt, left=0pt]
    \item Deployment Option 1: In this option, the network slice \glspl{MF}---namely the \gls{NSMF} and \gls{RAN} \gls{NSSMF}---are deployed within the \gls{SMO} framework, as shown in Figure~\ref{fig:Ref3GPP-MANOBasedSMODeployment}.
    \item Deployment Option 2: In this option, both the \gls{NSMF} and \gls{RAN} \gls{NSSMF} are decoupled from and deployed externally to the \gls{SMO} framework.
    \item Deployment Option 3: This deployment option integrates the \gls{NSMF} within the \gls{SMO} framework, while the \gls{RAN} \gls{NSSMF} is placed outside the \gls{SMO} framework.
    \item Deployment Option 4: In this deployment option, the \gls{NSMF} is deployed outside the \gls{SMO} framework, whereas the \gls{RAN} \gls{NSSMF} is integrated within the \gls{SMO}.
\end{itemize}

Within \gls{O-RAN}, the \gls{RAN} \gls{NSSMF}, including its interactions with the \gls{SMO} framework, is the primary area of focus for the \gls{O-RAN} Alliance \cite{O-RAN.WG1.StdSA}. During the creation and provisioning of \gls{NSSI}, the \gls{RAN} \gls{NSSMF}, in coordination with the \gls{SMO} framework, triggers the instantiation of essential \glspl{NF}, such as the \gls{Near-RT RIC}, \gls{O-CU-CP}, \gls{O-CU-UP}, and \gls{O-DU}, according to specific slice requirements. Following the establishment of \gls{RAN} \gls{NSSI}, the \gls{RAN} \gls{NSSMF} in coordination with \gls{SMO} framework, may execute procedures for \gls{NSSI} modification and termination \cite{O-RAN.WG1.SA}.

Each \gls{RAN} \gls{NSSI} is identified using the \gls{NSSAI}. The \gls{NSSAI} includes one or a list of \glspl{S-NSSAI} each serving as unique identifier for a \gls{RAN} slice \cite{3GPP-TS-38.300}. An \gls{S-NSSAI} is a combination of two values. The first value is the mandatory \gls{SST} field, which defines the type of network slice. The \gls{SST} is an 8-bit value ranging from 0 to 255 and may represent a standardized service type such as \gls{eMBB}, \gls{URLLC}, or a network-specific slice type. The second value is the optional \gls{SD} field, a 24-bit value used to distinguish among slices with the same \gls{SST}. According to \gls{3GPP} specifications \cite{3GPP-TS-38.300}, the \gls{NSSAI} may contain up to eight \glspl{S-NSSAI}, which means a single \gls{UE} can be connected with a maximum of eight \gls{RAN} \glspl{NSSI} simultaneously.

The architectural components of the \gls{NFV-MANO} framework integrated within the \gls{SMO} framework may consists of following core functional blocks.

\paragraph{\glsreset{NFVO}\Gls{NFVO}}
The \gls{NFVO} has two responsibilities { within the \gls{NFV-MANO} framework}: Firstly, it {executes resource} orchestration {by coordinating the allocation and management of} \gls{NFVI} resources across multiple \glspl{VIM}. Secondly, it performs network service orchestration, managing the lifecycle of network services by coordinating groups of \glspl{VNF} {that collectively deliver} complex {service functionalities}. The \gls{NFVO} {enables} joint instantiation and configuration of \glspl{VNF}, ensuring inter-\gls{VNF} connectivity, and manages dynamic service adaptation. The network service orchestration function relies on {collaborative interactions with} both the \gls{VNFM} and the resource orchestration function that enables abstracted access to the \gls{NFVI} resources irrespective of the underlying \glspl{VIM} implementations. Furthermore, it manages \glspl{VNF} that shares resources within the underlying \gls{NFVI}~\cite{3GPP-TS-28.533,ETSI-GS-NFV-006-R4}.

\paragraph{\glsreset{VNFM}\Gls{VNFM}}
The \gls{VNFM} is responsible for the lifecycle management of one or more \gls{VNF} instances within a network slice \cite{ROTSOS2017203}. These \glspl{VNF} {may belong to} the same type or different {functional categories}. In addition, the \gls{VNFM} is responsible for the \gls{FCAPS} management of the \glspl{VNF}, {and it also facilitates elastic scalability, that enables} \glspl{VNF} to be dynamically scaled up or down in its designated service region \cite{ETSI-GS-NFV-006-R4}.

\paragraph{\glsreset{NFVI}\Gls{NFVI}}
The \gls{NFV} {recognizes both} software and hardware accelerators as auxiliary resources capable of virtualization that can be exposed as virtual accelerators within the \gls{VNF} layer \cite{ETSI-GR-NFV-IFA-046}. The \gls{NFVI} includes all the underlying components of the infrastructure---hardware and software---necessary to host \glspl{VNF}. {It abstracts and presents these} resources in virtualized forms to be utilized by \glspl{VNF} and network services, {including virtualized} compute, storage, and networking capabilities \cite{ETSI-GS-NFV-006-R4}.

However, it is important to highlight that current \gls{NFV-MANO} specifications do not comprehensively address \gls{NFVI} management aspects, particularly regarding physical infrastructure within cellular networks. As a result, full support of complete \gls{IMS} functionality is not achievable under the current specifications \cite{ETSI-GR-NFV-IFA-046}.

\paragraph{\glsreset{VIM}\Gls{VIM}}
The \gls{VIM} is responsible for managing and controlling the compute, storage, and network resources of the \gls{NFVI} within the underlying telecommunication infrastructure \cite{ROTSOS2017203}. {While the} deployment and maintenance of the \gls{VIM} fall outside the formal scope of the \gls{NFV-MANO} framework, the interfaces {it exposes} are explicitly included within its scope \cite{ETSI-GS-NFV-006-R4}. The \gls{NFV-MANO} framework utilizes these interfaces to influence the decisions made regarding the three types of resources (compute, storage, and networking) within the underlying infrastructure.
 
\paragraph{\glsreset{EM}\Gls{EM}} The \gls{EM} is equialent to the \gls{NFMF} within the \gls{3GPP} management system. It is responsible for the \gls{FCAPS} managment of a \gls{VNF}, encompassing both functional and application layer perspectives. Notably, this functional block also manages the \gls{FCAPS} of a \gls{VNF}, but exclusively from a virtualization standpoint \cite{ETSI-GS-NFV-006-R4}.

In addition to these functional blocks, the \gls{ETSI} \gls{ISG} \gls{NFV} introduced five new \glspl{MF} {as part of Release 4. These additions aim to enhance the \gls{NFV-MANO}'s capability to support} containerized network functions and manage the transport aspects of virtualized infrastructures \cite{ETSI-GS-NFV-006-R4,ROTSOS2017203}. {Comprehensive details regarding these newly introduced \glspl{MF} can be found in the Release 4 specifications published by ETSI}.

\subsubsection{3GPP and ONAP-based SMO Deployment}
As discussed in Section~\ref{Sec:SecondChapter}, the \gls{ONAP} framework provides the necessary management, orchestration, and automation capabilities for \gls{E2E} network architecture. {Within} the \gls{OSC}, the \gls{SMO} leverages \gls{ONAP} along with other components {to enable standardized, modular orchestration functions}. Particularly noteworthy is the \gls{Non-RT RIC}, which {complements} the \gls{SMO} and utilizes \gls{ONAP} for efficient A1 policy management \cite{9367572}. \gls{ONAP} encompasses {predefined} workflows and \glspl{UI} for \gls{3GPP}-defined network slice orchestration functions---namely, the \gls{CSMF} and the \gls{NSMF}---along with an additional interface to external \glspl{NSSMF} for managing \gls{RAN}, \gls{CN}, and \gls{TN} domains. These slice \glspl{MF} empower the \gls{ONAP} framework to orchestrate and allocate an \gls{E2E} \gls{NSI}, comprising suitable \glspl{NSSI} across \gls{RAN}, \gls{CN}, and \gls{TN} to meet specific service and use case requirements \cite{OnapE2ENetworkSlicing}.

\gls{ONAP} proposes two deployment options for the \gls{SMO} framework, that emphasizes enhanced integration with the \gls{O-RAN} architecture, improved cloud-native \gls{NF} orchestration, and progressing towards intent-driven, closed-loop automation.

In the first option, the \gls{RAN} \gls{NSSMF} is deployed within \gls{SMO} and is responsible for the \gls{MO} of the \gls{RAN} network slice subnet, including \gls{O-RAN} \glspl{NF} and the associated \gls{TN} components. 
The \gls{RAN} \gls{NSSMF} determines the slice-specific configuration of \gls{O-RAN} \glspl{NF} based on slice profile received from the \gls{NSMF} and identifies the corresponding requirements for the \gls{FH} and \gls{MH} interface. It then communicated to the \gls{TN} \gls{MD}, which executes the actual configuration using \gls{ETSI} \gls{ZSM} based \gls{MD} approach \cite{O-RAN.WG1.SA,OnapE2ENetworkSlicing}.

The second option {assigns more comprehensive role} to the \gls{NSMF}, which not only determine the slice profile of \gls{RAN} \glspl{NF}, \gls{FH}, and \gls{MH} segments but also stitches together \gls{E2E} network slice instances across domains. This centralization enables the \gls{NSMF} for consistent orchestration across \gls{RAN} and \gls{TN} subnets. In both options, separate \gls{RAN} \glspl{NSST} are designed to support customized \gls{RAN} and \gls{TN} configurations \cite{OnapE2ENetworkSlicing}.

\begin{figure}[htbp!]
    \centering
    \includegraphics[width=\columnwidth]{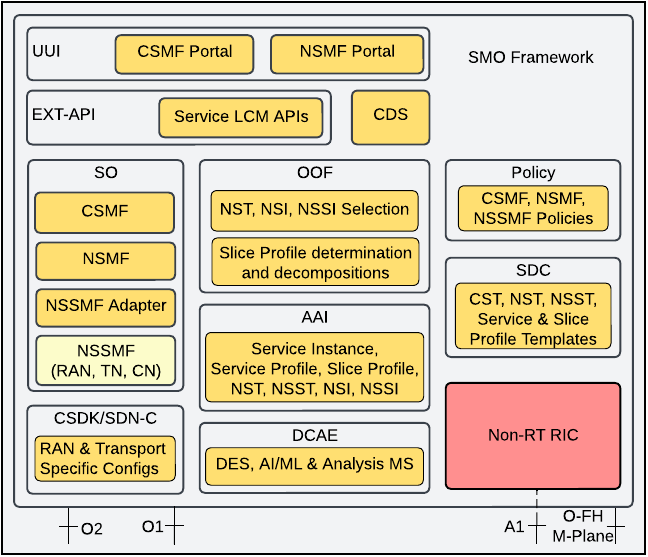}
    \caption{3GPP and ONAP-based SMO deployment option with a particular emphasis on \gls{O-RAN} slicing}
    \label{fig:RefONAPBasedSMODeployment}
\end{figure}

Figure~\ref{fig:RefONAPBasedSMODeployment} illustrates the \gls{ONAP}-{based \gls{SMO} framework incorporating \gls{3GPP}-defined} slicing management functions. The {realization of} \gls{5G} \gls{E2E} network slicing {depends on} the coordination and integration of multiple \gls{ONAP} functional modules, each described in detail in the following section.

\paragraph{\Gls{UUI}} The \gls{UUI} operations support an extensive range of lifecycle management actions through an simple point-and-click interface, thereby enabling network operators and service providers to execute tasks more easily \cite{OnapArchitectureWhitepaper}. Within the \gls{CSMF} portal, users can create service request forms to establish network services using specific \glspl{NSI}. These services can be viewed in a list and managed performing operations like activation, deactivation, or termination. In parallel, the \gls{NSMF} portal offers functionality to manage slicing-related tasks triggered by customers. Operators can monitor task status, take appropriate actions, and refine slice configurations as suggested by the \gls{OOF}. Additionally, the \gls{NSMF} portal includes a comprehensive network slicing resource management interface, which allows users to visulize and manipulate existing network slices, \glspl{NSI}, and \glspl{NSSI} \cite{onap_e2e_slicing_usecase}.

\paragraph{External API (EXT-API)} The EXT-API offers northbound interoperability for the \gls{ONAP} platform, serving as an access point for third-party frameworks. Upon receiving a service request, EXT-API responds with a \textit{Service Order ID}, which can be used to track the status of the service order. Subsequently, the EXT-API activates the \gls{SO} \gls{API} to initiate the actual service creation process. This action represents progress in establishing uniform external interfaces for automated network slice orchestration \cite{OnapE2ENetworkSlicing,onap_e2e_slicing_usecase}.

\paragraph{\glsreset{CDS}\Gls{CDS}} The \gls{CDS} framework provides blueprint definitions and archives for configuration management processes. It comprises a \gls{GUI} and run time components. The \gls{GUI} manages user input and displays both the design time and the run time activities. At run time, it allows users to direct the system to resolve the dynamic parameters in blueprint {and generate final configuration, which are subsequently} downloaded to \gls{VNF}. The major role of the \gls{CDS} is to generate and populate a controller blueprint, create a configuration file, and download it to \gls{VNF}/\gls{PNF} \cite{OnapArchitectureWhitepaper}.

\paragraph{\glsreset{SO}\Gls{SO}} The \gls{SO} automates sequences of activities, tasks, rules, and policies to execute specified processes required for the on-demand creation, modification, or removal of network, application, or infrastructure services and resources \cite{OnapE2ENetworkSlicing}. Within the \gls{SO}, distinct \gls{BPMN} workflows are established for the \gls{CSMF} and \gls{NSMF}. The \gls{CSMF} workflow manages service requests originating from the \gls{CSMF} portal and stores order information in a communication service instance within the \gls{AAI}. It then interacts with the \gls{NSMF} workflow to initiate slice requests.

The \gls{NSMF} {workflow is responsible for} generating service profiles, \gls{NSI}, and \gls{NSSI}, all of which can be reused or shared across multiple services \cite{OnapE2ENetworkSlicing}. {Furthermore, the \gls{SO} incorporates} an \gls{NSSMF} adapter that interacts with internal or external \glspl{NSSMF} for \gls{NSSI} orchestration. 
The \gls{NSSMF} functionality includes a common part for subnet capability queries from \gls{SO}, invoking domain-specific \gls{NSSMF} functions for the \gls{RAN}, \gls{CN}, and \gls{TN} domains. 
The specialized workflows of the domain-specific \gls{NSSMF} handle the essential tasks involved in creating or updating the \gls{NSSI} according to the guidance provided by the \gls{OOF}, particularly for new \gls{NSSI} creation or reuse \cite{OnapE2ENetworkSlicing}.

\paragraph{\glsreset{OOF}\Gls{OOF}} The \gls{OOF} offers a declarative and policy-driven method for developing and executing optimization applications such as placement and change management, scheduling optimization \cite{OnapE2ENetworkSlicing}. The \gls{SO} interacts with the \gls{OOF} to select the \gls{NST} and \gls{NSI}/\gls{NSSI}. The \gls{OOF} may recommend either creating new instances or reusing the existing ones. {In the case of} \gls{NSI}/\gls{NSSI} selection, the \gls{OOF} could return an existing \gls{NSI} if it is shareable and suitable, an existing \gls{NSSI} if shareable and no suitable \gls{NSI} exists, or a slice profile if the service request is non-shareable or no suitable \gls{NSI} or \gls{NSSI} exists. The recalibration of \gls{NSI} and \gls{NSSI} selection is managed by the orchestration task, which allows network operators to intervene manually through the \gls{NSMF} portal in \gls{UUI} \cite{onap_e2e_slicing_usecase}.

\paragraph{\glsreset{SDC}\Gls{SDC}} The \gls{SDC} offers tools, methods, and repositories for defining, simulating, and certifying system assets along with their corresponding processes and policies. These assets are categorized into four groups: resources, services, products, or offers. The \gls{SDC} environment {serves a diverse range of} users through shared services and utilities. Within the design studio, product and service designers can onboard, extend, or retire resources, services, and products \cite{OnapArchitectureWhitepaper}.    

\paragraph{\glsreset{AAI}\Gls{AAI}} The \gls{AAI} provides real-time and historical views of a system's resources, services, products, and their interrelationships. It serves as a dynamic registry, continuously updated by controllers in real-time to support the flexibility of \gls{SDN}/\gls{NFV}. The \gls{AAI} module introduces three additional nodes: Communication-service-profile, Service-profile, and Slice-profile, along with modifications to the service-instance nodes. {Furthermore, t}hree new nodes have been incorporated as attributes of the service-instance node. To align with \gls{SDC} templates such as \gls{CST}, Service Profile Template, Slice Profile Template, \gls{NST} , and \gls{NSST}, the run-time instances include \gls{CSI}, Service Profile Instance, Slice Profile Instance, \gls{NSI}, and \gls{NSSI}. The Slice Profile Instance for the all three subnets---\gls{RAN}, \gls{CN}, and \gls{TN}---are distinct \cite{OnapE2ENetworkSlicing, OnapKuhn}.

The \gls{AAI} offers query \glspl{API} to \gls{CSMF} and \gls{NSMF}, enabling them to retrieve various information such as communication service instances, service profile instances, \gls{NSI}, and \gls{NSSI}. Additionally, \gls{AAI} provides creation of \glspl{API} to \gls{SO}, allowing the creation of communication service profiles, service profiles, slice profiles, and the establishment of relationships between service instances \cite{OnapE2ENetworkSlicing}.

\paragraph{\glsreset{CCSDK}\glsreset{SDN-C}\Gls{CCSDK}/\Gls{SDN-C}} The \gls{CCSDK}/\gls{SDN-C} components manage specific configurations for both the \gls{RAN} and \gls{TN} subnets of a network slice. When requested by the \gls{SO} from the \gls{TN} \gls{NSSMF}, they set up and configure a new \gls{TN} \gls{NSSI}, including updating the \gls{TN} during \gls{NSI} reuse, {as well as during} activation, deactivation, and termination phases. Similarly, when invoked by the \gls{SO} from the \gls{RAN} \gls{NSSMF}, they (re)configure existing \gls{RAN} \glspl{NF} for \gls{RAN} \gls{NSSI} or \gls{NSI} reuse. Additionally, when policy triggers closed loop actions within the \gls{RAN} for \gls{RAN} \glspl{NSSI}, they send relevant configuration updates to the \glspl{Near-RT RIC} \cite{OnapE2ENetworkSlicing}.

\paragraph{\glsreset{DCAE}\Gls{DCAE}} In collaboration with other \gls{ONAP} runtime components, \gls{DCAE} provides closed loop automation. It introduces two new micro-services \cite{onap_e2e_slicing_usecase}. The first is
\textbf{\gls{DES}}, which offers a simplified interface for network operators, slice tenants, or other \gls{ONAP} component to query both current and historical \gls{PM}/\gls{KPI} data. The second is the \textbf{Slice Analysis MS}, which analyzes \gls{PM} data received from the \gls{RAN} through the PM-Mapper micro-service to detect any updates. When it receives configuration updates, it initiates a control loop by transmitting a suitable \gls{DMaaP} message to policy framework.

\paragraph{Policy Framework} At a granular level, policies consists of machine-readable rules that define actions triggered by specific events or requests, {based on given} conditions. This approach enables the rapid policy adjustments by updating rules, {allowing} the adjustment of technical behaviors without rewriting code. The Policy framework simplifies the management of complex mechanisms using abstraction \cite{OnapArchitectureWhitepaper}.

\vspace{-2.5mm}
\section{Slicing the Underlying Infrastructure within the O-RAN Architecture} \label{sec:UnderlyingInfrastructure}
\vspace{-1.5mm}
The transition from distributed to centralized architectures marks a significant shift in \gls{RAN} designs{. The} \gls{O-RAN} adopts {the} centralized model where major \glspl{NF}, such as the \gls{O-CU}, are located in \glspl{DC} at \gls{O-Cloud} sites. The \gls{O-DU} optionally resides either in \gls{DC} or at the cellular network site. The \gls{TN} {enables} data paths {across various \glspl{NF} between and within the} \gls{RAN} and \gls{CN} {domains}, thereby delineating distinct \gls{TN} segments, such as the \gls{FH}, \gls{MH}, and \gls{BH} \cite{lucena2021RolloutOfSlicing}. Network slicing integrated into the virtualized \gls{O-RAN} infrastructure boosts efficiency and unlocks unprecedented opportunities for innovation and service differentiation \cite{wu2020dynamic-RAN-Slicing}.

In the rest of this section, we delve into the underlying infrastructure within the \gls{O-RAN} {architecture}, including the components of the cellular network site, the \gls{O-Cloud} platform, and the Xhaul \gls{TN} {domain}, examining the network slicing aspects {in} these critical elements {and domains}.

\vspace{-2.5mm}
\subsection{O-RAN Cellular Network Site}
\vspace{-1.5mm}
In wireless networks, a cellular network site serves as a {fixed hub} for transmitting and receiving radio signals {and ensures consistent} coverage over a specified area. {A cellular network site} encompasses two primary components: first, one or more antennas {that transmit and receive} radio signals, and second, a supply unit that houses essential switching and control elements critical for managing antenna operations.

In {standard} design, cellular network sites are structured to support {multiple} sectors, thus inherently associated with {several} \glspl{O-RU}. The \gls{O-RU} serves as a fundamental {component} in establishing seamless \gls{PHY} layer connections with \glspl{UE} \cite{O-RAN.WG1.OAD,O-RAN-WG6-CADS}. {It integrates} antenna elements {with} important \gls{RF} components {such as} transceivers and amplifiers {to ensure efficient signal processing}. Additionally, the \gls{O-RU} manages lower-level \gls{PHY} tasks such as digital beamforming and \gls{FFT} operations \cite{Pw-OpenRan2020}. The Open \gls{FH} interface is crucial for connecting the \gls{O-RU} with the \gls{O-DU} to ensure seamless communication in \gls{O-RAN} architecture.

As {explained} in Section~\ref{sec:DeploymentOptions}, a balance between \gls{FH} latency and cost considerations is necessary to optimize {the deployment of \gls{O-RAN} components}. Consolidating all elements of the \gls{O-gNB} at cellular network site minimizes latency {but is the most} expensive option. Conversely, relocating control and connection anchors towards a centralized edge cloud facilitates resource management across multiple sites while preserving low-latency data processing. Strategically relocating processing functions to the edge cloud while retaining only the \gls{O-RU} at the cellular network site achieves {an optimal balance between optimized latency and cost efficiency}.

For services with less stringent time requirements, {moving} the \gls{Near-RT RIC} and \gls{O-CU} to a regional cloud may increase latency {to around} 50 milliseconds. However, {this approach} optimizes resource processing across various cellular network sites. By centralizing these functions, a single \gls{Near-RT RIC} can efficiently manage resource allocation while {keeping} critical processing units closer to end users.

\vspace{-2.5mm}
\subsection{O-RAN Cloud Platform}
\vspace{-1.5mm}
{One of the objectives} of the \gls{O-RAN} Alliance is to enhance the flexibility and deployment speed of the \gls{RAN} architecture while lowering both capital and operati{onal expenses} through {\gls{O-Cloud}-based implementations}. 
{The \gls{O-Cloud} platform comprises hardware and software components that deliver \gls{O-Cloud} capabilities and services to host \gls{O-RAN} \glspl{NF}.} The logical architecture of the \gls{O-RAN} {combined} with the \gls{O-Cloud} platform {and technologies} provide a fully open {and cloud-native} solution where software is decoupled from hardware.

Hardware and software {decoupling follows} a three-tiered approach: a hardware layer, an intermediary layer {with the} cloud stack and acceleration abstraction functions, and a top layer dedicated to virtual \gls{RAN} functions {(i.e., O-CU and O-DU)}. These layers {can be sourced from} different vendors and the decoupling ensures interoperability between a cloud stack and numerous hardware suppliers as well as accommodate \gls{RAN} \glspl{VNF} from various \gls{RAN} software providers \cite{O-RAN-WG6-CADS}.

\begin{figure}[!htbp]
    \centering
    \includegraphics[width=\columnwidth]{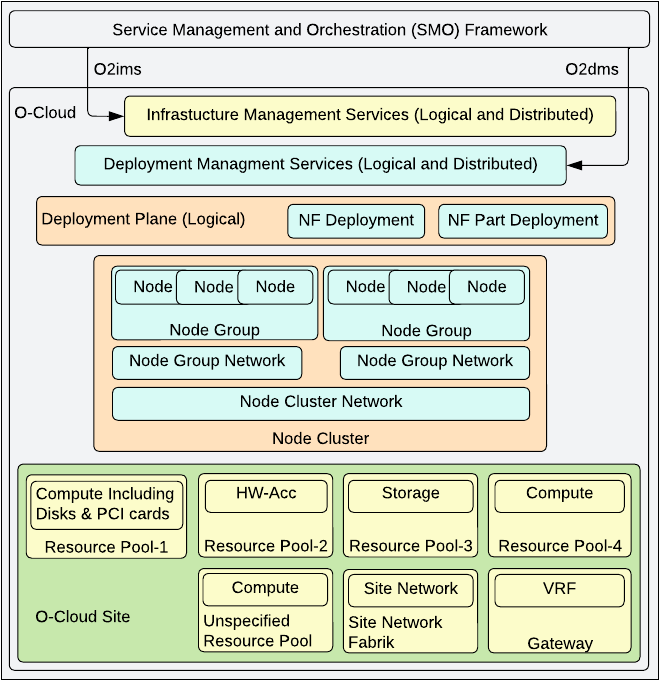}
    \caption{{Key components in an O-Cloud within the O-RAN architecture}}
    \label{fig:RefO-CloudComponents}
\end{figure}

An O-Cloud platform can automate and autonomously manage tasks with a certain level of complexity such as placing \gls{NF} deployment workloads on suitable \gls{O-Cloud} nodes, executing self-repair, and auto-scaling based on deployment artifacts, and policies, without \gls{SMO} intervention. {As illustrated in Figure~\ref{fig:RefO-CloudComponents}, an} \gls{O-Cloud}  includes \gls{O-Cloud} resources, Resource pools, and \gls{O-Cloud} services across multiple sites including the software that manages resource provisioning, nodes, clusters, and deployments {on them. It includes the functionality to support the deployment} and management services. The \gls{O-Cloud} provides a unified reference point for all the elements and services within its boundary and scope.

An \gls{O-Cloud} site refers to a collection of \gls{O-Cloud} resources at a specific geographical location. The resources are interconnected through \gls{O-Cloud} site network fabrics. Multiple \gls{O-Cloud} sites can be interconnected to form a distributed \gls{O-Cloud}, that requires bridging, routing, or stitching at networking layer in between each {\gls{O-Cloud}} site and its respective external transport network attachment point \cite{O-RAN-WG6-CADS}.

The O2 interface {facilitates connectivity to various} \gls{O-Cloud} services offered by the \gls{O-Cloud} platform in conjunction with the \gls{SMO} framework. These services are tailored to address specific functionalities and requirements within the \gls{O-Cloud} ecosystem \cite{O-RAN-WG6-CADS, oran_risk_analysis}. {The following sections provide a detailed explanation of each \gls{O-Cloud}} component within the \gls{O-RAN} architecture, as shown in Figure~\ref{fig:RefO-CloudComponents}.

\subsubsection{Infrastructure Management Services} 
Within the intricate framework of \gls{O-Cloud} {site} operations, \glsreset{IMS} \gls{IMS} is a crucial subset of O2 functions, entrusted with the deployment and {management} of {O-Cloud} infrastructure. The \gls{IMS} assumes an important role {in} provisioning by efficiently allocating and configuring resources for \gls{O-Cloud} node clusters \cite{O-RAN.WG6.O2IMS}.

{In addition,} the \gls{IMS} {provides} fault and performance management identifying issues and providing measurements {to \gls{SMO}} through the O2ims interface. It also provides \gls{O-Cloud} inventory reporting through O2ims containing details of \gls{O-Cloud} sites, deployment management services, node clusters, and resources \cite{O-RAN.WG6.O2IMS}. The O2ims inventory services enables the \gls{SMO} {framework} to understand the requested allocation and available \gls{O-Cloud} capabilities and capacities. The \gls{O-Cloud} lifecycle management involves registering, structuring, and configuring infrastructure services and resources. {Furthermore, this component performs} maintenance operations, such as switching \gls{O-Cloud} nodes to maintenance mode autonomously or on demand to ensure a seamless communication with various components within the \gls{SMO} framework.

In the following paragraphs, we explore the concepts and perspectives {related to} \gls{O-Cloud} \gls{IMS} and the \gls{O-Cloud} infrastructure as detailed in {reference} \cite{O-RAN-WG6-CADS}:

\paragraph{O-Cloud Resource} This is a defined unit comprising capabilities such as compute, hardware acceleration, storage, and gateway within {an} \gls{O-Cloud} {site}. These resources are provisioned and utilized for the \gls{O-Cloud} deployment plane, enabling efficient allocation and management of computing resources in cloud-based network infrastructure.

\paragraph{O-Cloud Resource Pool} The \gls{O-Cloud} resource pool consists of a grouping of \gls{O-Cloud} resources possessing similar capabilities and traits within an \gls{O-Cloud} environment. It comprises one or more such resources, each equipped with network connections and, optionally, internal hardware accelerators and storage devices. Additionally, it may include standalone servers lacking an associated \gls{O-Cloud} site network fabric, like infrastructure deployed at a {cellular network} site.

\paragraph{Unspecified O-Cloud Resource Pool}  It refers to a collection of \gls{O-Cloud} resources listed in the \gls{O-Cloud} \gls{IMS} inventory but not yet categorized or allocated to any specific \gls{O-Cloud} resource pool within an \gls{O-Cloud} site.

\paragraph{O-Cloud Site Network Fabric} It serves as an interconnecting resource within an \gls{O-Cloud} site, linking various resources within a site to enable seamless communication and data exchange between them. {This enhances the} overall functionality, interactions, and resource utilization.

\paragraph{O-Cloud Site Network} It represents a meticulously provisioned network resource, {which showcases} its defined capabilities and characteristics derived from an intricately configured \gls{O-Cloud} site network fabric.
    
\subsubsection{Deployment Management Services} The \gls{DMS} efficiently handles various tasks by leveraging information received over O2dms. These tasks encompass the strategically placing \gls{O-RAN} \gls{NF} deployment workloads within \gls{O-Cloud} node clusters. Additionally, \gls{DMS} manages the entire lifecycle of these workloads, including resource allocation, configuration adjustments, and the executing essential lifecycle management operations such as autonomous scaling, self-healing, and workload relocation within the same \gls{O-Cloud} node cluster to meet \gls{SLE}. It also supervises the cessation of \gls{NF} deployments based on directives from the \gls{SMO} framework. Furthermore, the \gls{DMS} ensures the \gls{O-Cloud} inventory is regularly updated with the latest status information on resources dedicated {(or allocated)} to \gls{NF} deployment workloads.

Below are the conceptual insights related to \gls{O-Cloud} \gls{DMS} and its interaction with \gls{O-Cloud} resources generated or modified via \gls{IMS} provisioning as demonstrated in \cite{O-RAN-WG6-CADS}.

\paragraph{O-Cloud Deployment Plane} It refers to a conceptual framework comprising \gls{O-Cloud} nodes, \gls{O-Cloud} networks, and \gls{O-Cloud} node clusters, which are pivotal components {for} \gls{NF} deployments. This framework is established by leveraging \gls{O-Cloud} resources provisioned via \gls{IMS}, derived from \gls{O-Cloud} resource pools and \gls{O-Cloud} site network fabrics.

\paragraph{O-Cloud NF Deployment} \gls{NF} deployment refers to deploying software on \gls{O-Cloud} resources to implement cloudified \glspl{NF}, either fully or partially. {This enables the deployment of} \gls{NFV} within cloud environments.

\paragraph{O-Cloud Node} It is a network connected computer or function, that can be provisioned into \gls{O-Cloud} node clusters by \gls{IMS}. The nodes comprises physical or logical components, and expose \gls{IMS}-assigned resources to form \gls{O-Cloud} deployment plane constructs. Additionally, an \gls{O-Cloud} node may operate independently as a standalone entity.

\paragraph{O-Cloud Node Cluster} It consists of a set of \gls{O-Cloud} nodes operating together via interconnected \gls{O-Cloud} node cluster networks. The operating system and cluster software of these nodes identify their capabilities and characteristics managed by \gls{IMS}.

\paragraph{O-Cloud Node Cluster Network} It denotes a dedicated network infrastructure tailored for an \gls{O-Cloud} site network allocated to an \gls{O-Cloud} node cluster.

\paragraph{O-Cloud Node Group} It refers to a subset of \gls{O-Cloud} nodes in an \gls{O-Cloud} node cluster treated equally, particularly by the \gls{O-Cloud} node cluster scheduler. These nodes are interconnected through \gls{O-Cloud} node cluster networks and optionally through \gls{O-Cloud} node group networks.

\paragraph{O-Cloud Node Group Network} It refers to the \gls{O-Cloud} site network designated for a specific grouping of \gls{O-Cloud} nodes within an \gls{O-Cloud} node cluster.

\begin{figure*}[!htbp]
    \centering
    \includegraphics[width=\textwidth]{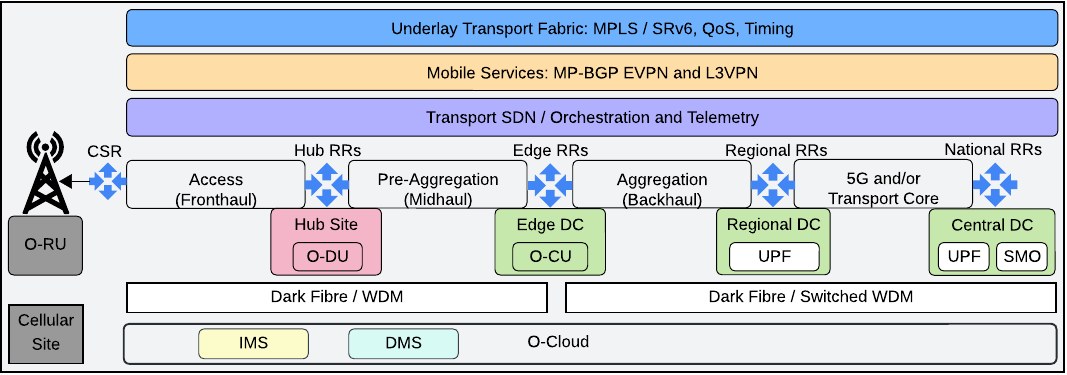}
    \caption{{Packet switched Xhaul TN architecture with a common underlay transport fabric overlaid with a mobile service layer}}
    \label{fig:XhaulArchitecture}
    \vspace{-6.0mm}
\end{figure*}

\vspace{-2.5mm}
\subsection{Xhaul Transport Network}
\vspace{-1.5mm}
The Xhaul \gls{TN} {serves as a} unified \gls{TN} {providing seamless} connectivity within and between \gls{RAN} and \gls{CN} components. {In the} \gls{O-RAN} architecture, {it integrates various} \gls{TN} segments across \gls{RAN} and \gls{CN} functions such as \gls{FH}, \gls{MH}, and \gls{BH}. The \gls{TN}, particularly the access {segments} like \gls{FH} and \gls{MH}, {can simultaneously manage} diverse transport flows, {which becomes especially important} when operators integrate mixed-use cases into their \gls{RAN} deployments.

An efficient network resource management strategy is essential due to the varying nature of these transport flows, each {with} distinct requirements for latency, throughput, and transmission reliability \cite{eBook-Keysight}. This is crucial to mitigate complexity and maintain optimal performance across the \gls{TN} \cite{9741386}. A {strategic approach involves} categorizing transport flows into transport slices according to shared service {requirements}, which enables more structured and efficient {TN} management. These slices {can be further} subdivided into sub-slices as needed enabling tailored support for diverse \gls{E2E} user applications \cite{O-RAN.WG9.XPSAAS}. 
{As network slicing integration advances}, discussions within \gls{O-RAN} focus on incorporating network slicing into the existing \gls{TN} infrastructure \cite{9165514}. {This includes determining} which mobile interfaces—FH, MH, BH, and N6—require network slicing, the structure of these slices, and the optimal number of slices needed at the {\gls{TN}} level.

In the subsequent sections, we explore the architecture of the \gls{TN}, focusing on the complexities of network slicing. The aim is to provide insights into the fundamental principles and practical considerations crucial for the successful deployment and operation of Xhaul \glspl{TN} {within the} \gls{O-RAN} {architecture}.

\subsubsection{Xhaul TN Architecture}
In the \gls{O-RAN} Xhaul \gls{TN} architecture, the \gls{FH} network connects the \gls{O-DU} and \gls{O-RU} with latency models based on \gls{eCPRI} reference points \cite{waypior2022OpenRan}. The \gls{MH} network enables communication between \gls{O-DU} and \gls{O-CU} with \gls{3GPP}-defined F1/E1 interfaces, while the \gls{BH} network connects the \gls{O-CU} to the \gls{CN} \cite{O-RAN.WG9.XPSAAS}.

The Xhaul \gls{TN} {architecture must be highly adaptable to accommodate varying} requirements based on the specific use case and \gls{RAN} designs. {It involves} accommodating numerous {next-generation network} services, multiple network slices \cite{9204600}, and diverse \gls{3GPP} interfaces  \cite{O-RAN.WG9.XPSAAS} across {different} segments of the physical transport network. The \gls{O-RAN} Alliance in its \gls{WG}9 transport requirements document \cite{O-RAN.WG9.XTRP-REQ} has meticulously outlined {key} prerequisites for the \gls{O-RAN} Xhaul \gls{TN}, encompassing bandwidth and {latency expectations} within the \gls{5G} network, as well as logical transport connectivity needs across \gls{FH}, \gls{MH}, and \gls{BH} and even the N6 interface.

The \gls{O-RAN} {introduces new} requirements for \gls{FH} networks, {particularly regarding} \gls{FH} latency and data rate \cite{O-RAN.WG9.XTRP-REQ}. To overcome these obstacles, \gls{WDM} has emerged as a promising solution, offering various architectural approaches such as passive, active, and semi-active \gls{WDM}. For more details on \gls{WDM}, please refer to \cite{o-ran-wg9-wdm}.

The deployment of an \gls{E2E} Xhaul \gls{TN} relies on packet switched transport solutions, {which is} influenced by various key factors. These encompass the extent of packet switching components spanning from cell sites to the transport core and the potential integration of other technologies {with}in the \gls{FH} network \cite{8722598}. {Additional} considerations include the nature of the underlying Layer 0/Layer 1 transport, the {choice of} network protocols implemented at the packet switching layer, and the framework for {implementing the} overlay services on the Xhaul {\gls{TN}} infrastructure.

Figure~\ref{fig:XhaulArchitecture} illustrates a unified \gls{E2E} packet switched {\gls{TN}} infrastructure. The architecture {consists of} a common underlay infrastructure overlaid with a service layer that utilizes the shared transport fabric to support mobile services. The \glspl{DC} are strategically distributed {and} integrated {into} the \gls{TN} {architecture}, which enables \glspl{VNF}/\glspl{PNF} essential for mobile and fixed communication services.

The underlay {infrastructure is designed to be} scalable which ensures that it meets the diverse service requirements of an \gls{O-RAN} \gls{TN}. In contrast, the service infrastructure, or overlay operates above the underlay {supporting} \gls{FH}, \gls{MH}, and \gls{BH} {segments} of the \gls{O-RAN} \gls{TN} \cite{8320765,O-RAN.WG9.XPSAAS}. {However,} the logical architecture shown in Figure~\ref{fig:XhaulArchitecture} may vary in physical implementation. For instance, some operators may adopt packet switched technology in the \gls{MH} and \gls{BH}, while using simpler physical networking for the \gls{FH} \cite{O-RAN.WG9.XPSAAS,9771187}.

\paragraph{Xhaul TN Underlay/Fabric Technologies}
Underlay networks form the physical infrastructure of a \gls{TN}, comprising Ethernet switches, routers, \gls{DWDM} equipments, and the fiber optic cabling that interconnects these components into a coherent topology. To support an Xhaul \gls{TN} environment, the packet switched network must handle both \gls{L2} and \gls{L3} services. Currently, \gls{L2} underlay networks predominantly rely on Ethernet, often utilizing \glspl{VLAN} for segmentation. Within \gls{O-RAN}, following two prevalent packet switched underlay technologies are discussed in \cite{O-RAN.WG9.XPSAAS}.

\begin{itemize}[noitemsep, topsep=0pt, left=0pt]
\item \textbf{\Gls{MPLS}:} \gls{MPLS} employs label switching in the data plane with multiple control plane technologies including \gls{SR} an extension to \gls{IGP} and \gls{BGP}. Regardless of the \gls{MPLS} control plane used, the service layer is independent and supports native Ethernet and \gls{L3} services \cite{rfc4364}.

\item \textbf{\Gls{SRv6}:} \gls{SRv6} is {built} on the \gls{SR} architecture and {operates using} an \gls{IP}v6 data plane, where segments are identified by \glspl{SID} embedded in the \gls{IP}v6 header \cite{rfc9252}. While it shares some similarities with \gls{SR}-\gls{MPLS}, key differences {exist, particularly} in the requirements for scaling the underlay infrastructure {to support} \gls{5G} environment.
\end{itemize}

The underlay is anticipated to offer a comprehensive set of tools necessary to deliver essential network services, encompassing functionalities such as universal connectivity, prioritization, isolation, scalability, rapid convergence, shortest path routing, traffic engineering, packet-based \gls{QoS}, and precise timing mechanisms \cite{O-RAN.WG9.XPSAAS}.

\paragraph{Xhaul TN Overlay/Services Infrastructure}
Overlay networks utilize network virtualization principles to create virtualized networks {composed} of overlay nodes such as routers. They leverage technologies like \gls{EVPN} and \gls{MP-BGP}-based \glspl{L3VPN} for tunneling encapsulation within the overlay service layer. This encapsulation enables data packets transmission over the underlying physical network while maintaining logical separation and isolation between different virtual networks or network segments \cite{O-RAN.WG9.XPSAAS}.

Both \gls{MPLS} and \gls{SRv6} packet switched underlays utilize \gls{EVPN} for \gls{L2} support and \gls{MP-BGP} for \glspl{L3VPN}. In the \gls{MP-BGP} architecture, the protocol is configured with suitable address-family support for both \gls{EVPN} and \gls{L3VPN}, facilitating the transmission of service connectivity information among \gls{PE} {devices} \cite{rfc4364}.

Ethernet services are provided by \gls{EVPN}, where \gls{EVPN} \gls{VPWS} acts as a transport service for open \gls{FH} and \gls{RoE} to ensure redundancy for open \gls{FH} interface. \gls{BGP} \glspl{L3VPN} support \gls{IP}v4 and \gls{IP}v6. {This approach presents} flexible connectivity models with default shortest path routing, along with the option for automatic steering into \gls{SR} policy \cite{rfc9256}.

Mobile \gls{IP} services are facilitated by \gls{MP-BGP}-based \glspl{L3VPN}, which establish \gls{L3} connectivity among various mobile components. With \gls{BGP} \gls{L3VPN} support, {network} operators {can deploy} both \gls{IP}v4 and \gls{IP}v6 \glspl{VPN}, thereby enabling adaptable connectivity models to {meet diverse service and} network requirements.

\subsubsection{Xhaul TN Slicing}
The packet switched \glspl{TN} provide a robust framework to support the network slicing. The Xhaul \gls{TN} infrastructure {is designed to support} the diverse transport demands. {They} require tailored solutions for control, management, and user plane interfaces. This includes \gls{5G} \gls{TN} segments, where \gls{FH}, \gls{MH}, and \gls{BH} interfaces require tailored resource allocation \cite{9410215,9165451}. Each interface has specific latency, bandwidth, and traffic demands, which slicing helps to address by enabling precise resource customization \cite{5GamericasTN45G}.

In \gls{TN}, {slicing is categorized into} \textbf{hard slicing} and \textbf{soft slicing}, which {determine the level} of isolation between network slices \cite{ITU-TR}. Hard slicing allocates resources exclusively to a particular \gls{NSI}, ensuring strict assignment {with} limited resource sharing. Conversely, soft slicing preserves the characteristics of a transport slice {while allowing} shared and reusable resources across different \glspl{NSI} \cite{O-RAN.WG9.XTRP-MGT,9410215}. This approach {enhances} flexibility and {improves} resource utilization. While hard slicing prioritizes exclusive allocation, soft slicing fosters {efficient} resource sharing, enhancing resource management.

\begin{figure}[!htbp]
    \centering
    \includegraphics[width=\columnwidth]{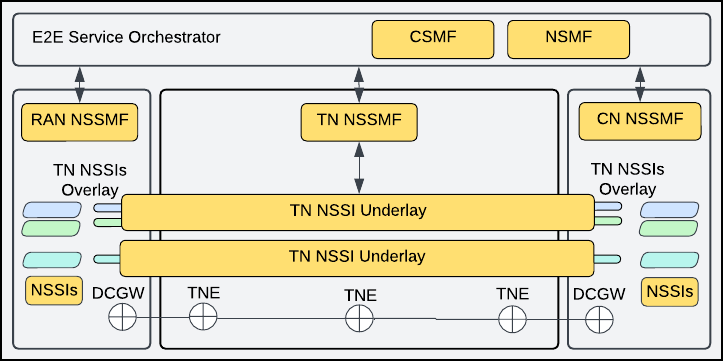}
    \caption{Functional architecture of TN slicing}
    \label{fig:RefTNSlicingArchitecture}
\end{figure}

Figure~\ref{fig:RefTNSlicingArchitecture} illustrates Xhaul \gls{TN} slicing, incorporating orchestration infrastructure {along with} the \gls{RAN}, \gls{CN}, and Xhaul \gls{TN}. It integrates \glspl{NSSI} at overlay with \gls{TN} underlays, demonstrating a comprehensive system design.

The seamless mapping of \glspl{NSI} to physical or logical TN instances is {essential for maintaining coherence between} the \gls{TN} architecture and the unique requirements of each network slice. This mapping {process} heavily depends upon the available deployment options within the Xhaul \gls{TN} \cite{ietf-teas-5g-ns-ip-mpls-02}. The following sections elucidate some key concepts as outlined in \gls{O-RAN} Alliance specification documents.

\paragraph{Transport Plane}
Within the \gls{O-RAN} \gls{TN} infrastructure, both \gls{L2} \gls{EVPN} and \glspl{L3VPN} leverage \gls{MP-BGP} to establish individual \glspl{NSI}. These \glspl{VPN} support numerous instances and endpoints while offering diverse connectivity models. Furthermore, four {distinct} approaches are outlined for constructing the underlay transport plane(s), each designed to {optimize} network performance and {address} specific {network slice operational} requirements \cite{O-RAN.WG9.XPSAAS}.

\begin{itemize}[noitemsep, topsep=0pt, left=0pt]
\item \textbf{Single transport plane for all slices:} In this configuration, a single transport plane serves as the backbone for all network slices and ensures a uniform distribution of traffic paths. {As a result}, each slice follows identical routes between network endpoints, fostering consistency and cohesion across the \gls{TN} infrastructure. This approach represents the softest form of slicing within the underlay transport plane, as it prioritizes resources sharing and harmonious coexistence among slices.
    
\item \textbf{Transport plane per \gls{5G} service type:} {The underlay network configuration to support various} \gls{5G} service types involves constructing dedicated transport planes, each customized to accommodate the distinct forwarding behaviors of different services. These transport planes are accessible to multiple customers and enables the deployment of \glspl{VPN} and traffic steering mechanism. {They} can adopt distinct topologies and optimizations depending on various criteria, such as service requirements and network conditions. For example, \gls{URLLC} service types {emphasize} reliability {by selecting} the most dependable links and optimizing paths based on link delay metrics. In contrast, \gls{eMBB} service types prioritize cost-effective, high-bandwidth links, {with paths selection} determined by \gls{IGP} metrics {that are} correlated with link capacity. For \gls{NB-IoT} service types, which do not require low latency or high capacity, a separate transport plane could be designated, with paths established based on \gls{TE} metrics that favor links {specifically suited to} these services.
    
\item \textbf{Transport plane per slice customer:} In this approach, rather than allocating a transport plane for each \gls{5G} service type, a separate transport plane is assigned to individual customers. While similar techniques are employed, scalability now depends on the number of customers utilizing the network, rather than the diversity of \gls{5G} service types. To tackle scaling challenges, a {hybrid approach can be implemented} combining the mappings per customer and per \gls{5G} service type. For example, the primary approach might entail mapping according to \gls{5G} service types {with a subset} of premium customers receiving dedicated mappings to individual transport planes.

\item \textbf{Transport plane per 5QI group:} In this {configuration}, the \gls{TN} supports the integration of traffic streams {with} different \gls{5QI} values into specific {network} slices. It enables the efficient allocation of {numerous} \glspl{5QI} values to a limited pool of \gls{TN} resources, such as queues within \glspl{TNE}. As a result, the network can effectively manage and prioritize various types of traffic within these designated {\glspl{NSI}}.
\end{itemize}

\paragraph{Quality of Service}
In \gls{TN}, \gls{QoS} {is essential to} ensure that various types of traffic receive appropriate levels of service, such as bandwidth, and latency, to meet specific performance requirements. {The importance} of \gls{QoS} becomes even more {pronounced} when slicing the transport infrastructure to accommodate {diverse} traffic types or services. {Similar to} the transport plane, various strategies can be used to provide varying levels of isolation between slices. These strategies may include traffic prioritization, bandwidth allocation, traffic shaping, and congestion management techniques \cite{ietf-teas-5g-ns-ip-mpls-02}. When addressing \gls{QoS} within the context of \gls{TN} slicing, it's crucial to consider both edge \gls{QoS} and core \gls{QoS} solutions.

\begin{itemize}[noitemsep, topsep=0pt, left=0pt]
\item \textbf{Edge \gls{QoS}:} Edge interfaces within packet transport networks often experience delays and congestion. In the context of network slicing, it's essential to {implement} traffic conditioning at the network's edge upon entry and scheduling upon exit to ensure {that} each slice retains its allocated bandwidth. Moreover, when mobile clients present traffic via \glspl{VLAN}, the \gls{PE} router {must posses} hierarchical \gls{QoS} capabilities to effectively manage both the overall allocated bandwidth at the \gls{VLAN} level and the designated class bandwidth within each \gls{VLAN}.
    
\item \textbf{Core \gls{QoS}:} In the \gls{CN}, various \gls{QoS} strategies {are employed to} manage bandwidth allocation and queue sharing across different slices to ensure optimal performance and efficient resource utilization. 
\end{itemize}

\paragraph{Service Models}
In addition to the physical infrastructure, it is essential to establish dedicated networks for managing and controlling various aspects of the \gls{3GPP} and \gls{O-RAN} frameworks. These management and control plane networks are crucial for regulating the functionality and performance of the overall {\gls{O-RAN} architecture} \cite{O-RAN.WG9.XPSAAS}. {Within the} \gls{TN} slice architectures, management networks {and frameworks} can function as independent \gls{VPN} entities, or in some cases, multiple \glspl{MF} may be consolidated into a single \gls{VPN}. This consolidation improves efficiency by centralizing tasks and control within a unified network structure, thereby simplifying \gls{MO} and enhancing overall coordination.

The \gls{TN} management network utilizes \gls{MP-BGP} \gls{L3VPN} technology to {interconnect} \glspl{TNE} through an any-to-any or hierarchical topology, with optional out-of-band networks for redundancy \cite{ietf-teas-ietf-network-slices-25}. \gls{DC} management networks provide centralized oversight of \glspl{DC}, particularly those supporting \gls{5G} and \gls{O-RAN} services, by integrating logical Ethernet interfaces with \glspl{TNE} via \gls{MP-BGP} \gls{L3VPN}. The \gls{O-RAN} \gls{FH} management network operates as a distinct \gls{VPN} {separate} from the control \gls{VPN}, ensuring effective management of mobile elements. Similarly, the \gls{O-RAN} control and management network consolidates A1, E2, and O1 interfaces into a unified \gls{VPN} that spans the \gls{TN}, while the \gls{3GPP} control plane network manages all control plane traffic and ensures seamless connectivity and operations for mobile components across {different network slice instances} and protocols \cite{O-RAN.WG9.XTRP-MGT}.

\vspace{-2.5mm}
\section{Exploring Use Cases Related to Network Slicing in O-RAN Architecture}\label{sec:UseCases}
\vspace{-1.5mm}
A use case is a concept that describes how a system can be utilized to achieve specific goals or tasks. It outlines the interactions between users or actors and the system to accomplish a particular outcome. The exploration of network slicing within \gls{O-RAN} encompasses deploying network slices for various use cases while also highlighting the diverse requirements of business customers {(also known as tenants)} seeking to realize their specific needs. These requirements may encompass ultra-reliable services, high-bandwidth communication, {massive machine-type communication,} and low latency, among many others. The \gls{O-RAN} Alliance has identified specific use cases for network slicing to showcase its potential in meeting the demands of business customers. In this section, we delve into several use cases outlined in \gls{O-RAN} Alliance specifications, expected to be supported within the {context of the} slicing-aware \gls{O-RAN} architecture. The requirements derived from these use cases will be integrated into \gls{O-RAN} architecture as network slicing requirements. Prioritizing and specifying support for these use cases by the \gls{O-RAN} community are essential, as not all of them have been realized {by the} specifications {of the \gls{O-RAN} Alliance} yet.

\vspace{-2.5mm}
\subsection{Slice Subnet Management and Provisioning Use Cases}
\vspace{-1.5mm}
The aspects related to the \gls{MO} of network slicing, including \gls{NSI} and \gls{NSSI}, are provided by the \gls{3GPP} {\gls{TSG} \gls{SA5}}. The \gls{NSI} refers to an instance of an \gls{E2E} network slice, while the \gls{NSSI} represents a part of an \gls{NSI}, such as \gls{NSSI} for the \gls{RAN} domain or \gls{NSSI} for the \gls{CN} domain. For a comprehensive detailed discussion on the lifecycle management and provisioning of both the \gls{NSI} and \gls{NSSI}, interested readers may refer to \cite{3GPP-TS-28.531,3GPP-TR-28801MgmtOrch}. In this subsection, we outline the most essential procedures for \gls{O-RAN} slice subnet management {and} Provisioning, ensuring alignment with the \gls{3GPP} slice management framework and requirements \cite{GSMA-E2ENSA}. The use case further {outlines} the steps involved in {its various phases, which include} \textit{Creation, Activation, Modification, Deactivation, Termination, Configuration, and Feasibility Check} within \gls{O-RAN} architecture. It encompass{es} a diverse array of actors {at every phase with their defined roles \cite{O-RAN.WG1.SA}. The actors include all \gls{O-gNB} components and \gls{MF} while their roles are designated} as \gls{NFMS-P}, \gls{NSSMS-P}, \gls{NSSMS-C}, and \gls{O-RAN} \glspl{NF}. Table~\ref{tab:actors&roles} outlines the actors, {their roles,} and the corresponding phases of involvement.

\begin{table}[!htbp]
\centering
\caption{Actors and their roles in different phases of O-RAN Slice Subnet Management and Provisioning use cases}
\label{tab:actors&roles}
\renewcommand{\arraystretch}{1.4}
\normalsize
\begin{tabular}{|p{1.9cm}|p{0.60cm}|p{0.35cm}|p{0.60cm}|p{0.35cm}|p{0.60cm}|p{0.35cm}|p{0.60cm}|}
\hline \hline
\rowcolor{lightgray}
\multicolumn{1}{|c|}{\textbf{{\rot{Actors}}}} &
\multicolumn{1}{c|}{\textbf{{\rot{Creation}}}} & 
\multicolumn{1}{c|}{\textbf{{\rot{Activation}}}} &
\multicolumn{1}{c|}{\textbf{{\rot{Modification}}}} & \multicolumn{1}{c|}{\textbf{{\rot{Deactivation}}}} & \multicolumn{1}{c|}{\textbf{{\rot{Termination}}}} & \multicolumn{1}{c|}{\textbf{{\rot{Configuration}}}} & \multicolumn{1}{c|}{\textbf{{\rot{Feasibility Check}}}}  \\ \hline \hline 
NSMF            & \multicolumn{7}{c|}{NSSMS\_C} \\         \hline
NSSMF           & \multicolumn{7}{c|}{NSSMS\_P} \\        \hline

NFMF            & \multicolumn{4}{c|}{NFMS\_P} &  & \multicolumn{2}{c|}{NFMS\_P} \\           \hline
SMO OAM         & \multicolumn{4}{c|}{NFMS\_P} &  & \multicolumn{2}{c|}{NFMS\_P} \\         \hline
O-gNB         & \multicolumn{7}{c|}{NF} \\      \hline
Near-RT RIC     & \multicolumn{7}{c|}{NF} \\      \hline
Non-RT RIC      & NF &  & NF &  & NF &  & NF \\         \hline
O-Cloud M\&O    & MOP &  & MOP &  & MOP &  & MOP \\     \hline

\multicolumn{8}{|l|}{MOP$\rightarrow$ O-Cloud M\&O provider within SMO} \\ \multicolumn{8}{|l|}{NF$\rightarrow$ O-RAN Network Functions} \\ \hline

\end{tabular}
\end{table}

\paragraph{Creation} The objective of this phase is to establish the \gls{O-NSSI} or initialize the existing one to meet the \gls{RAN} slice subnet requirements. {The phase assumes that the \gls{NSSMS-P} is already aware of \gls{O-Cloud} \gls{MO} and} begins when the request for an \gls{NSSI} is received by the \gls{NSSMS-P}. {The \gls{NSSMS-P} evaluates the feasibility of the request by analyzing the network slice subnet requirements. It then decides whether to modify an existing \gls{O-NSSI} or create a new slice subnet.} The \glspl{VNF} within \gls{O-RAN} will then be instantiated by a service request from \gls{NSSMS-P} to \gls{O-Cloud} \gls{MO}. The response will then be forwarded to \gls{O-NSSI}, which configures its constituents of \gls{O-NSSI} using \gls{O-RAN} \gls{NF} provisioning service \cite{O-RAN-WG6-ORCH-USE-CASES}. After that, the \gls{NSSMS-P} activates \gls{TN} Manager to establish necessary links such as A1, E2, as well as \gls{FH} and \gls{MH} connectivity. The network slice subnet requirements are forwarded to the \gls{Non-RT RIC}, and \gls{NSSMS-C} will be informed about the resulting status of this process. {Upon successful completion of all steps, the phase concludes with} the establishment of the necessary \gls{O-RAN} \glspl{NF} and \gls{O-NSSI}, along with the configuration of the \gls{Non-RT RIC} \cite{O-RAN.WG1.SA}. {The failure exception may occur due to a full or partial failure of any of the above identified steps.}

\paragraph{Activation} The goal of this phase is to activate the \gls{O-NSSI}. {It requires} that an \gls{O-NSSI} has already been created but is in an inactive state \cite{3GPP-TS-28.531}. This means that several \gls{O-RAN} \glspl{NF} may be contained in the \gls{O-NSSI} but not yet have been activated. To begin the procedure, \gls{NSSMS-C} sends a request to \gls{NSSMS-P} to activate the \gls{O-NSSI}. The \gls{NSSMS-P} then identifies and decides to activate the parts that are inactive. For example, consider the elements listed in Table~\ref{tab:preconditions}, where all the NFs are inactive since they are not shared with other O-NSSI, but Near-RT RIC is only activated for other services. \gls{NFMS-P} makes sure that all the constituents of NSSI are installed and activated on request of \gls{NSSMS-P}. When all {inactive constituents of \gls{O-NSSI}} are activated, \gls{NSSMS-P} receives a notification from \gls{NFMS-P} and notifies \gls{NSSMS-C} about the activation of the \gls{O-NSSI}. It also changes the administrative state of the \gls{O-NSSI} to unlocked. This {phase} ends {when all steps are successfully completed, activating the O-NSSI; otherwise, a failure exception is triggered} \cite{O-RAN.WG1.SA}.

\paragraph{Modification} The objective of this {phase} is to ensure compliance with \gls{O-RAN} slice subnet requirements by refining the existing \gls{O-NSSI}. The only prerequisite of this phase is that the \gls{VNF} packages for virtualized \gls{O-RAN} \glspl{NF} {intended} for the \gls{O-NSSI} have been previously incorporated \cite{O-RAN.WG1.SA}. The process initiates upon receiving a request to modify an existing \gls{O-NSSI} along with new requirements by the \gls{NSSMS-P}. Subsequently, feasibility is assessed, leading to two potential outcomes. Should the requirements prove unattainable, the \gls{NSSMS-P} informs the \gls{NSSMS-C} of the status along with \gls{O-NSSI} details. Conversely, the provided information is segmented into modification requests for each constituent of the \gls{O-NSSI}. If there are additional \glspl{O-NSSI} managed by other \glspl{NSSMS-P}, their respective \glspl{NSSMS-P} are notified of the modification via the primary \gls{NSSMS-P}, thereby activating their \glspl{O-NSSI}. Additionally, the \gls{NSSMS-P} can sequentially initiate various required aspects as below:

\begin{itemize}[noitemsep, topsep=0pt, left=0pt]
    \item A service modification request to \gls{O-Cloud} \gls{MO}, if the \gls{O-NSSI} contains virtualized parts.
    \item \gls{NF} provisioning service to reconfigure the \gls{O-NSSI} constituents, if the \gls{O-NSSI} contains \gls{NF} instances.
    \item \gls{O-RAN} \gls{TN} Manager coordination {process}, if the \gls{NSSI} contains \gls{TN} part, to set up or modify necessary {connectivity} such as A1, E2, \gls{FH}, and \gls{MH}.
\end{itemize}

Upon {successful} completion of the {above} steps, {the \gls{NSSMS-P} informs} the \gls{Non-RT RIC} of the revised network slice subnet requirements and \gls{O-NSSI} details. Subsequently, it notifies the \gls{NSSMS-C} regarding the process status, along with {relevant} \gls{O-NSSI} information. This phase concludes with {modification} to the \gls{O-NSSI} and associated \gls{O-RAN} \glspl{NF}, {as well as} the configuration of the \gls{Non-RT RIC} {to align} with the updated slice requirements and \gls{O-NSSI} specifics.

\paragraph{Deactivation} This {phase} deactivates a currently active \gls{O-NSSI}. {The prerequisite is that \gls{O-NSSI} exist, is active, and its constituent \gls{O-RAN} \glspl{NF} are not shared with other \glspl{O-NSSI}}. \gls{NSSMS-C} decides to deactivate the \gls{O-NSSI} {upon request of its authorized consumer and} sends a deactivation request to the \gls{NSSMS-P} to start the process. The \gls{NSSMS-P} identifies the active {constituents} of the {\gls{O-NSSI}} and proceeds to deactivate {them. For} example, {\gls{NSSMS-P} identifies following} active \gls{O-RAN} \glspl{NF} {that are not shared with other \gls{O-NSSI}.}

\begin{table}[!htp]
\centering
\caption{The O-RAN NFs, along with their current state, are required to satisfy the prerequisites for activation}
\label{tab:preconditions}
\renewcommand{\arraystretch}{1.4}
\normalsize
\begin{tabular}{|p{2.55cm}|p{2.5cm}|p{2.5cm}|}
\hline \hline
\rowcolor{lightgray}
\multicolumn{1}{|c|}{\textbf{{NF}}} & \multicolumn{1}{|c|}{\textbf{{Installed}}} & \multicolumn{1}{|c|}{\textbf{{Activated}}} \\ \hline \hline
Near-RT RIC & Yes & Yes \\ \hline
O-CU-CP & Yes & No \\ \hline
O-CU-UP & Yes & No \\ \hline
O-DU & Yes & No \\ \hline
O-RU & Yes (as \gls{PNF}) & No \\ \hline 
\end{tabular}
\end{table}

\begin{itemize}[noitemsep, topsep=0pt, left=0pt]
    \item The \gls{O-CU-CP} \gls{NF} constituent: {It calls the \gls{NF} provisioning service to request \gls{NFMS-P} to deactivate the \gls{O-CU-CP}. The \gls{O-CU-CP} terminates the E2 interface connecting to the \gls{Near-RT RIC} and releases the E1 interface between the \gls{O-CU-CP} and \gls{O-CU-UP}.}
    \item The \gls{O-CU-UP} \gls{NF} constituent: {It invokes the \gls{NF} provisioning service to request \gls{NFMS-P} to deactivate the \gls{O-CU-UP} and the \gls{O-CU-UP} terminates the E2 interface connection with the \gls{Near-RT RIC}}.
    \item The \gls{O-DU} \gls{NF} constituent: {It invokes the \gls{NF} provisioning service to request \gls{NFMS-P} to deactivate the \gls{O-DU}. The \gls{O-DU}} terminates the F1 interface connection with \gls{O-CU} and the E2 interface connection with \gls{Near-RT RIC}.
    \item The \gls{O-RU} constituent: {It invokes the \gls{NF} provisioning service to request \gls{NFMS-P} to deactivate the \gls{O-RU} and the \gls{O-RU} initializes to terminate} the M-Plane interface connecting to the \gls{O-DU}.
\end{itemize}

Once the {the \gls{NFMS-P} deactivates the requested constituents, it set the \textit{administrativeState} of that constituent to \textit{locked} and notify the \gls{NSSMS-P} that the constituent is successfully deactivated. The \gls{NSSMS-P} sets the \textit{administrativeState} of the \gls{O-NSSI} to \textit{locked}. Lastly, if all the steps above completes without triggering any exception,} this phase is concluded with the deactivation of the \gls{O-NSSI}.

\paragraph{Termination} This phase involves disassociating an existing but inactive \gls{O-NSSI} when it is no longer required. Upon receiving the termination request, the \gls{NSSMS-P} {takes one of two actions}. If the \gls{O-NSSI} is shared, it is disassociated {using the previously described} modification {phase}. If the \gls{O-NSSI} is non-shared, it is terminated. If there are constituent \glspl{NSSI} within the \gls{O-NSSI} that are not directly managed by the \gls{NSSMS-P}, {it requests the respective \glspl{NSSMS-P} to release them. It} also requests the \gls{O-Cloud} \gls{MO} to terminate the non-shared virtual \gls{O-RAN} \glspl{NF} that are no longer required \cite{O-RAN-WG6-ORCH-USE-CASES} and starts the \gls{TN} manager coordination process. {If all the above steps succeed, the \gls{O-NSSI} is terminated and \gls{NSSMS-P} notifies the \gls{Non-RT RIC} and the \gls{NSSMS-C} of the final status.}

\paragraph{Configuration} This phase involves (re-)configuring an existing \gls{O-NSSI}. {It assumes that \gls{NSSMS-P} is serving authorized customers and know the respective \glspl{NSSMS-P} and \glspl{NFMS-P} responsible for the management of \gls{O-NSSI} constituents and \glspl{NF}}. The \gls{NSSMS-C} initiate (re-)configuration of the \gls{O-NSSI} and its constituents {by sending the slice subnet (re-)configuration information} to the \gls{NSSMS-P}. {The \gls{NSSMS-P}} breaks down the received (re-)configuration information to prepare the configuration management for each constituent. The constituents managed directly by the \gls{NSSMS-P} is configured accordingly. If the constituents are managed by other \glspl{NSSMS-P}, {the respective \gls{NSSMS-P} is requested to configure them}. For the \gls{O-NSSI} with constituents \gls{O-RAN} \glspl{NF} managed by \gls{NFMS-P}, the \gls{NSSMS-P} sends configuration requests through the respective \glspl{NFMS-P}.

If any step fails partially or fully, an exception is triggered. Otherwise, the required (re-)configuration is successfully completed for the relevant constituent. The \gls{NSSMS-P} then sends the configuration results to the \gls{NSSMS-C}.

\paragraph{Feasibility Check} This phase assesses the possibility of provisioning an \gls{O-NSSI} and confirms whether its requirements are attainable. The precondition is that the \gls{NSSMS-C} has acquired or received the necessary requirements for the network slice subnet. To start the {feasibility check}, if an \gls{O-NSSI} meets the network slice subnet requirements, {\gls{NSSMS-C} sends a request to the \gls{NSSMS-P}}. The \gls{NSSMS-P} then identifies the involved constituents and {may consult} the \gls{SMO} and the \gls{Non-RT RIC} regarding the fulfillment of requirements. It then checks the availability of network constituents by submitting reservation requests to the \gls{O-Cloud} \gls{MO}.

In addition, the \gls{NSSMS-P} may request the \gls{TN} manager to collect information regarding the feasibility of the \gls{TN} links. {If all steps completed successfully without an exception, the \gls{NSSMS-P} provides} the feasibility check results, including details of reserved resources to the \gls{NSSMS-C}. {Subsequently, the feasibility check phase is concluded}.

\vspace{-2.5mm}
\subsection{RAN Slice SLA Assurance}
\vspace{-1.5mm}
The \gls{3GPP} standards provide a {flexible} \gls{5G} infrastructure that enables the creation and management of customized networks to meet diverse requirements across various applications and business verticals. These {standardized} requirements {define key} performance metrics such as throughput, energy efficiency, latency, and reliability \cite{O-RAN.WG1.Use-Cases}. Network slicing that extends across the \gls{CN}, \gls{TN}, and \gls{RAN}, {ensures the strict adherence to performance} criteria throughout the entire lifecycle of a network slice, with a particular emphasis placed on the \gls{RAN} architecture \cite{park_technology_2023}. 
However, the dynamic nature of the \gls{RAN} architecture {makes it challenging to maintain} consistent service quality for each \gls{RAN} slice within the complex multi-vendor \gls{O-RAN} environment \cite{brik2023survey,wu2020dynamic-RAN-Slicing}. Addressing this challenge {requires further research and} standardization efforts to establish the mechanisms and parameters for the \gls{RAN} slice \gls{SLA} assurance \cite{10329597}. The \gls{SLA} is a contract between the network service provider and the customer, defining responsibilities, performance standards, and service expectations \cite{10316961}.

{O-RAN with its} open interfaces and \gls{AI}/\gls{ML}-assisted architecture offers a promising {approach} for implementing the mechanisms that enable operators to {fully leverage the }opportunities of slicing \cite{GSMA-E2ENSA}. For instance, the \gls{O-RAN} architecture and interfaces empower operators to optimize spectrum resource utilization by dynamically allocating resources across slices {based on changing} usage patterns. The use case progresses through the following phases:

\subsubsection{Creation and Deployment of RAN slice SLA Assurance Models and Control Apps} In this phase, the training and deployment of the model begin with the activation of an \gls{O-RAN} slice. {The prerequisites include an established A1 interface between the \gls{Near-RT RIC} and the \gls{Non-RT RIC}, as well as an O1 interface between the \gls{SMO} and the \gls{Near-RT RIC}}. {The phase starts with \gls{RAN} slice activation}. The \gls{Non-RT RIC} retrieves a \gls{RAN} slice \gls{SLA} from the \gls{SMO} framework, specifically the \gls{NSSMF}, then collects performance measurements {(e.g., CSI, latency)} via the O1 interface and enrichment information {(e.g., public safety apps, location-based information)} from external applications. The \gls{Non-RT RIC} {then} analyzes {collected performance measurements} and/or enrichment information over an extended monitoring period, which contributes in the model training process \cite{park_technology_2023}.

The \gls{Non-RT RIC} {performs} model training and {obtains} \gls{RAN} slice \gls{SLA} assurance models using {either} an \gls{AI}/\gls{ML} model or a control app \cite{tripathi2025fundamentals}. {If} an \gls{AI}/\gls{ML} model is used, it can be deployed internally for slow loop optimization or sent to the \gls{Near-RT RIC} via the O2 interface for fast loop optimization. Conversely, if a control app is {chosen}, the \gls{SMO} deploy it to the \gls{Non-RT RIC} for slow loop optimization or transfer it to the \gls{Near-RT RIC} via O2 interface for fast loop optimization. The \gls{Non-RT RIC} {updates the \gls{RAN} slice \gls{SLA} assurance model and control app based on received feedback, either} internally or from the \gls{Near-RT RIC} via the A1 interface. The phase terminates with the deactivation of the \gls{RAN} slice.

\subsubsection{Slow Loop RAN Slice SLA optimization} This phase achieves slow loop \gls{RAN} slice \gls{SLA} optimization. The preconditions for this phase mirror those of the Creation and Deployment phase, with the addition that the \gls{RAN} slice \gls{SLA} assurance model or control apps are already deployed. The \gls{Non-RT RIC} {has} two options for slow loop optimization. It can adjust the \gls{RAN} configuration in accordance with long-term trends, {using data from the} O1 interface or develop A1 policies tailored to the requirements of the \gls{RAN} slice \gls{SLA}. {The A1 policies} incorporate inputs such as A1 feedback, O1 long-term trends, and operator-defined \gls{RAN} intents.

In the second option the \gls{SMO} framework updates the slice configuration of the \gls{Near-RT RIC} or \gls{RAN} nodes based on instructions from the \gls{AI}/\gls{ML} model or control app. After the update, two outcomes are possible: either the \gls{Near-RT RIC} and the \gls{RAN} nodes implement the updated configuration, or the \gls{Near-RT RIC} receives the updated A1 policies, take control of the \gls{RAN} nodes and provide feedback to the \gls{Non-RT RIC}.

\subsubsection{Fast Loop RAN Slice SLA optimization} In this phase, the \gls{Non-RT RIC} evaluates the necessity {to generate} a policy to ensure slice \gls{SLA} assurance for the \gls{Near-RT RIC}. The evaluation is based on the \gls{RAN} slice \gls{SLA} requirements and operator-defined \gls{RAN} intents. It also considers feedback from the \gls{Near-RT RIC} via the A1 interface or long-term trends observed through the O1 interface, as well as enrichment information from external application servers.

Afterwards, the \gls{Near-RT RIC} is furnished with slice-specific O1 configurations from \gls{SMO} and A1 policies from the \gls{Non-RT RIC}. It proceeds to collect performance measurements via the E2 interface. {The collected performance measurements}, combined with the A1 policies from the \gls{Non-RT RIC}, and analyzed by to guide the \gls{RAN} nodes to meet the slice \gls{SLA}. The phase concludes with the deactivation of the \gls{RAN} slice.

\vspace{-2.5mm}
\subsection{Managing Multi-vendor Network Slices}
\vspace{-1.5mm}
This use case involves {managing} multiple network slices, each incorporating the \gls{RAN} components from different vendors. For example, network slice 1 uses \gls{O-DU} and \gls{O-CU} from vendor A, while network slice 2 employs components from vendor B, with \gls{O-RU} from vendor C being shared between both slices \cite{O-RAN.WG1.UseCases-AR}. This {enables the use} of different slices for {specific} application scenarios, as each component offers unique specifications. While the implementation may vary, they all involve a single \gls{O-RU} connected to one or more \glspl{O-DU}. To support multiple slices, the schedulers of the \gls{vO-DU} and \gls{vO-CU} must manage each \gls{NSI} separately \cite{brik2023survey}.

The vendor providing \gls{vO-DU} and \gls{vO-CU} functionalities must have a robust {service-specific} customized scheduler. Moreover, effective coordination between the \gls{vO-DU} and \gls{vO-CU} is essential {for seamless allocation} of radio resources in multi-vendor slices, preventing conflicts. The coordination is evaluated based on service objectives and their {impacts on} the \gls{O-RAN} architecture \cite{O-RAN.WG1.Use-Cases}. For instance, the following three potential coordination approach could be explored:

\par \textbf{Case 1:} The resource allocation between the \gls{vO-DU} and \gls{vO-CU} is {managed} with loose coordination through the O1/A1/E2 interface. Each \gls{vO-DU} and \gls{vO-CU} pair is responsible for allocating radio resources to individual {business} customers within the radio resources allocated by {both} the \gls{Near-RT RIC} or the \gls{Non-RT RIC}.
\par \textbf{Case 2:} A moderate {level of} coordination where the resource allocation can be negotiated between slices or between the \gls{vO-DU}/\glspl{vO-CU} via the X2 and F1 interfaces, {after managed} through the O1/E2/A1 interface. The negotiation period is extended to several seconds, influenced by the periodic exchange of the X2 and F1 messages between the \glspl{vO-CU}.
\par \textbf{Case 3:} A tight coordination through a new interface between the \glspl{vO-DU} for adaptive resource allocation, which needs a more frequent negotiation.

The utilization of multi-vendor {network} slices is applicable in scenarios involving \gls{RAN} sharing. In such cases, two {network} operators possess their respective \gls{vO-DU} and \gls{vO-CU} components from distinct vendors while jointly utilizing the \gls{O-RU} component. However, the scenario with \gls{O-DU} and \gls{O-CU} components from different vendors within a single slice requires further examination \cite{O-RAN.WG1.UseCases-AR}.

Adopting a multi-vendor approach cultivates a resilient and adaptable network ecosystem, benefiting operators and end-users alike. Upon the successful implementation of multi-vendor scenarios, the anticipated benefits include:

\paragraph{Flexibility and time-to-market deployment} Numerous vendors offer virtualized \gls{RAN} components like the  \gls{vO-DU}, \gls{vO-CU}, and schedulers for different network slices. Network operators can thus select the most suitable components for each network slice, whether they prioritize high data rates or low latencies. This flexibility also enables network operators to introduce new services effortlessly, with the option to implement additional functions from different vendors without changing their existing setups and configurations \cite{s23218792}.

\paragraph{Flexible deployment for RAN equipment sharing} In scenarios where multiple vendors aim to share \gls{RAN} equipment and resources, challenges may arise concerning vendor selection and the placement of \gls{RAN} functions. However, by addressing these challenges through collaborative use cases, network operators can reach agreements on shared \gls{RAN} equipment and resources, thereby optimizing \gls{CAPEX} and \gls{OPEX} \cite{groen_implementing_2023,9812489} and potentially opening doors to further business investment opportunities.

\paragraph{Supply chain risks reduction} In scenarios where a vendor discontinues support for certain \gls{vO-DU} and \gls{vO-CU} functions due to business circumstances, network operators retain the ability to implement substitute \gls{vO-DU} and \gls{vO-CU} functions from different vendors within a multi-vendor framework. This proactive approach serves to alleviate potential risks to network operators' ongoing business operations, bolstering their resilience amidst market dynamics \cite{O-RAN.WG1.UseCases-AR}.

\vspace{-2.5mm}
\subsection{NSSI Resource Allocation Optimization}
\vspace{-1.5mm}
The increasing complexity of {the existing} \gls{5G} {and emerging \gls{6G}} networks, marked by the proliferation of millimeter-wave small cells and diverse services like \gls{eMBB}, \gls{URLLC}, and \gls{mMTC}, poses {significant} challenges in dynamically and efficiently allocating resources among network nodes \cite{9627832}. These services, realized as \glspl{NSI}, exhibit varying characteristics such as high-speed data, ultra-low latency, and sporadic traffic patterns influenced by factors such as time, location, \gls{UE} distribution, application types, {and others}.

To tackle the aforementioned challenges, the optimization of resources allocated to \gls{NSSI} is crucial. Various scenarios, such as \gls{IoT} applications running during off-peak hours or weekends and large events causing a surge in data flow, are considered. The data collected from the \gls{O-RAN} nodes serves as input to train an \gls{AI}/\gls{ML} model embedded within the \gls{NSSI}, enabling proactive determination of traffic demand patterns for different times and locations across network slices. This approach facilitates the automatic {and intelligent} reallocation of resources ahead of network issues, optimizing resource utilization, and ensuring flexibility in responding to diverse service requirements \cite{brik2023survey}.

Implementing resource quota policies within \glspl{NF}, notably E2 nodes within their respective \glspl{NSSI}, facilitates efficient management of resource allocation across diverse slices \cite{O-RAN.WG1.Use-Cases}. This flexibility enables the prioritization of resource distribution based on service importance, fostering effective resource sharing during periods of both abundance and scarcity. Premium service slices within an \gls{NSSI} may receive a more substantial allocation of resources compared to standard or best-effort service slices, while emergency services also benefit from additional resource allocation during critical situations \cite{9003208}. Acting as constraints for resource allocation, these policies aim to optimize resource utilization across slices. They are adaptable and can be tailored to specific requirements, such as analyzing past resource allocation failures evident in \gls{RAN} node measurements. This ensures optimal utilization, mitigates historical trends, and minimizes resource inefficiencies.

The \gls{O-RAN} {components} involved in this use case are the \gls{SMO} framework, the \gls{Non-RT RIC}, and the \gls{O-RAN} nodes. The \gls{SMO}  establishes the default \gls{NSSI} resource quota policy, which acts as a parameter for optimizing resource allocation. Meanwhile, the \gls{Non-RT RIC} gathers performance metrics from the \gls{O-RAN} nodes, employs the \gls{AI}/\gls{ML} models to analyze historical data, predicts traffic demand patterns, and determines appropriate resource adjustments for each \gls{NSSI} \cite{O-RAN.WG1.SA,O-RAN.WG1.Use-Cases}. Subsequently, the \gls{Non-RT RIC} optimizes the \gls{NSSI} resource allocation by adjusting attributes and updating cloud resources through the O1 and O2 interfaces, respectively. The \gls{O-RAN} nodes facilitate performance data collection and configuration updates regarding the \gls{NSSI} resource allocation via the O1 interface. They also facilities management data collection.

The process of the \gls{NSSI} Resource Allocation Optimization on the \gls{Non-RT RIC} may encompass the following steps:

\paragraph{Monitoring} The \gls{Non-RT RIC} monitors the \gls{RAN} to collect data through the O1 interface and gathers \gls{RAN} performance measurements from the \gls{RAN} nodes.

\paragraph{Analysis \& decision} The \gls{Non-RT RIC} leverages an appropriate \gls{AI}/\gls{ML} {models} to analyze measured data and forecast future traffic demand for each \gls{NSSI} within a specified time interval and geographical location. Based on this analysis, {the AI/ML model} determines the necessary actions to adjust resources such as the \gls{VNF} resources and slice subnet attributes for the \gls{RAN} \glspl{NF} specifically the E2 Nodes within their respective \gls{NSSI} at the designated time and location.

\paragraph{Execution} The \gls{Non-RT RIC} executes operations through two sequential steps guided by model inference. Firstly, it adjusts slice subnet attributes via the \gls{OAM} functions in \gls{SMO} framework, utilizing O1 interface to configure E2 nodes \cite{O-RAN.WG10.OAM}. Secondly, it triggers a request to the \gls{O-Cloud} \gls{MO} to update resource allocation via the O2 interface. The \gls{SMO} framework coordinates these operations following recommendations from the \gls{Non-RT RIC}.

\vspace{-2.5mm}
\section{Key Lessons Learned}\label{Sec:LessonsLearned}
\vspace{-1.5mm}
In this section, we summarize {a number of} key insights and lessons learned from our study on the ongoing research, development, {and deployment efforts} of slicing-aware \gls{O-RAN} architecture. Drawing on a comprehensive review of the literature within the research community, as well as documents from various \glspl{SDO} {(mainly the \gls{O-RAN} Alliance)}, we identify several critical observations and valuable lessons. Below, we present a list of these major lessons learned.

\vspace{-2.5mm}
\subsection{Lessons Learned Related to the Architecture of O-RAN}
\vspace{-1.5mm}
We discussed on several occasions that the \gls{O-RAN} architecture is composed of key components and interfaces that support its open, intelligent, and modular design. We gained several valuable insights from our study on these components and interfaces. The key takeaways are summarized below.

\textbf{Fully Disaggregated Cloud-based RAN:} The adoption of \gls{O-RAN} marks a pivotal shift towards fully disaggregated \gls{RAN} architectures, where key \gls{O-gNB} functions are modularized into distinct components such as the \gls{O-CU}, \gls{O-DU}, and \gls{O-RU}. The key contribution of \gls{O-RAN} Alliance is the definition of \gls{O-FH} interface that splits the \gls{DU} into \gls{O-DU} and \gls{O-RU}. \textbf{We learned that such a disaggregation can play a significant role in the isolation of \gls{O-RAN} slices. To further enhance isolation, resource efficiency, and {support for} cloud-native solutions, it {may be beneficial for} network operators and vendors to further split the \gls{O-CU} and \gls{O-DU} into micro \glspl{VNF}, such as the \glspl{VNFC} defined by the \gls{ETSI} \gls{ISG} \gls{NFV}.}

\textbf{RAN Intelligent Controller:} Within the context of the \gls{O-RAN} architecture, the \gls{RIC} is pivotal in managing and optimizing \gls{RAN} functions through various control loops. These control loops, are designed to operate at various time scales to enhance network performance. For example, near-real-time control loops operating in milliseconds to seconds allow the \gls{RIC} to manage dynamic tasks such as load balancing, interference management, and resource allocation. In contrast, non-real-time control loops operating above second support broader network optimization goals, such as policy-based configurations and performance tuning. Depending on the control loops, the \gls{RIC} is classified as \gls{Near-RT RIC} and \gls{Non-RT RIC}. The \glspl{RIC} are equipped with specialized software tools, known as xApps and rApps, to enhance \gls{RAN} automation and intelligence. \textbf{The key takeaway from these intelligent controllers is that by enabling {their} distinct control loops, the \gls{RIC} facilitates continuous, adaptive decision-making that enhances the efficiency and responsiveness of the \gls{RAN}. Therefore, this approach contributes to a more intelligent and automated environment in \gls{O-RAN}}.

\textbf{Service Management \& Orchestration:} The \gls{SMO} centralizes the \gls{MO} of the resources in \gls{O-RAN}. It plays a vital role in optimizing and automating the network {by monitoring the health, performance, and \gls{QoS} within the \gls{O-RAN} ecosystem}. The integration of rApps within the \gls{SMO} enables a more granular level of control and adaptability in the \gls{RAN} by supporting various time-sensitive and {critical functionalities}. This setup not only enhances real-time decision-making capabilities but also fosters \gls{E2E} automation and self-optimization across the {\gls{O-RAN} architecture}. The \gls{SMO} could further evolve by integrating additional management functions, such as slicing management from other \glspl{SDO}, including \gls{3GPP} and \gls{NFV-MANO}, for the orchestration of \gls{O-RAN} slicing. \textbf{{Throughout our study,} we learned that this integration would support a more robust, flexible, and standardized approach to network management within \gls{O-RAN}. Combined with xApps and rApps, the \gls{SMO} offers network operators a comprehensive management tool, {which enables} seamless orchestration, automation, and enhanced adaptability across {the \gls{O-RAN} architecture}.}

\vspace{-2.5mm}
\subsection{Lessons Learned Related to Standardization and Interfaces}
\vspace{-1.5mm}
In \gls{O-RAN}, a major advancement lies in the {development and promotion of open standards. One major obstacle has been the absence of standardized interfaces and protocols. This makes it difficult for hardware and software from different vendors to work seamlessly together. The} adoption of open interfaces, which include both {the interfaces} inherited {from} \gls{3GPP} and additional new open interfaces defined by the \gls{O-RAN} {Alliance} \cite{spirent2023}. This approach facilitates a broader ecosystem of interoperability, moving beyond traditional proprietary interfaces to enable a more diverse integration of hardware and software solutions from multiple vendors. \textbf{{One} key lesson is that open interfaces in cellular networks reduce vendor lock-in, {which enables} flexible component selection and fostering vendor diversity. {However, this} interoperability supports customized configurations, {which drives} innovation and efficiency by allowing seamless multi-vendor integration. {This ultimately makes} the {\gls{O-RAN} architecture} more adaptable to {the} changing demands {in 5G, 6G, and beyond cellular networks}.}

\vspace{-2.5mm}
\subsection{Lessons Learned Related to Vendor Diversity and Collaboration Across the O-RAN Ecosystem}
\vspace{-1.5mm}
Alongside open interfaces, a key advantage of \gls{O-RAN} is vendor diversity. By decoupling hardware from software, operators are no longer tied to a single vendor. This approach allows network operators to choose the best of breed solutions \cite{YGNEC}. This fosters a competitive multi-vendor ecosystem, driving innovation, reducing costs, and accelerating the deployment of new technologies. \textbf{Based on this, another key lesson is the role of \gls{O-RAN} in {defining the O-Cloud reference architecture}, which emphasize the cloudification and automation of \gls{RAN} functions}. This transformation has attracted prominent \gls{IT} companies, historically focused on infrastructure and software, into the telecommunication sector. These players are not only contributing essential infrastructure like servers and \gls{CaaS} platforms but are also entering the realm of telecommunication by developing critical \gls{RAN} components such as the \glspl{RIC} and \gls{O-CU}/\gls{O-DU} units {\cite{10056724}}. \textbf{Their involvement is accelerating innovation, strengthening standards development, and expanding global testing capabilities. This trend underscores a major shift towards software-driven \gls{RAN} functions, with cloud and automation technologies shaping the next generation of mobile networks, setting a foundation for more adaptive, efficient, and scalable wireless communication infrastructures.}

\vspace{-2.5mm}
\subsection{Lessons Learned Related to Network slicing}
\vspace{-1.5mm}
{The integration of network slicing with \gls{O-RAN} allows network operators to deliver more tailored network services while maximizing commercial benefits \cite{101145}. In \gls{O-RAN}, network slicing} leverages its disaggregated {architecture} and the separation of user and control plane traffic, {enhancing flexibility and efficiency.} This approach {enables} \gls{RAN} \glspl{NF} to be deployed across edge {, regional,} or central clouds {depending} on the specific {requirements of the} use case and application. {This flexibility allows} network resources to {be dynamically adapted to diverse} needs. The \gls{TN} slicing in \gls{O-RAN} is implemented through \glspl{VPN}, categorizing transport flows into distinct transport slices represented by \glspl{NSI}. \textbf{A key lesson learned is that this approach makes the mapping of \glspl{NSI} to specific physical or logical transport networks a critical aspect of \gls{TN} slicing, as precise mapping ensures that each slice receives the required network resources and isolation for the respective use case.} Currently, as of \gls{O-RAN} slicing phase-3, \gls{O-RAN} supports slicing over the \gls{MH} and \gls{BH} segments, while \gls{FH} slicing remains unsupported but is anticipated to be introduced in later releases of \gls{O-RAN} specifications.

\vspace{-2.5mm}
\subsection{Lessons Learned Related to Transport Network in O-RAN}
\vspace{-1.5mm}
In \gls{O-RAN}, the \gls{TN} is divided into distinct segments—\gls{FH}, \gls{MH}, and \gls{BH}—and is primarily supported by a packet-switched architecture. This architecture relies on an underlay fabric, typically based on \gls{MPLS} and \gls{SRv6}, to ensure reliable and efficient data transport across each network segment {in both upstream and downstream directions}. Overlay services, provided by L2/L3 \glspl{VPN} such as \gls{EVPN} and \gls{L3VPN}, offer flexible, \gls{E2E} connectivity and enable service isolation. \textbf{{One} key lesson learned from our study with respect to \gls{TN} is that this segmented, packet-switched approach enhances data flow and scalability across \gls{O-RAN}, enabling efficient and adaptable connections throughout a cellular network.}

\vspace{-2.5mm}
\section{Existing Major Research Challenges}
\label{Sec:ResearchChallenges}
\vspace{-1.5mm}
{The} \gls{O-RAN} {architecture and} technologies hold substantial potential to drive the evolution of mobile networks toward next-generation solutions. {They offer} a flexible, disaggregated, and multi-vendor architecture {for cellular networks}. Although the key principles and specifications for \gls{O-RAN} are established, the technology remains in its early development stages. As with any emerging technology, the widespread adoption of \gls{O-RAN} presents significant challenges. This section outlines some key {research and enginnering} challenges, highlighting several obstacles as identified throughout this research work that must be addressed to ensure the successful deployment and integration of \gls{O-RAN} in cellular networks.

\vspace{-2.5mm}
\subsection{Challenges in Multi-Vendor Interoperability}
\vspace{-1.5mm}
{The} \gls{O-RAN} {architecture} strives to foster seamless {multi-vendor} interoperability. {This approach empowers cellular network} operators to evade vendor lock-in and stimulate a more competitive market. However, ensuring seamless integration of components {(both software and hardware)} from diverse vendors remains one of the most significant research challenges. \textbf{For full multi-vendor interoperability, the E2 interface must undergo more extensive testing, with detailed test definitions and profiling}. This testing involves ensuring that components from different suppliers can work harmoniously without compromising performance, security, manageability, {or other \glspl{KPI}}. \textbf{Experts also doubt whether the industry will unite around a single set of standards}, as vendor-specific interpretations, rapid technological changes, and integration with legacy cellular systems complicate the goal of true ``plug-and-play" interoperability \cite{peter101108}. {Therefore}, \textbf{some {experts} see full interoperability within the context of \gls{O-RAN} as more aspirational than realistic}.

\vspace{-2.5mm}
\subsection{Optimizing Performance and Resource Management}
\vspace{-1.5mm}
Transitioning from traditional, vertically integrated \gls{RAN} to a cloud-based {RAN architectures}, multi-vendor \gls{O-RAN} presents a complex environment where network resources must be allocated and managed effectively to ensure seamless, reliable, and real-time communication across a wide range of applications {and use cases}. \textbf{One of the primary obstacles in achieving optimal network performance in \gls{O-RAN} is the allocation and utilization of network resources}. While current cellular networks already struggle with managing diverse traffic flows against network capacity, the challenge is magnified in an \gls{O-RAN} {architecture}, which must support a variety of services {and applications} with unique demands \cite{aryal03967401}. {In addition, the} \gls{O-RAN} {architecture} should be adaptable to meet the demands of dynamic resource management, enabling support for applications and services like network slicing that cater to distinct use cases or user groups.

The \glspl{RIC}, with {their} xApps and rApps, play a crucial role in enabling intelligent, targeted resource allocation. {This approach allows the} \gls{O-RAN} systems and technologies to adapt dynamically to changing network conditions. In reference \cite{aryal03967401} numerous studies have been presented that highlight xApp-based solutions for optimized resource allocation, congestion management, and enhanced network performance. {This demonstrates a} strong potential for maintaining high service quality in {the practical implementations of} \gls{O-RAN}.

\textbf{\gls{QoS} is another pivotal aspect for \gls{O-RAN} performance, especially for latency-sensitive applications and services like autonomous driving and telemedicine that require \gls{URLLC}} \cite{9814435}. Achieving ultra-low latency is a significant {research and engineering} challenge in \gls{O-RAN}, especially in defining Open \gls{FH} interface requirements between the \gls{O-DU} and \gls{O-RU} of an \gls{O-gNB} \cite{en15072429}. Additionally, \textbf{determining the most effective placement of functionalities across the network and ensuring scalability are crucial to meeting performance objectives of \gls{O-RAN}}. While high performance often demands a more complex architecture, simpler designs may constrain system capabilities {\cite{YGNEC}}. Balancing these factors is key to optimizing both efficiency and adaptability within the {context of \gls{O-RAN} architecture}.

To achieve {the aforementioned objective}, \textbf{critical factors like dynamic service chaining, virtualized operating systems, deployment strategies, and functional grouping require thorough analysis} \cite{aryal03967401}. For example deploying network functions closer to the edge {of a cellular network} can help reduce congestion at key interfaces, and improving data flow. Additionally, numerous studies suggest using \gls{RIC} applications to optimize network performance metrics, implementing power regulation to enhance throughput, and continuously monitoring \gls{AI}/\gls{ML} models to sustain high-quality network services and prevent performance declines.

\vspace{-2.5mm}
\subsection{Leveraging Automation and AI for O-RAN Management}
\vspace{-1.5mm}
{Network and service automation} in \gls{O-RAN} presents both opportunities and challenges, as it goes beyond merely introducing open interfaces to drive the cloudification and automation of \gls{RAN} operations. \textbf{To fully harness the potential of \gls{O-RAN} and address the increasing complexity, network operators can heavily invest in cloud computing, edge computing, automation, and orchestration technologies}. As networking controllers {and orchestrator} have evolved from hardware-based solutions to software-defined controllers and are now moving towards \gls{AI}-driven networks {and services}, this shift introduces new layers of intricacy.  

The \gls{RIC} and the integration of rApps and xApps will pave the way for a highly automated \gls{OAM} of \gls{O-RAN}, achieving the vision of a \gls{ZSM}, defined within the \gls{ETSI} framework \cite{Cisco5GOranDeployment}. However, \textbf{integrating these standards into an automated \gls{O-RAN} system while balancing operational efficiency with fault tolerance requires substantial technological advancements and collaboration across the ecosystem}. The increasing reliance on \gls{AI}/\gls{ML} in future networking systems highlights the inevitable need for robust automation solutions that can handle the complexities of \gls{O-RAN} effectively.

\vspace{-2.5mm}
\subsection{Conflict Mitigation Between the Applications of RICs}
\vspace{-1.5mm}
{The applications of \gls{RIC}}, including xApps and rApps, are designed to optimize and manage \gls{O-RAN} operations {and maintenance} by using advanced \gls{AI}/\gls{ML} algorithms. These include training models on live data, maintaining low-latency performance, and ensuring that \gls{AI}/\gls{ML} models operate transparently and without bias. However, \textbf{conflicts between these applications can arise when multiple applications are working to optimize the same network parameters simultaneously}, leading to potential performance degradation \cite{10439167}. Effective mitigation strategies {for conflict detection, resolution, and avoidance} are essential to ensuring seamless operation among the applications of both types of \glspl{RIC}.

To address conflicts that arise between xApps and rApps within the \gls{Near-RT RIC} and \gls{Non-RT RIC}, a conflict mitigation module is introduced {to both RICs}. These conflicts may relate to specific users, bearers, or cells, and often stem from \gls{RIC} configurations or actions. According to the WG3 specifications {of the \gls{O-RAN} Alliance}, conflicts fall into three categories: direct, indirect, and implicit.

Direct conflicts, which are easily identifiable by the conflict mitigation module, occur when multiple xApps apply conflicting configurations to the same control target or request more resources than are available. In such cases, the conflict mitigation module resolves the issue by determining which xApp takes precedence and limiting the control action accordingly. On the other hand, indirect and implicit conflicts are less obvious and harder to detect, as they do not present a direct relationship between the conflicting xApps. For instance, one app might optimize the network for certain user groups while unintentionally degrading performance for others. \textbf{These types of conflicts are identified and managed through ongoing verification and system monitoring} after control policies are implemented within the \gls{Near-RT RIC} and \gls{Non-RT RIC}.

\vspace{-2.5mm}
\subsection{Challenges Related to Network Slicing and Orchestration}
\vspace{-1.5mm}
For \gls{O-RAN} slicing, different split configurations are essential to accommodate various slice types, fulfilling their unique performance and resource requirements effectively. Additionally, \textbf{in the \gls{TN}, slicing is not yet {fully} supported on \gls{FH} components due to the lack of slicing capabilities in \glspl{O-RU} and \glspl{O-DU}}. This functionality is anticipated to be introduced in phase-5. A key use case is the "shared \gls{O-RU} scenario", where a single \gls{O-RU} is expected to serve multiple slices and multiple \glspl{O-DU}. In this scenario, the system should be capable of mapping \gls{PLMN} ID information to the corresponding \gls{VLAN} and optional \gls{IP} pair on the control and user planes of the Open \gls{FH} interface.

\textbf{Developing effective mechanisms for the \gls{MO} of various types of slices within \gls{O-RAN} is another pressing challenge}. The intricacies of assigning resources to different network slice instances, guaranteeing \glspl{SLA} which are tailored to the requirements of diverse use cases. Achieving this while balancing resource efficiency and avoiding over-provisioning is {a critical research challenge} for optimizing network slice performance and {therefore requires} further research and study.

Moreover, \textbf{managing network slice instances across an \gls{E2E} cellular network—from the \gls{RAN}, \gls{TN}, to the \gls{CN}—adds another layer of complexity}. Each network segment must be able to adapt dynamically to fluctuating traffic demands and maintain seamless coordination with the \gls{SMO} framework to enable real-time control of resources. The \gls{O-RAN} architecture, with its disaggregated and multi-vendor environment, \textbf{further complicates the \gls{MO} process as it requires precise synchronization among multiple components and vendors}.

Additionally, the integration of \gls{AI}/\gls{ML} into the orchestration process holds promise for automating and optimizing network slice management. Intelligent applications, such as rApps and xApps running on the \glspl{RIC}, can predict traffic demands, optimize resource allocation in real-time, and enforce \gls{SLA} policies. However, \textbf{implementing these \gls{AI}/\gls{ML}-driven optimizations across a multi-vendor \gls{O-RAN} {architecture} remains an open research area} due to the need for standardized interfaces and seamless data sharing across different vendors.

Network slicing within {the} \gls{O-RAN} architecture \textbf{involves a multi-dimensional challenge that spans resource management, \gls{SLA} enforcement, orchestration, real-time optimization, security, and multi-vendor coordination}. Addressing these {major} research challenges requires not only advanced algorithms and \gls{AI}/\gls{ML} integration but also the development of standardized frameworks that support interoperability, unification, and automation across the diverse \gls{O-RAN} components and interfaces {in next-generation of wireless networks}.

\vspace{-2.5mm}
\subsection{Technical and Standardization Gaps}
\vspace{-1.5mm}
While \gls{O-RAN} offers clear benefits, and build upon \gls{3GPP} architecture and protocol, also collaborate with \gls{ETSI}, there are gaps in standardization and technical implementations, especially related to the full integration of network slicing capabilities and automation. \textbf{Harmonizing \gls{O-RAN} standards with existing \gls{3GPP}, \gls{ETSI}, and other \glspl{SDO} is an ongoing standardization and research challenge}. There is a need for unified frameworks that can accommodate the diverse requirements of \gls{O-RAN} and traditional network infrastructures. To advance the growth of open cellular networks, these organizations need to work together on establishing common standards, conducting interoperability tests, supporting open source development, and advocating for effective policies.

\vspace{-2.5mm}
\subsection{Security Challenges and Risk Mitigation}
\vspace{-1.5mm}
{The} \gls{O-RAN} {architecture} introduces several security challenges due to its openness and disaggregation {principles}, which expand potential threats to network and user data {\cite{10807044}}. \textbf{These challenges include vulnerabilities in the global supply chain and increased attack surfaces}. Network operators need advanced monitoring systems to detect and prevent threats, while also leveraging automation and distributed security analytics. Virtualized environments enable quick deployment of security patches, but compliance with relevant security standards and certifications e.g., \gls{3GPP}, \gls{ETSI} is essential. Additionally, \textbf{the use of open source protocols, third-party interfaces, and cloud services requires careful security management to ensure network resilience and reliability}.

\vspace{-2.5mm}
\subsection{Challenges in Managing the O-RAN Ecosystem}
\vspace{-1.5mm}
The decentralized nature of \gls{O-RAN} presents this significant research challenge, \textbf{as no single vendor is responsible for the entire \gls{E2E} \gls{RAN} implementation}. Although the \gls{OSC} has established standard interfaces to facilitate interoperability among vendors, challenges persist regarding operations, administration, and maintenance. For example, alarm handling, system commissioning, fault resolution, and performance monitoring in a multi-vendor environment. In the event of unexpected alarms or \gls{KPI} degradation, effective troubleshooting relies on collaborative efforts among all involved vendors to pinpoint the root cause and implement corrective measures. \textbf{Without well-defined processes for engineering support and escalation, such situations can lead to operational delays or conflicts between network operators and suppliers} within the \gls{O-RAN} {architecture}.

In such a multi-vendor \gls{O-RAN} environment, \textbf{accurately diagnosing issues and tracing them to their root cause is essential to avoid miscommunication and delays in fault resolution}. This requires advanced systems capable of real-time diagnostics, alongside the integration of \gls{AI}/\gls{ML} applications that can predict faults and failures before they happen. Such predictive capabilities enable operators and vendors to take timely, preemptive action, ensuring network stability and minimizing downtime. However, \textbf{developing \gls{AI} models that can adapt to the complex and dynamic nature of \gls{O-RAN} presents a significant challenge}.

Furthermore, \gls{O-RAN} must remain adaptable to the continually evolving cellular network landscape. {This necessitates} regular updates from third-party providers to maintain compatibility. Managing these adjustments within a multi-vendor ecosystem adds complexity, with additional operational costs. While \gls{O-RAN} seeks to lower overall \gls{RAN} expenses through enhanced interoperability, the ongoing setup and maintenance across multiple vendors may lead to higher expenditures over time. \textbf{This highlights the need for robust cost-monitoring practices to effectively manage expenses and mitigate potential financial strain after deployment}.

\vspace{-2.5mm}
\section{Concluding Remarks and Future Outlook}\label{sec:conclusion}
\vspace{-1.5mm}
In conclusion, the exploration of \gls{O-RAN} illuminates its transformative potential {within the context of cellular communications systems}. As the {wireless industry} evolves to meet the demands of \gls{5G}, \gls{6G}, and beyond, \gls{O-RAN} emerges as a promising paradigm shift, offering flexibility, interoperability, and cost-efficiency in telecommunications networks deployment and management. Through our comprehensive analysis in this paper, it becomes evident that \gls{O-RAN}'s disaggregated approach to wireless network elements, enabled by open interfaces, automation, intelligence, and \gls{SDN} principles, fosters innovation and competition among vendors while reducing vendor lock-in. This approach not only spurs the development of diverse and specialized \glspl{NF} but also empowers operators to tailor their networks to specific use cases and environments with greater agility and granularity through the deployment of network slicing at both network and management domains. In essence, while the journey towards realizing the full potential of network slicing in \gls{O-RAN} may be fraught with {several} challenges, the destination promises a network {architecture} that is more open, agile, and responsive to the evolving needs of wireless communication in the next decade.

To explore the topic of network slicing within \gls{O-RAN} in a detailed manner, we presented its several aspects in this paper, including the architectural framework, network slice deployment options, \gls{MO} procedures, and underlying infrastructure, among many others. We began by exploring the ongoing standardization activities within various \glspl{SDO} and the efforts of the \gls{OSC} with respect to the realization of \gls{O-RAN}. Then, we discussed the \gls{O-RAN} architecture with a {particular} emphasis on network slicing, covering its \gls{SMO} {framework}, \gls{O-gNB} functionalities, and underlying infrastructure. Next, we studied a number of deployment options for \glspl{O-gNB} and {various types of} network slice {instances}, as well as several deployment options for the MFs and management systems within the \gls{SMO} {framework}. We then surveyed network slicing associated with the underlying infrastructure within \gls{O-RAN}, covering slicing in the cellular network sites, O-Cloud sites, and transport networks. Finally, we addressed several use cases related to the deployment of \gls{O-RAN} slicing.

Looking ahead, future research endeavors could extend the current work by exploring the potential of xApps and rApps {in} \gls{O-RAN}, delving into their capabilities for enhancing network intelligence, service orchestration, and resource optimization. The xApps and rApps may employ advanced \gls{ML} algorithms to dynamically allocate resources, predict traffic patterns, and optimize performance for each network slice. Additionally, integrating {\gls{AI}/\gls{ML}} models into various optimization functions within \gls{O-RAN} presents a promising avenue for improving network efficiency, performance, and user experience. By harnessing the power of advanced analytics and automation, future research initiatives can further unlock the transformative potential of \gls{O-RAN}, propelling the evolution of {wireless network} infrastructure into a new era of connectivity and innovation. We hope that the insights, together with the deep dive into the \gls{O-RAN} slicing specifications, architecture, and interfaces, will provide more flexibility for \gls{O-RAN} slicing deployment by using {advanced \gls{AI}/\gls{ML}} models{, as well as various types of} xApps and rApps.

In addition to the above research directions, the exploration of a unified \gls{SMO} architecture that integrates \gls{NFV-MANO} and \gls{ONAP}, alongside the decoupled \gls{SMO} use case defined by {the \gls{O-RAN} Alliance}, presents a promising avenue for improving network management efficiency. Tackling the {research} challenges outlined in Section \ref{Sec:ResearchChallenges} will be crucial for enhancing the capabilities and reliability of \gls{O-RAN} deployments, {which ultimately foster} the development of more resilient and adaptable {cellular network infrastructures}.

\vspace{-2.5mm}
\section*{Acknowledgment}
\vspace{-1.5mm}
This {research} was partially supported by the German Federal Ministry of Education and Research (BMBF) through the project 6G-Terafactory under Grant no. 16KISK186 and partially within the project Open6GHub under Grant no. 16KISK003K \& 16KISK004. {The authors are grateful to the anonymous reviewers for their insightful comments, which have significantly improved the quality of this article.}

\vspace{-2.5mm}
\bibliography{bibliography} 
\bibliographystyle{ieeetr}

\end{document}